\documentclass[a4paper,twocolumn,fleqn]{article}
\usepackage{amsmath}             % recommended for advanced math
\usepackage{txfonts}               % free Times-based fonts for text and math
\usepackage{bm}                  % \bm{..} for bold math (italic)fig:stages_fire_freq-pzeta
\usepackage{cite}                  % Ref. [1--3] instead of [1,2,3]
\usepackage{graphicx}
\usepackage{pfr}                   % PFR style

\usepackage[normalem]{ulem} % sout
\usepackage{units} % nicefrac
\usepackage{float} % H
\usepackage{hyperref}
\hypersetup{
	colorlinks=true,
	linkcolor=blue,
	urlcolor=blue,
	citecolor=blue,
	pdfpagemode=FullScreen,
}

\AtBeginDocument{\mathcode`v=\varv} % makes "v" distinguishable from "nu"

\def\nablab{{\bm \nabla}}

\usepackage{color}
\definecolor{gray}{rgb}{0.5,0.5,0.5}
\definecolor{lgray}{rgb}{0.8,0.8,0.8}
\definecolor{dgray}{rgb}{0.6,0.6,0.6}
\definecolor{dred}{rgb}{0.5,0.0,0.0}
\definecolor{dgreen}{rgb}{0.0,0.5,0.0}
\definecolor{dblue}{rgb}{0.0,0.0,0.5}
\definecolor{violet}{rgb}{0.7,0.0,0.5}

\definecolor{lred}{rgb}{1.0,0.5,0.5}
\definecolor{lgreen}{rgb}{0.5,1.0,0.5}
\definecolor{lblue}{rgb}{0.5,0.5,1.0}

\def\O{{\mathcal O}}

\def\A{{\mathcal A}}

\begin{document}
	
	%\preprint{}
	
	\title{\vspace{-0.3cm}
		Study of Low-Frequency Core-Edge Coupling in a Tokamak: \\ II. Spatial Channeling \& Focusing In Antenna-Driven MHD}%\vspace{-0.1cm}}
% Study of Low-Frequency Core-Edge Coupling in a Tokamak: \\ {I. Antenna}-Driven {MHD} Model with Fixed Boundary

\author{Andreas~BIERWAGE\sup{1,2,\ast}, Wonjun~LEE\sup{3,4}, Young-chul~GHIM\sup{3}, Panith~ADULSIRISWAD\sup{1}, Nobuyuki~AIBA\sup{2}, Seungmin~BONG\sup{3}, Gyungjin~CHOI\sup{3}, Matteo~FALESSI\sup{5,6}, Philipp~W.~LAUBER\sup{7}, Masatoshi~YAGI\sup{1}}
\affiliation{
	\sup{1}National Institutes for Quantum Science and Technology (QST), Rokkasho Institute for Fusion Energy, Aomori 039-3212 \\
	\sup{2}QST, Naka Institute for Fusion Science and Technology, Ibaraki 311-0193, Japan \\
	\sup{3}Department of Nuclear and Quantum Engineering, Korea Advanced Institute of Science and Technology (KAIST), Daejeon 34141, Republic of Korea \\
	\sup{4}Max-Planck-Institut f\"{u}r Plasmaphysik, Wendelsteinstrasse~1, D-17491 Greifswald, Germany \\
	\sup{5}Center for Nonlinear Plasma Science and C.R. ENEA Frascati, Via E. Fermi 45, 00044 Frascati, Italy \\
	\sup{6}Istituto Nazionale di Fisica Nucleare (INFN), Sezione di Roma, Piazzale Aldo Moro 2, 00185 Roma, Italy \\
	\sup{7}Max-Planck-Institut f\"{u}r Plasmaphysik, Boltzmannstrasse~2, D-85748 Garching, Germany
}

\date{}

\email{bierwage.andreas@qst.go.jp}

\begin{abstract}
	Motivated by evidence for core-edge coupling in the form of double-peaked fishbone-like low-frequency modes ($\lesssim 20\,{\rm kHz}$) in KSTAR, which exhibit synchronized Alfv\'{e}nic activity both in the central core and near the plasma edge [Lee {\it et al}, {\it Phys.\ Plasmas} {\bf 30} (2023) 022502], we study the nonlocal response of a tokamak plasma in a visco-resistive full MHD simulation model using the code {\tt MEGA}. The waves are driven by an internal ``antenna'' that is localized both radially and azimuthally in the poloidal $(R,z)$ plane and has a sinusoidal form $\exp(in\zeta - i\omega t)$ with Fourier mode number $n=\pm 1$ in the toroidal angle $\zeta$ and fixed angular frequency $\omega$ in time $t$. By flattening the safety factor profile $q(r)$ at suitable locations in the minor radius $r$, we created plateaus in the low-frequency Alfv\'{e}n continua that act as wave ``receivers''. First, we confirm that such continuum plateaus respond with a coherent quasi-mode even when the driving antenna is located at a distant radius. Second, by varying the antenna location, we confirm the expectation of inward drive being more efficient than outward drive, which we attribute to volumetric focusing. Third, we find that the central core also responds well at frequencies below the central Alfv\'{e}nic continuum plateau, which could facilitate chirping. Our results show that a core-localized low-frequency response does not necessarily require core-localized drive nor an exactly matching continuum, but may be driven from the edge and sub-resonantly. It remains to be seen to what extent the examined effects play a role in double-peaked fishbone-like activity. Other possible contributing mechanisms are discussed to motivate further study. Our analyses also elucidate the mode structure formation process, from transients to quasi- or eigenmodes, here in the realm of MHD, and to be followed by a verification study against kinetic models.{\looseness=-1}%\vspace{0.5cm}
\end{abstract}

%\keywords{tokamak, fast particles, Hamiltonian guiding center simulation, phase space dynamics, chirping}

%\DOI{xx.xxxx/pfr.x.xxx}

\maketitle

\thispagestyle{empty} % no page number 1
\pagestyle{plain} % page numbers 2,3,...
\everypar{\looseness=-1} % squeeze space

\tableofcontents

% =============================================================================
\section{Introduction and motivation}
\label{sec:intro}

\begin{figure}
	[tb]\vspace{-1.9cm}
	\centering
	\includegraphics[width=0.48\textwidth]{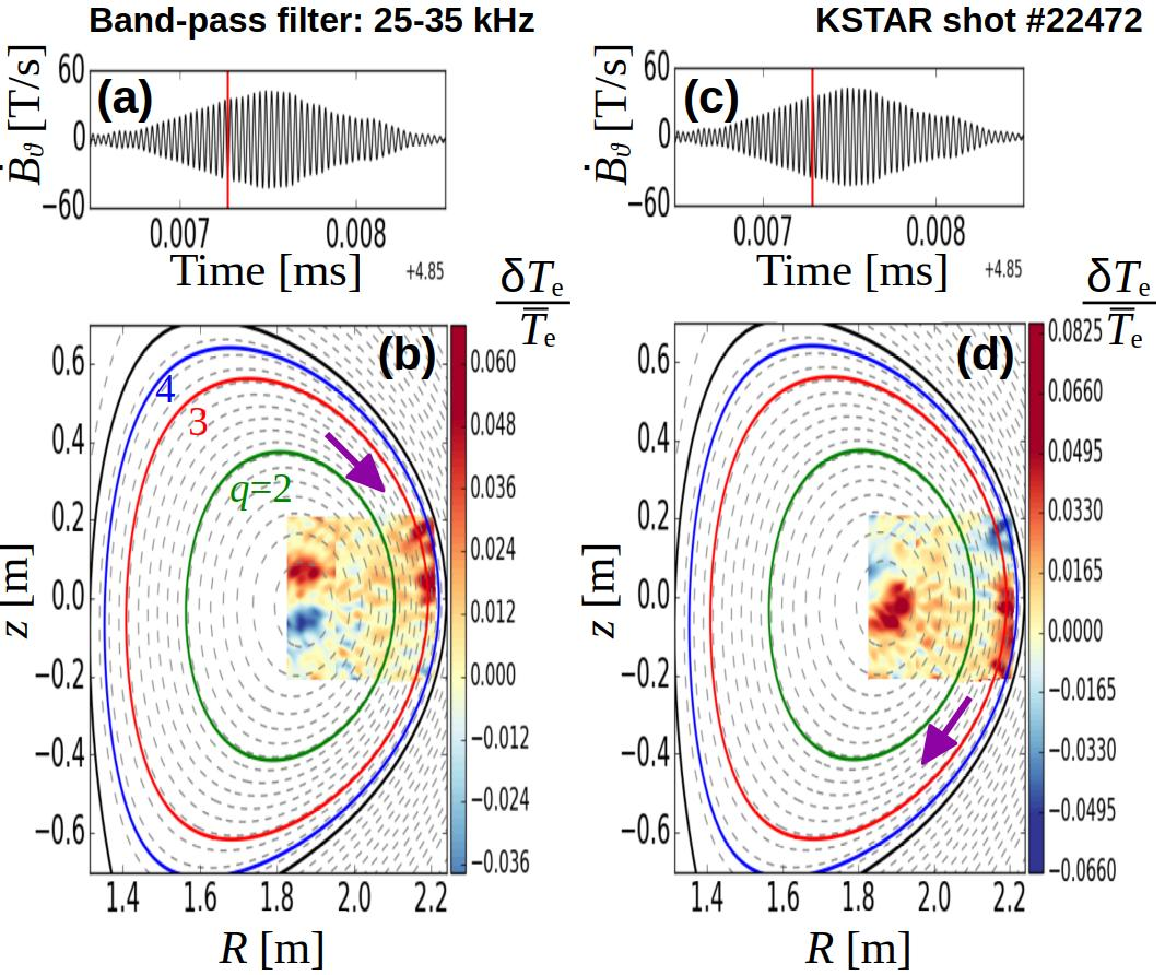}\vspace{-0.3cm}
	\caption{Example of a moderate-strength double-peaked fishbone-like mode in KSTAR similar to those analyzed in Ref.~\protect\cite{Lee23} (adapted from Fig.~5 of Ref.~\protect\cite{Lee26}). Panels (a) and (c) show time traces of the magnetic fluctuation signal $\dot{B}_\vartheta \equiv {\rm d}B_\vartheta/{\rm d}t$ measured by Mirnov coils. The vertical red lines indicate the time at which electron cyclotron emission imaging (ECEI) snapshots (b) and (d) were taken. For orientation, the ECEI data are overlaid with contours of {\tt EFIT}-reconstructed \protect\cite{Lao85} magnetic flux surfaces (gray dashed). Rational magnetic surfaces with safety factor values $q=2,3,4$ are indicated by solid colored lines. The last closed flux surface is drawn black. The inner and outer portions of ECEI were taken by different detectors that, at the time of these experiments (post-2018), had a toroidal angular distance of $\Delta\zeta = 18.5^o$. To be precise, panels (b) and (d) show $\delta Y/\overline{Y} \equiv (Y-\overline{Y})/\overline{Y}$, where $Y(R,z,t)$ is the measured signal subject to $25$-$35\,{\rm kHz}$ band-pass filtering, and $\overline{Y}(R,z)$ is its time-average. In an optically thick plasma, one can assume that $Y \approx T_{\rm e}\times C(R,z)$, with a scaling factor $C(R,z)$ that differs between detector channels. Since $\delta T_{\rm e}/\overline{T}_{\rm e}$ is a relative amplitude with a radially nonuniform reference profile satisfying $\overline{T}_{\rm e}^{\rm core} \gg \overline{T}_{\rm e}^{\rm edge}$, it is clear that, in absolute terms, we have $\delta T_{\rm e}^{\rm edge} \ll \delta T_{\rm e}^{\rm core}$. It would have been interesting to estimate also the radial displacement $\delta\xi_r \approx \delta T_{\rm e}/T_{\rm e}'$ (or the displacement in flux space), but unfortunately the electron temperature profile measurement (Thomson scattering) was not accurate enough to yield reliable values for its radial gradient $T_{\rm e}' \equiv {\rm d}T_{\rm e}/{\rm d}r$, especially near the edge.}\vspace{-0.55cm}
	\label{fig:01_kstar_ecei}%
\end{figure}

Measurements made in neutral-beam-driven KSTAR tokamak experiments exploring high-confinement mode (H-mode) physics revealed synchronized chirping bursts of Alfv\'{e}nic activity both in the plasma's central core and near its surface (edge) \cite{Lee23}. The authors of that study originally set out to look for so-called edge harmonic oscillations (EHO) in H-mode plasmas and utilized divertor Langmuir probes (DLP) to measure particle losses. After the DLP signals were found to exhibit repeated frequency chirping, the same patterns were also confirmed in magnetic signals from external Mirnov coils and in internal measurements using electron cyclotron emission (ECE) radiometry and beam emission spectroscopy (BES). Spectrograms and analyses of coherence between different diagnostics were reported in Ref.~\cite{Lee23}. Two snapshots of the mode's spatial structure in the poloidal $(R,z)$ plane as seen by KSTAR's ECE imaging (ECEI) diagnostic systems \cite{Yun14} are shown in Fig.~\ref{fig:01_kstar_ecei} during the course of a millisecond-long Alfv\'{e}nic burst with moderate amplitude. In the toroidal direction, the dominant Fourier mode number is thought to be $n=1$.

These modes have low frequencies of a few dozen kHz, and their chirping and bursty behavior is reminiscent of fast-ion-driven modes known as ``fishbone-like oscillations'', or simply ``fishbones''. Historically, fishbones were first seen in core-localized (on-axis) form, in the region where the tokamak safety factor $q$ is close to unity ($q\sim 1$) \cite{McGuire83}. Later, so-called ``off-axis fishbones'' or ``energetic-particle driven wall-stabilized modes (EWM)'' were also observed, which are thought to peak in the region where $q\sim 2$, while causing significant fluctuations across the entire plasma, from the core all the way to the boundary \cite{Matsunaga09, Okabayashi09, Matsunaga10, Okabayashi11, Heidbrink11}. In some cases, interactions with edge-localized modes (ELM) were reported. The radial structure of off-axis fishbones inferred from numerical models (Fig.~8 of Ref.~\cite{Matsunaga10}) and ECE measurements (Fig.~12 of Ref.~\cite{Okabayashi11}) differ significantly from the double-peaked mode in Fig.~\ref{fig:01_kstar_ecei}, which has a distinct minimum in the region around the $q \sim 2$ surface that is indicated by a green contour in Fig.~\ref{fig:01_kstar_ecei}(b,d). Moreover, the spectrograms in Figs.~3 and 4 of Ref.~\cite{Lee23} showed essentially identical downward chirps in the mode frequency in the core and near the edge, whereas no chirping signal could be detected at normalized minor radius $r/a \sim 0.7$ in the intermediate region, where $1 < q < 3$.

We interpret these observations at KSTAR \cite{Lee23, Lee26} as evidence for the existence of yet another type of fishbone, which we refer to as ``double-peaked fishbone mode''.

A few points should be noted concerning the observed mode structure. First, our present knowledge about the outer peak's structure in KSTAR is limited to the region near the outer midplane that is monitored by ECE, BES (cf.~Fig.12 of Ref.~\cite{Lee23}) and ECEI (Fig.~\ref{fig:01_kstar_ecei}). We shall assume that it covers the entire plasma circumference and consists of dominant poloidal harmonics around $m \sim 3,4$. Second, it must be noted that Fig.~\ref{fig:01_kstar_ecei}(b,d) shows {\it relative} amplitudes $\delta T_{\rm e}/\overline{T}_{\rm e} \equiv (T_{\rm e}-\overline{T}_{\rm e})/\overline{T}_{\rm e}$, where the time-averaged reference profile $\overline{T}_{\rm e}(r)$ is nonuniform in $r$, with $\overline{T}_{\rm e}^{\rm core} \gg \overline{T}_{\rm e}^{\rm edge}$. In absolute terms, the outer peak is always much smaller. On the other hand, if one assumes that the outer component occupies the entire magnetic surfaces in that region, its total energy content could be comparable to and potentially even larger than that of the fishbone's inner component.

Comparable magnitudes in the central core and near the edge, separated by a deep minimum at intermediate radii, are also obtained if one examines the correlations between the amplitude envelopes of Mirnov and ECE signals, $\dot{B}_\vartheta$ and $\delta T_{\rm e}$. Such an analysis is reported in Ref.~\cite{Lee26} (Part I of this paper series) using a large database of fishbone burst events under various conditions. It was then found that the radial width of the annular minium in the amplitude correlation profile varies with the intensity or ``strength'' of the fishbone burst that, in turn, depends on the H-mode performance controlled by external magnetic perturbations.

Finally, it is worth noting that KSTAR's double-peaked fishbones seem to be able to withstand significant differential toroidal rotation: Due to tangential beam injection, the core of these KSTAR plasmas rotated faster than the plasma edge, and the toroidal rotation frequency difference $\Delta\nu_{\rm rot} = \nu^{\rm core}_{\rm rot} - \nu^{\rm edge}_{\rm rot} \sim 10\,{\rm kHz}$ (Fig.~12 of Ref.~\cite{Lee23}) is comparable to the mode frequency $\nu \sim (10...30)\,{\rm kHz}$ in the lab frame. This rotation shear may contribute to the decorrelation of fluctuations at intermediate radii that we discussed in the previous paragraph. This raises the question how the mode's inner and outer components are still coupled into one coherent entity that is able to perform synchronized core-localized and edge-localized chirps with identical frequencies in the lab frame.

Taking into account various pieces of existing evidence, both experimental and theoretical, four plausible explanations are currently being considered and explored:
\begin{enumerate}
\item[1.] Outer peak appears in BES and ECE through the displacement of steep gradients in the H-mode edge.

\item[2.] Double-peaked global eigenmode.

\item[3.] Pair of modes (quasi- or eigenmodes, possibly different parities, polarizations) coupled by MHD waves,

\item[4.] Pair of modes coupled by beam ions on wide orbits.
\end{enumerate}

{\bf Option 1} would perhaps be the simplest explanation: Even if the fishbone's plasma displacement profile $\delta{\bm \xi}(r)$ had only a single peak (like a classical $q=1$ fishbone), if that peak is sufficiently broad to also perturb the edge region, BES and ECE-based diagnostics may show a second peak in H-mode plasmas because the steep gradients in electron density $n_{\rm e}$ and temperature $T_{\rm e}$ will amplify the observed fluctuations $\delta n_{\rm e} \sim \delta{\bm \xi}\cdot\nabla n_{\rm e}$ and $\delta T_{\rm e} \sim \delta{\bm \xi}\cdot\nabla T_{\rm e}$ in the edge transport barrier (ETB) region, as illustrated in Fig.~12 of the companion paper \cite{Lee26}. The implied influence of edge gradients on the modes' apparent properties is supported by the experimentally observed dependence of the double-peaked fishbones' strength on external magnetic perturbations that control the H-mode performance: The fishbones are stronger and the outer peak is more pronounced in cases where the magnetic perturbations enhance the H-mode, and they are weakened when the magnetic perturbations cause density pump-out \protect\cite{Lee26}.

The mechanism underlying Option 1 is likely to contribute to the observation of double-peaked structures in BES and ECE profiles as well as ECEI in H-mode plasmas. It can thus be a sufficient condition, though perhaps not a necessary one. Indeed, the necessity of Option 1 can probably be ruled out due to several reasons. First, double-peaked fishbones were found also in KSTAR data from low-confinement mode (L-mode) plasmas \cite{Lee26}. Second, the fishbone mode's outer component often seems to be broader than the expected width of an ETB. Third, the fact that the mode's bursts and chirps are also observable via DLPs indicates that the mode itself does have a significant amplitude near the edge. Fourth, the non-observation of chirping at intermediate radii is difficult to reconcile with the assumption of a mode consisting of only a single broad peak. All this suggests that Option 1 may not be a complete explanation for observations like those in Fig.~\ref{fig:01_kstar_ecei}, so it is worth considering other contributing factors.

{\bf Option 2} assumes that differentially rotating H-mode plasmas may support multi-peaked eigenmodes. The notion of an eigenmode implies that its spatial structure is robust as long as nonlinear effects are not too strong. That is, its structure should show only little or no sensitivity with respect to the radial location of the resonant drive, and all its components are expected to chirp in unison, regardless of whether the driving resonance is located in the central core or near the edge.

One may thus expect that a global eigenmode should not only have a uniform coherence profile (as in Fig.~7 of Ref.~\cite{Lee23}) but also a uniform amplitude correlation profile. The latter expectation, however, seems to contradict results reported in Ref.~\cite{Lee26}, which show that the amplitude correlation profiles have a distinct off-axis minimum.

There are also theoretical aspects that challenge Option 2 of a conventional eigenmode. First, geometric couplings due to toroidicity ($m \rightarrow m\pm 1$ \cite{Cheng85}) and ellipticity ($m\rightarrow m\pm 2$ \cite{Betty91}) are optimized at higher frequencies (here $90...360\,{\rm kHz}$), where corresponding continuum gaps and discrete global eigenmodes are known to form \cite{Cheng85, Betty91}. The underlying quantization conditions are not satisfied at the low plasma-frame frequencies ($\lesssim 10\,{\rm kHz}$) that are relevant for fishbones, so while the geometric couplings remain active in that they influence the form of low-frequency modes, they do not induce discrete eigenmodes.

Second, the plasma conditions in the central core and edge differ substantially, so it seems likely that double-peaked fishbones comprise components of different parity or differently weighted polarizations.

For instance, near the plasma edge, the thermal pressure is relatively low, so that the beta-induced Alfv\'{e}n eigenmodes (BAE) \cite{Heidbrink93} and beta-induced Alfv\'{e}n-acoustic eigenmodes (BAAE) \protect\cite{Gorelenkov07a}, which respectively reside near the top and bottom of the so-called kinetic thermal ion (KTI) gap \cite{Zonca96, Curran12, Chavdarovski14}, can have sufficiently low frequencies to fall into the range of our double-peaked fishbones.\footnote{Readers familiar with continuous spectra in tokamaks can readily verify these claims by inspecting Figs.~\ref{fig:04_kstar_continua} and \ref{fig:b10_kstar-1.051_scan-hi-f_n1} below.}
In contrast, the higher plasma pressure in the central core implies that the core-localized component of a double-peaked fishbone falls into the domains of diamagnetic frequencies and low-frequency shear Alfv\'{e}n continua, so it may acquire features of drift- and drift-Alfv\'{e}n waves \cite{Coppi86,Chen91,Du24}, kink-like perturbations \cite{Chen84,Chen94} or both \cite{Zhang89}. Moreover, when the magnetic shear is weak, so that low-frequency Alfv\'{e}n continua are relatively flat --- as is often the case in the central core of a tokamak plasma and possibly in the KSTAR plasmas at hand --- one may also encounter so-called Global Alfv\'{e}n Eigenmodes (GAE) \cite{Appert82}.

A concrete example of a compound mode with multiple distant peaks that have different physical properties was seen in numerical studies using the {\tt MARS-K} code including kinetic thermal ion effects \cite{Liu22}, following experimental observations of ``multiscale chirping modes'' in high-temperature DIII-D tokamak plasmas \cite{Du21}. These modes contained tearing-parity components that, in rare instances, were also seen in KSTAR's double-peaked fishbones \cite{Lee26}.

We anticipate that the structure of such compound modes can be influenced by the spatial distribution of sources and sinks to such a degree that their interpretation as eigenmodes becomes questionable. In fact, some part of a compound mode may not be identifiable as any known discrete eigenmode at all, so it is worth looking beyond Option 2 and consider the possibility of a quasi-modes.

{\bf Options 3 \& 4} refer to the possibility of a compound mode whose inner and outer components are linearly independent but synchronized by (Option 3) the spontaneous emission and absorption of MHD waves, or (Option 4) fast ions on barely mirror-trapped orbits that pass through both the core and the edge due to magnetic drifts.

The drift-orbit-based core-edge coupling mechanism underlying {\bf Option 4} was proposed in Fig.~12 of Ref.~\cite{Lee23}. This possibility will be examined in more detail in a separate forthcoming paper of this series, using recently developed methods to analyze resonances of arbitrarily wide orbits in the presence of an ambient radial electric field \cite{Bierwage25}.\footnote{Let us already note that Option 4 suffers from several caveats: (i) Based on KSTAR's beam geometry and the small size of the relevant portion of fast ion phase space, we expect the number of particles populating such wide orbits to be relatively small. (ii) The coupling strength between the kinetic compression of resonant particles and the electric field of the mode should depend on the mode amplitude. Thus, the drift-orbit-based coupling between the inner and outer mode components should be weaker when the fluctuation amplitude is low; namely, at the beginning and end of a fishbone burst, so one may expect threshold-like behavior or at least time-varying phase lags between the two peaks. This, however, was not seen in the detailed phase analyses reported in Ref.~\protect\cite{Lee26}.\label{fn:option4}}

The purpose of the present paper is to explore the physical possibility of {\bf Option 3}. To illustrate the underlying idea, envision an edge-localized primary mode (e.g., a stable external kink) that is resonantly excited by wave-particle interactions. That primary mode could effectively act as an antenna that emits MHD waves into the plasma core, where they are absorbed by another mode. Such a core-localized parasitic secondary mode could, for instance, arise on a flat piece of the shear Alfv\'{e}n continuum that commonly exists in steady-state tokamak plasma scenarios whose central safety factor is flat and remains close to unity ($q \sim 1$). This is precisely the situation that we will study here using numerical simulations of the MHD plasma response in the presence of an artificial antenna.

For simplicity, differential plasma rotation will be ignored in the present study, because it may do more than merely adding a Doppler shift between the antenna and the receiver in the plasma frame, so it should be examined with proper scrutiny in a dedicated study. In particular, we expect that differential rotation will alter, compete or synergize with the mechanisms that couple mode components at different radii \cite{Kolesnichenko20}, causing a shearing of the mode's radial phase profile, possibly to the point of cyclic phase slippage. Differential rotation may thus turn eigenmodes into quasi-modes whose spatial structure varies periodically, with bonds being transiently broken and restored in a cyclic fashion. Our above distinction between Option 2 (eigenmode) and Options 3 \& 4 (compound mode) may thus become vague. Theoreticians have found ways to describe such behavior in terms of a dense spectrum of Floquet-type (generalized) eigenmodes with complex eigenvalues and time-dependent wave vectors \cite{Cooper88, Waelbroeck91}, where an unstable modelet is coupled to an infinite number of stable ones \cite{Furukawa05}. In the present case, our compound modes are subject to the observation-based constraint that, throughout the fishbone burst, coherent phase-alignment should be preserved near the outer midplane, which was monitored by diagnostics as in Fig.~\ref{fig:01_kstar_ecei}. In other regions, cyclic phase slippage and realignment is allowed.

Compared to Options 1 \& 2, the realization of Options 3 \& 4 requires a more complex interplay between multiple factors, which implies sensitivity to a larger number of parameters, such as the shapes of the safety factor and plasma rotation velocity profiles, $q(r)$ and ${\bm V}(r)$. By virtue of Occam's razor, these options may thus appear less likely. However, they are physically interesting and --- if not for double-peaked fishbones --- may find applications elsewhere. The present study is also useful for gaining a deeper understanding of how the low-frequency plasma response is captured by the visco-resistive full MHD simulation model in our adapted version of the code {\tt MEGA} \cite{Todo98, Todo05, Todo25}. A comparison with results from kinetic models is in preparation.

One intriguing implication of Options 2 (eigenmode) and 3 (compound mode) is the possibility of resonances driving only one component of the mode. For instance, it is possible that the outer peak is the primary recipient of direct drive while the inner peak is secondary (parasitic).\footnote{Of course, one would have to explain how the population of fast ions in the plasma's edge region grows to the level needed to destabilize such a mode. However, this may still be easier than addressing the caveats of Option 4 outlined in Footnote~\protect\ref{fn:option4} above. Continuous outward transport of fast ions may suffice, and the outer plasma also has a larger population of trapped particles that may drive the mode via precession resonances.}
Such ``inward coupling'' appears energetically favorable based on the principle of volumetric focusing: the energy density should increase when fluctuations propagate from the edge to the core and occupy a progressively smaller volume. This expectation will be numerically confirmed in this work. An edge resonance is also more likely to find a matching low-frequency core response because the central $q$ profile tends to stay near unity in steady-state tokamak plasmas. As mentioned above, such configurations tend to have weak central magnetic shear and a fairly flat low-frequency Alfv\'{e}n continuum, which we will exploit here. Such a continuum plateau facilitates resonant absorption of Alfv\'{e}n waves without much radial phase mixing, so their conversion to kinetic Alfv\'{e}n waves via ion Larmor radius and electron inertia effects \cite{Chen74,Hasegawa75} is not considered here. However, we do inspect non-ideal effects associated with visco-resistive dissipation.

It must be noted that our nonchalantly constructed physical pictures involving ``{\it volumetric focusing}'', ``{\it emitted and absorbed MHD waves}'' and ``{\it coherent global quasi-modes forming on continuum plateaus}'' are papering over complex processes and a still incomplete understanding. Thus, in addition to taking steps towards an explanation of KSTAR's double-peaked fishbones, the numerical experiments reported in this paper are also meant to provide deeper insight into mode formation processes at a fundamental level. The present paper should be seen as a snapshot of ongoing research and evolving understanding.

This paper is organized as follows. In Section~\ref{sec:history}, we outline our current theoretical understanding of mode structure formation physics in a tokamak. Section~\ref{sec:setup} describes the simulation model and methods. Simulation results are presented in Section~\ref{sec:results} and discussed in Section~\ref{sec:discuss}. We conclude with a summary and outlook in Section~\ref{sec:summary}. Extensive appendices contain supplementary material.\vspace{-0.25cm}

% =============================================================================
\section{Theoretical underpinnings}
\label{sec:history}\vspace{-0.15cm}

This section revisits several theoretical concepts that are relevant for the present work. Some parts await a more rigorous treatment but are helpful for constructing a tentative physical picture. Readers interested primarily in the simulation results may skip this and proceed to Section~\ref{sec:setup}.\vspace{-0.15cm}

\subsection{Excitation of eigen- \& quasi-modes}
\label{sec:history_mode}

Our simulations will show in some detail how the coherent structure of a mode gradually emerges after the antenna has been active for some time. In the long run, the plasma's nonuniformities (magnetic geometry, density and pressure profiles, and numerical boundary conditions) determine the spectrum of long-lived eigen- and quasi-modes.

Elaborating somewhat on the rough definition given in Section~\ref{sec:intro}, we speak of a (conventional) {\it eigenmode} when a solution has a well-defined oscillation frequency and robust spatial structure, in the sense that its form is insensitive with respect to the radial location of resonant drive. Eigenmodes can be viewed as standing waves, a physical picture that becomes mathematically explicit in certain limiting cases, where the governing equations can be compactly written in Schr\"{o}dinger form, $(\tfrac{{\rm d}^2}{{\rm d}\theta^2} + V)\delta\psi = 0$, as in Eq.~(22) of Ref.~\cite{Chen95}. The form of the effective potential $V(\theta,\omega)$ is determined by the said spatial nonuniformities ($\theta$) and the mode frequency ($\omega$). Eigenmodes are then solutions satisfying characteristic quantization conditions given by $V$. Any damping mechanisms (including tunneling losses through potential barriers) must be much slower than the radial mode structure formation time.\footnote{In the spatial channeling theory discussed in Section~\protect\ref{sec:history_coupling}, the mode structure formation time is given by Alfv\'{e}n-magnetoacoustic coupling.}

Conversely, we speak of a {\it quasi-mode} whenever we observe a long-lived mode that does not seem to classify as a discrete eigenmode as defined above.\footnote{Our notion of a quasi-mode may encompass generalized eigenmode concepts like those mentioned in Section~\protect\ref{sec:intro}.}
For instance, the spatial structure of a quasi-mode may change noticeably when the spatial location of the drive changes. Quasi-modes with similar spatial structures may be also seen over a relatively broad range of driving frequencies.

The quasi-modes studied in the present work are found in the radial domain of an Alfv\'{e}nic continuum plateau. We will show that our quasi-modes can be excited near and below the plateau's resonant frequency. Moreover, we will drive them from a distance (the antenna will not overlap with the quasi-mode) and demonstrate that inward drive (towards the geometric $1/r$ singularity) is more effective. It is thus instructive to revisit in the following sections the concepts of volumetric focusing and spatial channeling, and the intricacies of weak magnetic shear.

\begin{figure}
	[tb]\vspace{-0.3cm}
	\centering
	\includegraphics[width=0.48\textwidth]{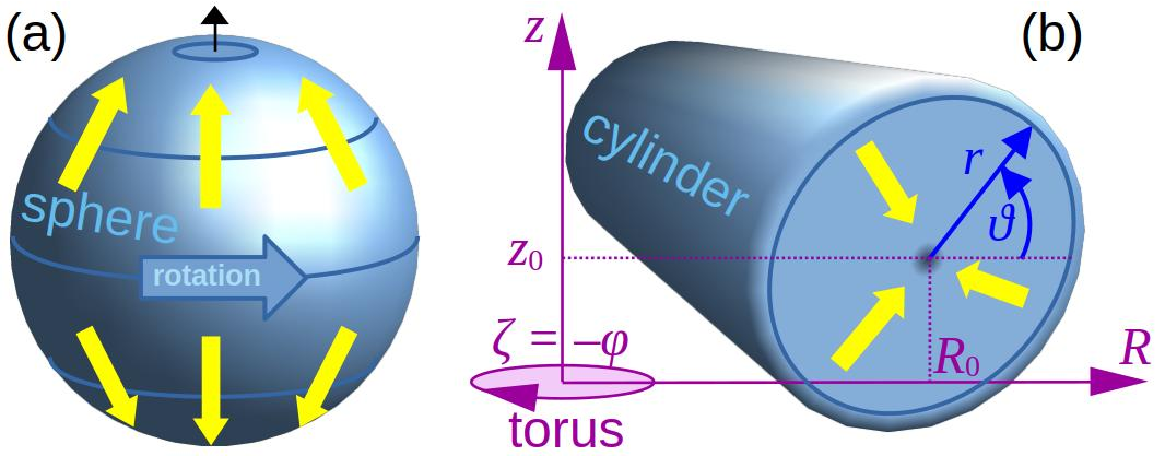}%\vspace{-0.2cm}
	\caption{Volumetric focusing of material or energy fluxes along the yellow arrows towards the poles of a rotating sphere (a) or the axis of a cylinder (b). The cylindrical case also represents the situation in the poloidal cross-section of a torus, where the singular axis is located at $(R,z) = (R_0,z_0)$ of the coordinate system $(R,z,\zeta)$ drawn magenta. Within a poloidal cross-section of the cylinder or torus, we also use the poloidal coordinates $(r,\vartheta)$ drawn blue, where $0 \leq r \leq a$ ranges from the axis $0$ to the boundary radius $a$.}
	\label{fig:02_focus_schematic}\vspace{-0.3cm}
\end{figure}

\subsection{Volumetric focusing \& spatial channeling}
\label{sec:history_focuschannel}

Consider a gas or fluid on a rotating sphere as in Fig.~\ref{fig:02_focus_schematic}(a), which may resemble a planetary atmosphere. If energy is transported from the equator to the poles as indicated by the yellow arrows, the energy density $\Delta W/\Delta V$ increases. The same happens in a cylindrical configuration as sketched in Fig.~\ref{fig:02_focus_schematic}(b) when energy or material is transported radially towards the central axis. Here, we refer to this effect as ``volumetric focusing'' and examine it in the context of a plasma that is confined by a magnetic field in a toroidal volume. The situation in such a torus is equivalent to that of the (periodic) cylinder in Fig.~\ref{fig:02_focus_schematic}(b) when the central axis is taken to be the magnetic axis, located at major radius $R = R_0$ and height $z = z_0$. Distance along the torus is measured by $R\zeta$ with toroidal angle $\zeta = -\varphi$.

Volumetric focusing of waves in such a cylindrical or toroidal magnetized plasma requires mechanisms that facilitate the transfer of energy across magnetic surfaces along the minor radial coordinate $r$. Here, we shall refer to such mechanisms collectively as ``spatial channeling'', a term we adopted from Kolesnichenko {\it et al}.~\cite{Kolesnichenko10, Kolesnichenko20} who, among many other things, used it to describe the transfer of energy from stable to unstable radial domains via global eigenmodes. In the case of shear Alfv\'{e}n eigenmodes, these are collections of shear Alfv\'{e}n waves that are radially coupled to form coherent global modes. When driven by a local resonance at a certain radius, the oscillations will be distributed (with some temporal delay) across the radial domain covered by the eigenmode. This idea underlies {\bf Option 2} in the introductory Section~\ref{sec:intro} above, and we conjecture that similar processes are at work in {\bf Option 3}, where two mode components with different physical properties are coupled into a coherently oscillating entity.

\subsection{Alfv\'{e}n-magnetoacoustic coupling}
\label{sec:history_coupling}

Kolesnichenko {\it et al}.\ also addressed the question how shear Alfv\'{e}n waves are coupled radially to facilitate the said spatial channeling. We know that the propagation of shear Alv\'{e}n waves as such is primarily guided by the magnetic field, even in strongly nonuniform plasmas, as has been demonstrated experimentally \cite{Borg96}. Nevertheless, information about a shear Alfv\'{e}n wave's presence travels also radially, which is why nonlocal Alfv\'{e}n eigenmodes exist, and antennas and external Mirnov probes are useful. For this, Kolesnichenko {\it et al}.\ invoke the coupling between shear Alfv\'{e}n and fast (magnetoacoustic)\footnote{Here, the attribute ``magnetoacoustic'' should be understood as emphasizing the aspect of magnetic field line compression in the cold-plasma limit. These magnetoacoustic effects do not require coupling to sound waves, which are, in fact, absent in Eq.~(\protect\ref{eq:imhd_disp}). This is often a reasonable approximation in tokamaks due to their low plasma beta (few \%).}
waves in a nonuniform plasma, which can be expressed in the zero-beta ideal MHD limit as (see Appendix B of Ref.~\cite{Kolesnichenko18b})
\begin{subequations}
\begin{align}
	(\omega^2/v_{\rm A}^2 - k^2)D^+ =& +iQD^-,
	\label{eq:imhd_disp_saw} \\
	(\omega^2/v_{\rm A}^2 - k_\parallel^2)D^- =& -iQD^+,
	\label{eq:imhd_disp_fw}
\end{align}	
\label{eq:imhd_disp}
\end{subequations}\vspace{-0.5cm}

\noindent with $Q \equiv \mu_0 J_\parallel k_\parallel/B$ and $D^\pm \equiv k_{\rm r}\xi_{\rm r} \pm k_{\rm b}\xi_{\bm b}$, where $B \equiv |{\bm B}|$ is the background magnetic field strength, $J_\parallel \equiv {\bm J}\cdot{\bm B}/B$ is the parallel current density, ${\bm k}$ and ${\bm \xi}$ are the wave and displacement vectors, and the subscripts ``r'' and ``b'' identify their radial and binormal components. It is evident that the shear Alfv\'{e}n (\ref{eq:imhd_disp_saw}) and fast wave (\ref{eq:imhd_disp_fw}) branches are coupled provided that $k_\parallel \equiv (m/q -n)/R \neq 0$ and at least one of the other two wave vector components, $k_{\rm r}$ or $k_{\rm b}$, is nonzero.\footnote{Although dispersion relations like Eq.~(\protect\ref{eq:imhd_disp}) are formulated in terms of well-defined harmonic frequencies $\omega$ and wave numbers ${\bm k}$, it appears reasonable to assume that the essential physics, such as the Alfv\'{e}n-acoustic coupling, remain active for more general forms of fluctuations, like those constituting fast transients emerging from an abrupt perturbation. This is confirmed in Appendix~\protect\ref{apdx:modeform_transients}.}
Using the ideal MHD equations of motion to write the magnetic fluctuations as
\begin{equation}
	\hspace{-0.5cm} \delta{\bm B} = \nablab\times\delta{\bm A} = -{\bm B}\nablab\cdot{\bm \xi}_\perp + ({\bm B}\cdot\nablab){\bm \xi}_\perp -({\bm \xi}_\perp\cdot\nablab){\bm B},
\label{eq:imhd_db}
\end{equation}

\noindent Ref.~\cite{Kolesnichenko18b} goes on to show that the magnetic compression associated with shear Alfv\'{e}n waves can be estimated as
\begin{equation}
\hspace{-0.5cm}\omega |\delta B_\parallel| \approx |\Omega_\parallel| |\delta\Phi|, \quad \text{with} \quad \Omega_\parallel^2 \equiv \left(k_\parallel^2 - \frac{\omega^2}{v_{\rm A}^2}\right) k_\perp^2
\label{eq:imhd_dbpar}
\end{equation}

\noindent where $\mu_0$ is the vacuum permeability, and where $\delta\Phi$ is the perturbed scalar potential. Note that, the last term of Eq.~(\ref{eq:imhd_db}) was ignored assuming weak magnetic nonuniformity. Moreover, since it has been derived in the MHD limit, Eq.~(\ref{eq:imhd_dbpar}) is valid only for wavelengths much larger than the ion Larmor radius: $k_\perp \varrho_{\rm Li} \ll 1$. The short-wavelength nonzero-beta limit was treated by Tang {\it et al}.\ \cite{Tang80} using the (kinetic) ballooning formalism \cite{Connor78, Dewar83} and Scott \cite{Scott24_arxiv} using modern gyrokinetic theory. Both arrive at the estimate (see Eq.~(3.23) of Ref.~\cite{Tang80}),
\begin{equation}
\hspace{-0.5cm}\omega\delta B_\parallel \approx -\Omega_{\rm p}\delta\Phi \quad \text{with} \quad \Omega_{\rm p} \equiv -\mu_0{\bm k}_\perp\cdot\frac{{\bm B}\times\nablab P}{B^3},
\label{eq:gk_dbpar}
\end{equation}

\noindent where we have taken the ideal MHD limit $k_\parallel\delta\Phi \approx \omega \delta A_\parallel$. The factor $\Omega_{\rm p}$ measures the difference between curvature and $\nablab B$ drift, with the gradient of the scalar pressure $P$ arising from the constraint of perpendicular MHD force balance, $\nablab_\perp P = {\bm J}\times{\bm B}$. The wavelength ordering underlying Eq.~(\ref{eq:gk_dbpar}) is $k_\perp/k_\parallel \sim \Omega_{\rm Li}/\omega$, where $\Omega_{\rm Li}$ is the ion Larmor frequency (see Eq.~(66) of Ref.~\cite{Scott24_arxiv}). For long-wavelength modes in finite-beta plasmas, we expect some combination of Eqs.~(\ref{eq:imhd_dbpar}) and (\ref{eq:gk_dbpar}). Graves {\it et al}.\ \cite{Graves19} suggested concrete corrections for reduced MHD and gyrokinetic models.

Kolesnichenko {\it et al}.\ \cite{Kolesnichenko18b, Kolesnichenko20} also provided recipes for estimating the radial energy flux. They state that, in the presence of sources and sinks, the net energy flux $\Gamma^{\rm (m)}$ across a radial slice $\delta r$ of a mode is
\begin{equation}
	\hspace{-0.3cm} \Gamma^{\rm (m)} = v_{\rm g}^{\rm (m)} W^{\rm (m)} = \left|v_{\rm g}^{\rm (t)}\right| (W_+^{\rm (t)} - W_-^{\rm (t)}) = \Gamma_+^{\rm (t)} + \Gamma_-^{\rm (t)}.
	\label{eq:fluxm}
\end{equation}

\noindent Eq.~(\ref{eq:fluxm}) determines the mode slice's group velocity $v_{\rm g}^{\rm (m)}$ in terms of the total energy density $W^{\rm (m)} = W_+^{\rm (t)} + W_-^{\rm (t)}$ in the slice, the energy densities $W_\pm^{\rm (t)}$ of waves traveling radially outward (+) or inward (--), and the traveling waves' local group velocity $v_{\rm g}^{\rm (t)}$. The latter is given by
\begin{equation}
	\hspace{-1.0cm} v_{\rm g}^{\rm (t)} = \frac{\partial\omega_{\rm loc}}{\partial k_r} = \frac{\Gamma^{\rm (t)}}{W^{\rm (t)}} = \frac{\omega k_r}{k_\perp^2}\frac{|\delta B_\parallel|^2}{8\pi W^{\rm (t)}} = \omega k_r \frac{B^2}{8\pi} \frac{(\nablab\cdot{\bm \xi}_\perp)^2}{k_\perp^2 W^{\rm (t)}},
	\label{eq:imhd_vt}
\end{equation}

\noindent where we used Eq.~(\ref{eq:imhd_db}). The authors emphasize that Eq.~(\ref{eq:imhd_vt}) is valid for both shear Alfv\'{e}n and fast magnetoacoustic waves. Taking Eq.~(\ref{eq:imhd_dbpar}) and $W^{\rm (t)}_{\rm A} = k_\perp^2|\delta\Phi|^2/(8\pi v_{\rm A}^2)$ for shear Alfv\'{e}n waves in cylindrical geometry, and $W^{\rm (t)}_{\rm F} = k^2 |\delta B_\parallel|^2 / (8\pi k_\perp^2)$ and $\omega = k v_{\rm A}$ for fast waves gives%\vspace{-0.2cm}
\begin{equation}
	v_{\rm g,A}^{\rm (t)} = \frac{k_r v_{\rm A}^2}{\omega k_\perp^2} \left(k_\parallel^2 - \frac{\omega^2}{v_{\rm A}^2}\right), \qquad
	v_{\rm g,F}^{\rm (t)} = \frac{k_r}{k} v_{\rm A}.
	\label{eq:imhd_vt_saw}
\end{equation}%\vspace{-0.4cm}

On the one hand, the vanishing of $v_{\rm g,A}^{\rm (t)}$ in Eq.~(\ref{eq:imhd_vt_saw}) at shear Alfv\'{e}n continua is consistent with our expectation that wave energy transfer should stagnate in the domain of a continuum plateau, which is one of the rationales underlying the simulation setup used in the present work. On the other hand, it is currently not clear to what extent the framework outlined above is applicable to the case where information must be transferred between two distant Alfv\'{e}nic peaks, as in a double-peaked fishbone.

It is also worth noting that, in the case of a shear Alfv\'{e}n mode, $\delta B_\parallel$ in Eqs.~(\ref{eq:imhd_dbpar}) and (\ref{eq:gk_dbpar}) varies on the shear Alfv\'{e}n wave time scale. In fact, fast waves are entirely excluded from the gyrokinetic theory underlying Eq.~(\ref{eq:gk_dbpar}). This means that, within the scope of the above low-frequency model, spatial channeling and the resulting radial coupling of shear Alfv\'{e}n waves into coherent global structures is not mediated by the propagation of fast waves themselves. Instead, only the {\it fast wave physics} is at work here, in the sense that the shear Alfv\'{e}n waves are not pure but modified by magnetic compression.

We will observe later that fast waves and magnetoacoustically modified shear Alfv\'{e}n waves coexist in our simulations based on the single-fluid full MHD model. Fast waves are most easily detectable immediately after activating the antenna, and we assume that they continue to be emitted in all directions at each time step and each grid point throughout the simulation. They are reflected by the numerical non-slip boundaries and fill the plasma incoherently but at very small amplitudes (here about 3 orders of magnitude smaller than the shear Alfv\'{e}n wave field).

An open question is whether fast waves contribute directly to spatial channeling in the region between the peaks of a compound mode, where the fluctuation amplitude is small. The coupling evidenced by Eq.~(\ref{eq:imhd_disp}) suggests that low-frequency shear Alfv\'{e}n modes continuously emit and absorb fast waves, whose amplitude will thus be modulated by the low-frequency mode. Although the fast wave amplitudes are very small, their radial velocity is high, so they could play a role for the transfer of signal information (e.g., frequency chirping) across the weakly perturbed region between the two peaks of a double-peaked fishbone.

\subsection{Intricacies of weak magnetic shear}
\label{sec:history_flat}

The peculiarities of weak magnetic shear and shear reversal regions in tokamaks are an active area of research. For instance, it has recently been revisited using methods borrowed from studies of topological phase transitions \cite{Parker20}. Particularly intriguing are regions of weak magnetic shear near a magnetic axis and near low-order rational values of the safety factor $q$, which is the subject of the present work.

On the one hand, the weakness of magnetic shear and near constancy of $k_\parallel v_{\rm A}$ suggests that such regions exhibit a high degree of degeneracy in the sense that many forms of perturbations are supported on various time scales. We suspect that this is one reason why Grad \cite{Grad69} referred to such regions as ``pathological''. The concrete electromagnetic structure of such a region tends to be sensitive to external drive, one illustrative example being so-called ``multiple flux tubes'' that can form in domains with $q \approx 1$ under the influence of electron cyclotron current drive (ECCD) \cite{Yun12, Choe15, Bierwage15a}. Moreover, radial phase mixing is slow, so in the absence of non-ideal effects, the structure of a perturbation can be expected to have a long life time, irrespective of its shape (coherent or corrugated).

On the other hand, we imagine that wave energy emitted radially by an undulating flux tube at one radius $r$ can be readily reabsorbed at a neighboring radius $r + \delta r$ since weak shear implies that neighboring flux tubes have a similar geometry and are in that sense resonant with each other.

We are thus faced with the following conundrum: While weakness of magnetic shear appears to facilitate radial coherence geometrically, it also weakens the radial coupling required to establish and sustain such coherence.

This problem can be resolved by noting that, with the vanishing of magnetic shear, other symmetry-breaking and radial coupling mechanism take over. Our expectation that a globally coherent quasi-mode should eventually form on or near a continuum plateau rests on the intuition that a higher degree of radial coherence should minimize the relative shearing between neighboring undulating flux tubes. For instance, if one starts from radially highly incoherent flux tube undulations the associated highly nonuniform electric currents can, given enough time, be expected to be smoothed by processes that facilitate non-ideal diffusion. The diffusion-mediated mode structure formation process may require a relatively long time, but this may not be a serious obstacle for modes at low frequencies, such as the fishbone-like modes considered here.

\subsection{Non-ideal effects}
\label{sec:history_nonideal}

Our simulations include non-ideal effects, which in the present study enter mainly in the form of an electric resistivity $\eta_{\rm e}$ and an ion viscosity $\eta_{\rm i}$. In the main part of this paper, we utilized a moderately large value $10^6$ of the Lundquist number $S$ (inverse normalized resistivity) and Reynolds number ${\rm Re}$ (inverse normalized viscosity),%\vspace{-0.1cm}
\begin{equation}
\hspace{-0.5cm} S \equiv \hat{\eta}^{-1}_{\rm e} \equiv \frac{\mu_0 v_{\rm A0} R_0}{\eta_{\rm e}} \equiv \frac{\tau_\eta}{\tau_{\rm A0}}, \quad
{\rm Re} \equiv \hat{\eta}_{\rm i}^{-1} \equiv \frac{v_{\rm A0}R_0}{\eta_{\rm e}},
\label{eq:dissipation}
\end{equation}%\vspace{-0.4cm}

\noindent where $\tau_{\rm A0} \equiv \omega_{\rm A0}^{-1} \equiv R_0/v_{\rm A0}$ is the Alfv\'{e}n time and $\tau_{\eta_{\rm e}} \equiv R_0^2/(\mu_0\eta_{\rm e})$ the resistive diffusion time. For $S = {\rm Re} \gtrsim 10^6$, one may expect that radial diffusion on the spatial scale of our macroscopic modes ($L \sim a/3 \sim R_0/10$) should take $\gtrsim 10^4\,\tau_{\rm A0}$. Indeed, we will confirm in Appendix~\ref{apdx:dissipation_sensitivity} that these dissipation terms have a negligible effect on the overall structure of our low-frequency quasi-modes with $\omega/\omega_{\rm A0} \sim 0.03$, whose radial mode structures form on the time scale of $\pi/\omega \sim 100\, \tau_{\rm A0}$. Nevertheless, as anticipated in Section~\ref{sec:history_flat} above, we will also see in Appendix~\ref{apdx:dissipation_sensitivity} that the amplitude of an antenna-driven quasi-mode needs more time to settle at a constant value when both $S$ and ${\rm Re}$ are increased to $10^7$.

\begin{figure}
	[tb]\vspace{-1.2cm}
	\centering
	\includegraphics[width=0.48\textwidth]{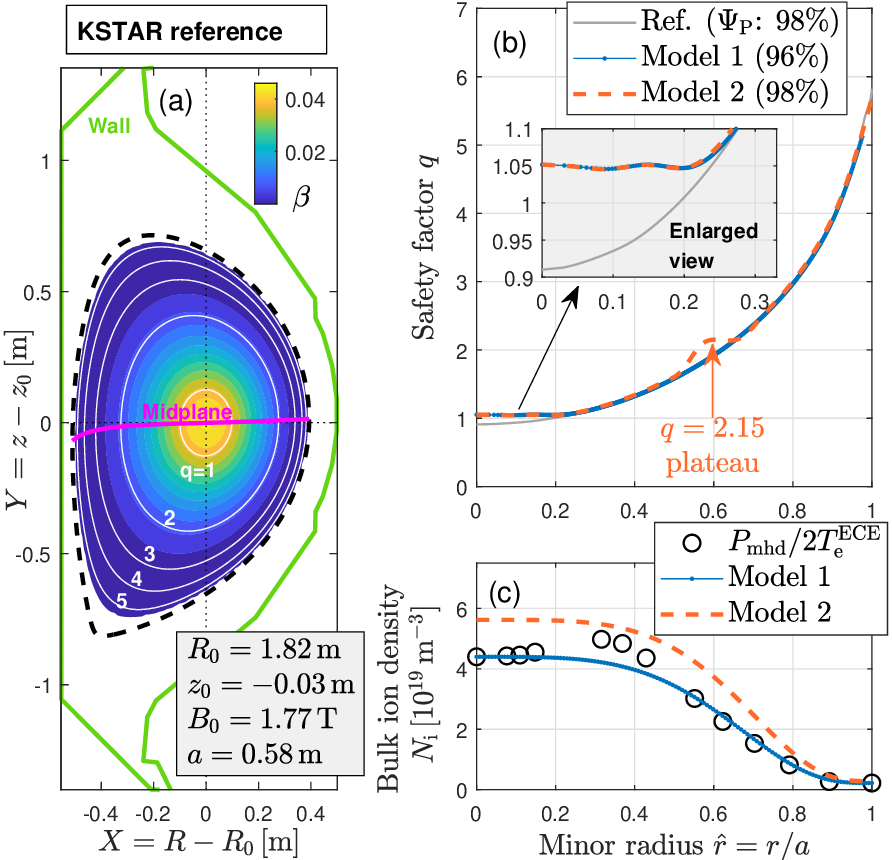}\vspace{-0.2cm}
	\caption{Simulation setup based on KSTAR shot 18567 @ $8.85\,{\rm s}$ \protect\cite{Lee23}. Panel (a) shows the plasma's poloidal $(R,z)$ cross-section, with the toroidal beta $\beta$ (color contours), magnetic flux surfaces with $q=1,2,3,4,5$ (white), last closed flux surface (dashed black), magnetic midplane (magenta), and wall (green). The coordinates of the magnetic axis $(R_0,z_0)$, the on-axis magnetic field strength $B_0$, and the mean minor radius $a$ are also shown. Panel (b) shows the safety factor profiles $q(r)$ as functions of the normalized minor radius $\hat{r} = r/a$ for the original reference case (gray) and for the two model cases (blue, orange) that we consider here. Panel (c) shows the bulk ion density profiles $N_{\rm i(r)}$ that are used in these two models.}\vspace{-0.3cm}
\label{fig:03_kstar_profiles}%
\end{figure}

\begin{figure}
[tb]\vspace{-1.2cm}
\centering
\includegraphics[width=0.48\textwidth]{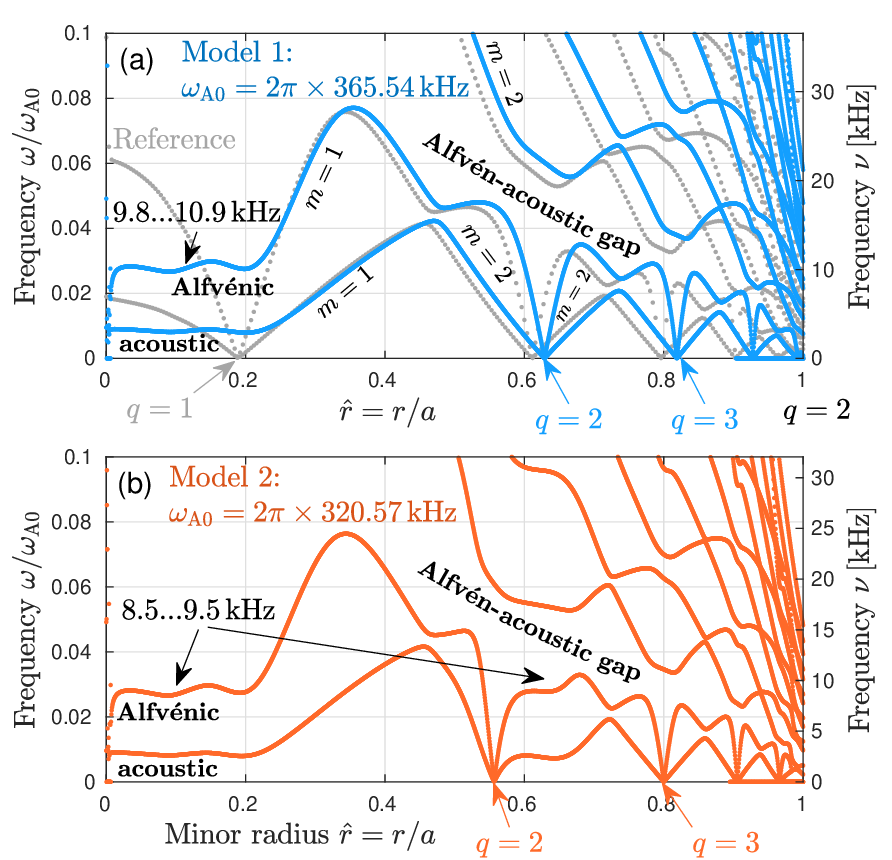}\vspace{-0.2cm}
\caption{MHD wave continua for the two model cases in Fig.~\protect\ref{fig:03_kstar_profiles}, computed by solving the equations in Appendix~A of Ref.~\protect\cite{Deng12} for toroidal mode number $|n|=1$. Arrows indicate flat portions of Alfv\'{e}nic low-frequency continua with dominant poloidal mode numbers $m=1$ and $m=2$, whose frequencies (a) $9.8...10.9\,{\rm kHz}$ and (b) $8.5...9.5\,{\rm kHz}$ are close to the plasma-frame frequency of the core-localized component of the double-peaked fishbone modes studied in Ref.~\protect\cite{Lee23}.}\vspace{-0.3cm}
\label{fig:04_kstar_continua}%
\end{figure}

% =============================================================================
\section{Simulation model and methods}
\label{sec:setup}

Our working examples are based on the Korean Superconducting Tokamak Advanced Research (KSTAR) device \cite{Kwon11}. Our reference case is KSTAR shot \#18567 (year 2017), which was studied in Refs.~\cite{Lee23, Lee26, Bierwage25}. For the chosen time slice ($8.85\,{\rm seconds}$), KSTAR modelers using the {\tt TRANSP} code \cite{TRANSP} obtained the Grad-Shafranov equilibrium shown in Fig.~\ref{fig:03_kstar_profiles}(a). The toroidal beta $\beta \equiv 2\mu_0 P_{\rm mhd}/B^2$ has a central value of about $\beta_0 \approx 4.7\%$, where $\mu_0$ is the vacuum permeability, $P_{\rm mhd}$ is the MHD pressure, and $B = |{\bm B}|$ is the magnetic field strength. The magnetic axis is located at $(R_0,z_0) = (1.82\,{\rm m},-0.03\,{\rm m})$, where $B_0 \equiv B(R_0,z_0) = 1.77\,{\rm T}$. The plasma current was about $0.5\,{\rm MA}$. The safety factor, whose profile $q(r)$ is plotted gray as a function of volume-averaged minor radius $\hat{r} = r/a$ in Fig.~\ref{fig:03_kstar_profiles}(b), turned out to have a central value of $q_0 \equiv q(0) = 0.91$. The magnetic surfaces where $q=1,2,3,4,5$ are shown as white contours in Fig.~\ref{fig:03_kstar_profiles}(a).

The shape of $q(r)$ near the axis is difficult to measure experimentally and thus highly uncertain, so we regard it as a free parameter. In this work, we used the equilibrium solver {\tt CHEASE} \cite{Luetjens96} to create equilibria with the two modified $q(r)$ profiles that are shown as blue and orange curves in Fig.~\ref{fig:03_kstar_profiles}(b). We designed both profiles to be nearly flat and slightly elevated above $1$ in the central region $\hat{r} \lesssim 0.2$.\footnote{This imitates ITER case 101006r50, which was studied in Ref.~\protect\cite{Bierwage25}. When constructing our Grad-Shafranov equilibria, we also removed a small portion of flux space ($2\%$ in the reference case and Model 2, $4\%$ in Model 1) in order to alleviate convergence problems that {\tt CHEASE} had when initialized with the original {\tt EQDSK} file from {\tt TRANSP}.}
By elevating $q_0$, we avoid the resistive internal kink instability (cf.~Appendix~\ref{apdx:kink}) that would otherwise cause dynamics similar to so-called sawtooth crashes in the visco-resistive MHD model that is used here. In reality, the internal kink may be stabilized by mirror-trapped fast ions \cite{Porcelli91} and sheared rotation \cite{Graves00} even when $q_0 \lesssim 1$, but simulating such a scenario faithfully is difficult and expensive. For the purpose of the present study, it is more efficient and sufficient to stabilize the plasma by letting $q \gtrsim 1$. We note that configurations with centrally flat $q \gtrsim 1$ bear relevance to the so-called ``hybrid scenario'' that is one of the attractive operational regimes for ITER. Under some conditions, such a configuration may bifurcate into a non-axisymmetric state known as helical core \cite{Cooper10, Cooper11, Wingen18, Adulsiriswad25}, but this was not seen in the cases studied here.

The centrally flattened $q$ profiles of our Models 1 (blue) and 2 (orange) imply that the continuous spectra of MHD waves are flat in the same region $\hat{r} \lesssim 0.2$. This can be seen in Fig.~\ref{fig:04_kstar_continua} for waves with  toroidal Fourier mode number $|n|=1$ at low frequencies $\nu \equiv \omega/(2\pi) \lesssim 30\,{\rm kHz}$. For comparison, the continua for the reference case, with a $q=1$ surface at $\hat{r} \approx 0.2$, are plotted gray in Fig.~\ref{fig:04_kstar_continua}(a).

As we explained in the introductory Section~\ref{sec:intro}, we will utilize these continuum plateaus as resonant ``receivers'' in our antenna-driven simulations. In Model 2, whose continua are shown in Fig.~\ref{fig:04_kstar_continua}(b), we have introduced an additional receiver plateau in the outer core, around $0.6 \lesssim \hat{r} \lesssim 0.65$. Fig.~\ref{fig:03_kstar_profiles}(b) shows that the safety factor has a corresponding plateau around $q \approx 2.15$, with the side effect of $q(r)$ having a steep gradient (i.e., large magnetic shear) at the nearby $q=2$ surface. We will see later that this has the consequence of Model 2 being subject an unstable tearing mode with dominant poloidal/toroidal Fourier mode numbers $m/n=2/1$. However, in contrast to the rapidly growing $m/n = 1/1$ internal kink in the reference case (gray), the $2/1$ tearing mode in Model 2 (orange) is sufficiently benign and forms sufficiently slowly to still allow us to study the antenna-driven plasma response. Model 1 (blue) is stable for our simulation parameters.

Our KSTAR model plasma consists only of deuterium (Carbon impurities are ignored). Direct measurements of particle density profiles were not available. Instead, the circles in Fig.~\ref{fig:03_kstar_profiles}(c) show the bulk ion density profile $N_{\rm i}(r)$ that one obtains by assuming that the ion temperature equals the electron temperature measured by ECE radiometry, $T_{\rm i} \approx T_{\rm e}^{\rm ECE}$, and letting $N_{\rm i} = P_{\rm i}/T_{\rm i} \approx N_{\rm e} \approx P_{\rm mhd}/(2T_{\rm e}^{\rm ECE})$. The off-axis peak is thought to be the result of {\tt TRANSP} modelers having used a simplified pressure profile model: $P_{\rm mhd}(r)/P_{\rm mhd,0} = (1 - \psi_{\rm P}(r))^2$, which is probably inconsistent with the sharply peaked $T_{\rm e}^{\rm ECE}(r)$ profile. We have thus chosen to construct simplified monotonic density profiles using the parametric model
\begin{equation}
	N_{\rm i}(\psi)/N_{\rm i0} = (1 - \hat{N}_{\rm a}) (1 - \psi_{\rm P}^{\hat{c}_{\rm 1}})^{\hat{c}_{\rm 2}} + \hat{N}_{\rm a},
	\label{eq:setup_ni}
\end{equation}

\noindent where $\psi_{\rm P}(r) \equiv (\Psi_{\rm P}(r) - \Psi_{\rm P0})/(\Psi_{{\rm P}a} - \Psi_{\rm P0}) \in [0,1]$ is the normalized poloidal flux with $\Psi_{\rm P0} \equiv \Psi_{\rm P}(0)$ and $\Psi_{{\rm P}a} \equiv \Psi_{\rm P}(a)$. Using the parameter values shown in Table~\ref{tab:setup_ni}, we constructed the $N_{\rm i}(r)$ profiles for Models 1 and 2 that are plotted blue and orange in Fig.~\ref{fig:03_kstar_profiles}(c).

\begin{table}[tbp]
	\centering\vspace{-1.0cm}
	\begin{tabular}{c|c@{\hspace{0.25cm}}c|c@{\hspace{0.35cm}}c@{\hspace{0.35cm}}c@{\hspace{0.35cm}}c}
		\hline\hline Model & $B_0\,[{\rm T}]$ & $R_0\,[{\rm m}]$ & $N_{\rm i0}\,[{\rm m}^{-3}]$ & $\hat{N}_{\rm a}$ & $\hat{c}_1$ & $\hat{c}_2$ \\
		\hline 1: & $1.79$ & $1.81$ & $4.4\times 10^{19}$ & $0.05$ & $3.8$ & $3.5$ \\
		\hline 2: & $1.78$ & $1.82$ & $5.6\times 10^{19}$ & $0.05$ & $4.2$ & $3.5$ \\
		\hline\hline
	\end{tabular}%\vspace{-0.1cm}
	\caption{On-axis field strength $B_0$, major radius $R_0$, and parameters for the bulk ion density profile model in Eq.~(\protect\ref{eq:setup_ni}).}
	\label{tab:setup_ni}\vspace{-0.4cm}
\end{table}

In the present study, the primary role of the ion density $N_{\rm i}$ is to determine the Alfv\'{e}n frequency $\omega_{\rm A0} = v_{\rm A0}/R_0$, where $v_{\rm A0} \equiv B_0/\sqrt{\mu_0 M_{\rm i} N_{\rm i0}}$ is the central Alfv\'{e}n velocity and $M_{\rm i}$ is the ion mass. This, in turn, determines the frequencies of the MHD wave continua in Fig.~\ref{fig:04_kstar_continua}. In Model 1, the central value $N_{\rm i0} \equiv N_{\rm i}(0)$ is set equal to $P_{\rm mhd,0}/(2 T_{\rm e0}^{\rm ECE}) \approx 4.4\times 10^{19}\,{\rm m}^{-3}$. In this case, the central plateau of the Alfv\'{e}nic low-frequency continuum with $n=1$ and dominant poloidal Fourier mode number $m=1$ has a frequency around $9.8...10.9\,{\rm kHz}$ as shown in Fig.~\ref{fig:04_kstar_continua}(a). In Model 2, we raised $N_{\rm i0}$ to about $5.6\times 10^{19}\,{\rm m}^{-3}$ and slightly broadened the density peak. With this, we were able to align the on-axis ($m/n = 1/1$) and off-axis ($m/n = 2/1$) continuum plateaus in Fig.~\protect\ref{fig:04_kstar_continua}(b) to have similar frequencies around $8.5...9.5\,{\rm kHz}$. The receiver plateau frequencies in both models are thus close to the expected plasma-frame frequency $\nu_{\rm pla} \approx \nu - n\nu_{\rm rot}(\hat{r}=0) \sim 10\,{\rm kHz}$ of a typical double-peaked fishbone mode's core-localized component before its downward chirp (i.e., its ``seed'' frequency) \cite{Lee23}.

\begin{table}[tbp]
	\centering\vspace{-1.0cm}
	\begin{tabular}{c|c@{\hspace{0.15cm}}c@{\hspace{0.15cm}}c@{\hspace{0.15cm}}c}
		\hline\hline $R_{\rm ant}\,{\rm [m]}$ & 1.98, & 2.04, & 2.10, & {\bf 2.18} \\
		($r_{\rm ant}/a$) & (0.13), & (0.45), & (0.613), & ({\bf 0.85}) \\
		\hline $\nu_{\rm ant}\,{\rm [kHz]}$ & \multicolumn{4}{c}{4.5,\; {\bf 9},\; 18,\; 27} \\
		\hline $w_{\rm ant}\,{\rm [m]}$ & \multicolumn{4}{c}{{\bf 0.02},\; 0.04,\; 0.06} \\
		\hline\hline
	\end{tabular}%\vspace{-0.1cm}
	\caption{Values of scanned parameters in our antenna model (\protect\ref{eq:setup_ant}). Default values are printed bold. The minor radii in parentheses are the approximate values of $r_{\rm ant} \equiv r(R_{\rm ant},z_{\rm ant})$. The values of the other parameters in Eq.~(\protect\ref{eq:setup_ant}) remain fixed: $\hat{A}_{\rm ant} = 2\times 10^{-8}$, $z_{\rm ant} = z_0 = -0.03\,{\rm m}$, $|n_{\rm ant}| = 1$.}
	\label{tab:setup_ant}\vspace{-0.4cm}
\end{table}

In our simulations, electromagnetic fluctuations are driven using an internal antenna model of the form
\begin{subequations}
\begin{align}
	\hspace{-0.8cm} \delta{\bm B}_{\rm ant} =&\; \varrho_{\rm A0} B_0 \hat{A}_{\rm ant} \nablab\times\left(\delta\hat{E}_{\rm ant}(R,z,\zeta,t) \hat{\bm e}_\zeta\right),
	\label{eq:setup_ant_dB}
	\\
	\hspace{-0.8cm} \delta \hat{E}_{\rm ant} =&\; {\rm cos}(n_{\rm ant}\zeta - \omega_{\rm ant} t)\times \left\{\begin{array}{l@{\hspace{0.15cm}}c@{\hspace{0.15cm}}l} {\rm cos}^3(\chi) & : & \chi < \frac{\pi}{2}, \\ 0 & : & {\rm else},\end{array}\right.
	\label{eq:setup_ant_phase}
	\\
	\hspace{-0.8cm} \chi \equiv&\; \frac{\pi}{2} \sqrt{\left(R - R_{\rm ant})^2 + (z - z_{\rm ant})^2\right)/w_{\rm ant}^2},
	\label{eq:setup_ant_env}
\end{align}\vspace{-0.4cm}
\label{eq:setup_ant}
\end{subequations}

\noindent where the on-axis field strength $B_0$ and the Alfv\'{e}nic Larmor radius $\varrho_{\rm A0} = M_{\rm i} v_{\rm A0} / (Z_{\rm i}e)$ are used to normalize ${\bm B}$ and $\nablab$, respectively. We will fix the antenna's amplitude $\hat{A}_{\rm ant} = 2\times 10^{-8}\,{\rm m}$ and vertical position $z_{\rm ant} = z_0 = -0.03\,{\rm m}$. As indicated in Fig.~\ref{fig:05_kstar-1.051_decay-ant_early}(a), we will vary the radial position $R_{\rm ant}$ of a narrow antenna ($w_{\rm ant} = 0.02\,{\rm m}$) with frequency $\nu_{\rm ant} \equiv \omega_{\rm ant}/(2\pi) = 9\,{\rm kHz}$. We will also vary the antenna's frequency $\nu_{\rm ant}$ and its spatial width $w_{\rm ant}$ in the range of values shown in Table~\ref{tab:setup_ant}. For the toroidal mode number we use $n_{\rm ant} = \pm 1$, whose sign is arbitrary since fast ions and plasma rotation are absent here. Most simulations reported in this paper were performed with $n_{\rm ant} = -1$ (for no particular reason). In a few cases, we use $n_{\rm ant} = +1$, because that will resemble to propagation of KSTAR fishbones in the $+\zeta$ direction; that is, clockwise and out-of-plane in Fig.~\ref{fig:01_kstar_ecei}(c,d).

In Eq.~(\ref{eq:setup_ant_dB}), the antenna's electric field is chosen to point in the toroidal direction. Within the scope of single-fluid MHD this choice can be considered to be arbitrary. Tests with $\delta\hat{E}_{\rm ant}$ pointing in the $R$- or $z$ directions on the outer midplane gave similar mode structures and spectra. Only the energy conversion efficiency varied significantly.\footnote{In our tests with a narrow antenna located near the plasma edge on the outer midplane, the energy of the $|n|=1$ fluctuations driven by the toroidally polarized antenna (\protect\ref{eq:setup_ant_dB}) was $25$ times larger than with vertical (poloidal) polarization and nearly $6000$ times larger than with horizontal (radial) polarization. These number may, however, depend on the spatial resolution because the coarse mesh ($N_R\times N_z\times N_\zeta = 200\times 200\times 16$ grid points) only poorly resolved the narrow antenna profile in the $(R,z)$-plane.}

Our simulations were performed using an adapted version of the MHD-PIC hybrid code {\tt MEGA} \cite{Todo98, Todo05,Todo25}, albeit solving only the full MHD equations without invoking the kinetic PIC module. The equations with all source terms shown explicitly can be found, for instance, in Section 2.2 of Ref.~\cite{Bierwage17a}. The antenna field is added to the total magnetic field as ${\bm B}(t) = {\bm B}_{\rm eq} + \delta{\bm B}(t) + \delta{\bm B}_{\rm ant}(t)$, where ${\bm B}_{\rm eq} = \Psi_{\rm P}\times\nablab\zeta + I(\Psi_{\rm P})\nablab\zeta$ is the equilibrium field, whose covariant toroidal component $I(\Psi_{\rm P}) = R B_{\rm tor}$ and poloidal flux $2\pi\Psi_{\rm P}(R,z)$ together with $P_{\rm mhd}$ satisfy the Grad-Shafranov equation. {\tt MEGA} then computes the total current density and electric field as $\mu_0{\bm J} = \nablab\times{\bm B}$ and $\delta{\bm E} = {\bm B}\times\delta{\bm u} + \eta_{\rm e}({\bm J} - {\bm J}_{\rm eq})$, where $\delta{\bm u}$ is the MHD velocity vector field and $\eta_{\rm e}$ the resistivity. The continuity equation is solved for the fluctuating part of the mass density $\rho_{\rm mhd} = M_{\rm i} N_{\rm i} + \delta\rho_{\rm mhd}$, and an equation of state with specific heat ratio $\Gamma = 5/3$ is solved for $\delta P_{\rm mhd}$.

The parameter space was first explored using relatively coarse resolution with $N_R\times N_z \times N_\zeta = 200\times 200\times 16$ grid points in right-handed cylinder coordinates $(R,z,\zeta) = (R,\varphi,z)$. The tentative results were subsequently confirmed with up to $400\times 400\times 128$ or $1024\times 1024\times 32$ grid points. This is sufficient for our purposes since we focus primarily on the long-wavelength $n = 1$ component at amplitudes that are small enough for MHD nonlinearities to be ignorable. The electric resistivity, ion viscosity and thermal diffusivity are all spatially uniform and fixed at the same normalized value: $\eta_{\rm e}/\mu_0 = \eta_{\rm i} = \chi = 10^{-6} v_{\rm A0} R_0$. This is expected to keep numerical dissipation at a reasonably low level \cite{Bierwage16c}. The sensitivity of the main results with respect to $\eta_{\rm e}$ and $\eta_{\rm i}$ is examined in Appendix~\ref{apdx:dissipation_sensitivity}.\vspace{-0.2cm}

% =============================================================================
\section{Simulation results}
\label{sec:results}\vspace{-0.2cm}

% -----------------------------------------------------------------------------
\subsection{Structure formation process \& time scales}
\label{sec:results_transients}

The purpose of this section is to illustrate the effect of our antenna and the process of mode structure formation, while also introducing the diagnostic methods that we use throughout this paper. Our KSTAR Model 1 with a narrow $9\,{\rm kHz}$ antenna located near the plasma center ($\hat{r}_{\rm ant} = 0.13$, $w_{\rm ant} = 0.02\,{\rm m}$) shall serve as a first concrete example.

\begin{figure}
	[tbp]\vspace{-0.8cm}
	\centering
	\includegraphics[width=0.48\textwidth]{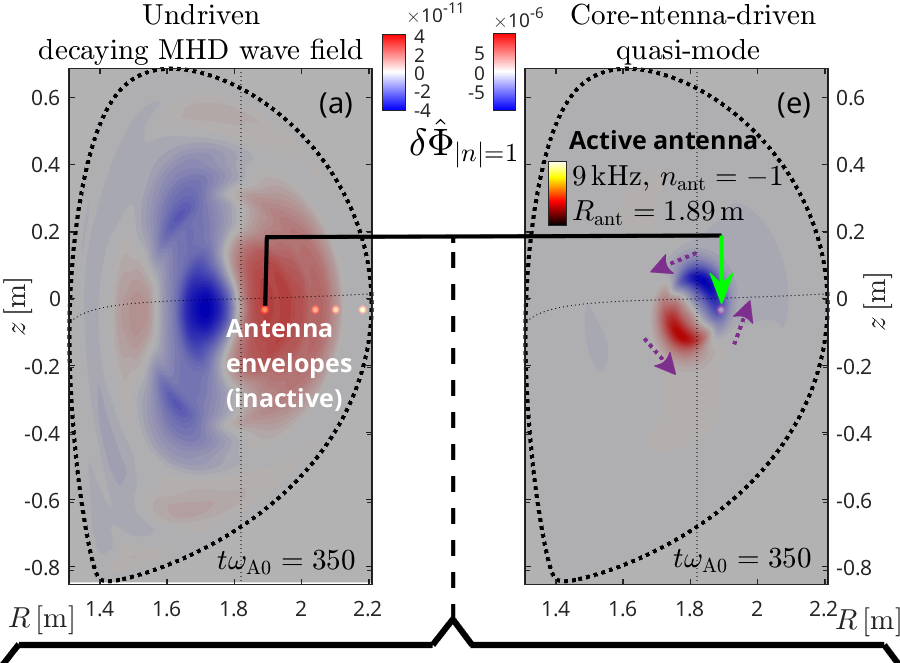}\vspace{-0.02cm}
	\includegraphics[width=0.48\textwidth]{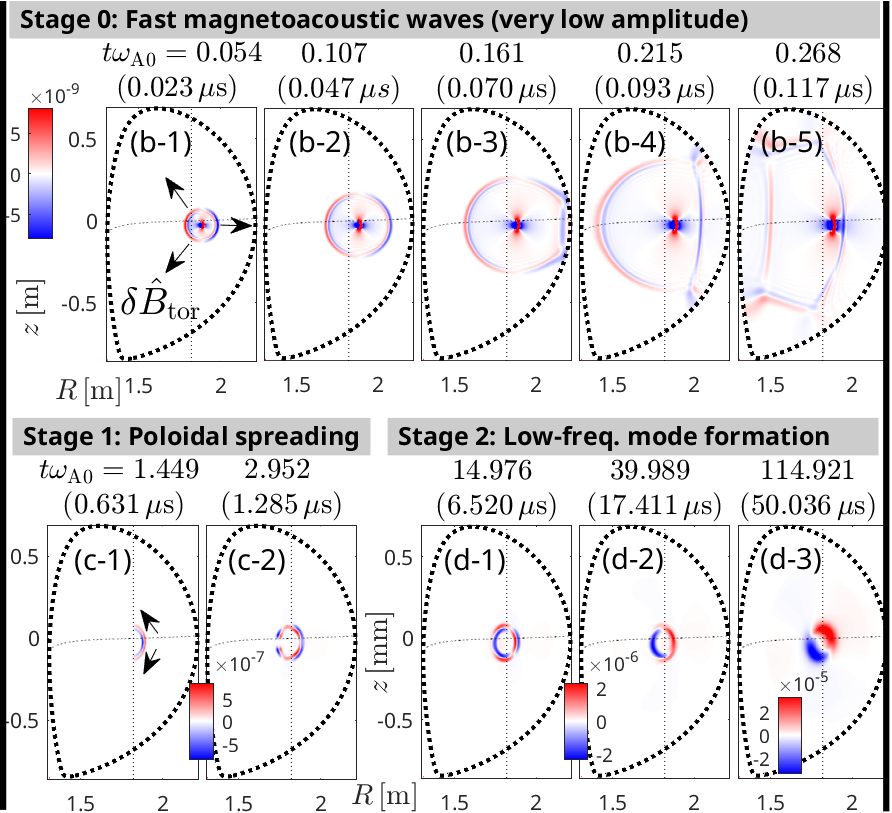}%\vspace{-0.3cm}
	\caption{Illustration of our antenna model and its effect on the plasma in the poloidal cross-section of our KSTAR Model 1. The four bright spots in panel (a) represent the antenna envelopes $\delta\hat{E}_{\rm ant} \propto \cos^3\chi$ of Eq.~(\protect\ref{eq:setup_ant}) at the four radii $R_{\rm ant}$ listed in Table~\protect\ref{tab:setup_ant}. In the background, we show a snapshot $\delta\hat{\Phi}_{|n|=1}(R,z)$ of the decaying electrostatic potential fluctuations after initializing a {\tt MEGA} simulation with an arbitrary perturbation and running it for $\hat{t}\equiv t\omega_{\rm A0} = 350$ Alfv\'{e}n times ($0.15\,{\rm ms}$) without antenna. After activating the innermost antenna ($R_{\rm ant} = 1.98\,{\rm m}$, $\hat{r}_{\rm ant} \approx 0.13$), and running the simulation for the same duration of $350$ Alfv\'{e}n times, we obtained the wave field $\delta\hat{\Phi}_{|n|=1}(R,z)$ shown in panel (e): a coherent core-localized response with dominant Fourier component $|m/n|=1/1$, which rotates counter-clockwise as indicated by the three violet arrows. As described in the text, this antenna-driven quasi-mode forms in four stages, which we label 0, 1, 2 and 3. The main features of Stages 0--2 are shown in panel groups (b)--(d) in terms of snapshots of the fluctuating component $\delta\hat{B}_{\rm tor}(R,z)$ of the toroidal magnetic field. The arrows in (b-1) and (c-1) indicate wave front propagation directions. The dotted curves indicate the plasma boundary, which can be seen to act as a reflector in (b). Note that the colorbar in snapshot set (b) has been scaled and clipped at a low amplitude in order to reveal the fine structure of the earliest transients in $\delta\hat{B}_{\rm tor}$. In contrast, the color scale in (c) and (d) is chosen to display the full range of $\delta\hat{B}_{\rm tor}$. The final Stage 3 is shown in Fig.~\protect\ref{fig:06_kstar-1.051_decay-ant_spec}(c,d).}%\vspace{-0.2cm}
	\label{fig:05_kstar-1.051_decay-ant_early}%
\end{figure}

The upper part of Fig.~\ref{fig:05_kstar-1.051_decay-ant_early} shows snapshots of the MHD plasma response (a) without and (e) with antenna drive by plotting the spatial structure of the $|n|=1$ component of the fluctuating electrostatic potential $\delta\Phi$. This quantity is normalized as $\delta\hat{\Phi} = \delta\Phi/(a v_{\rm A0} B_0)$ and estimated from the Fourier components of the radial MHD velocity as\vspace{-0.1cm}
\begin{equation}
	\delta\hat{\Phi}_n(r|m\neq 0) \approx \frac{\hat{r}}{m}[\delta\hat{u}_{\rm rad}]_n(r|m\neq 0),
	\label{eq:epot}\vspace{-0.5cm}
\end{equation}

\noindent where $\delta\hat{u}_{\rm rad} \equiv \delta{\bm u}\cdot\hat{\bm e}_r/v_{\rm A0}$. (In the few cases where the $m=0$ component is also needed, we will examine $\delta\hat{u}_{\rm rad}$ directly.) The brackets $[...]_n$ are used to indicate the $n$-component of a quantity that already has other subscripts.

\begin{figure}
	[tb]\vspace{-1.0cm}
	\centering
	\includegraphics[width=0.48\textwidth]{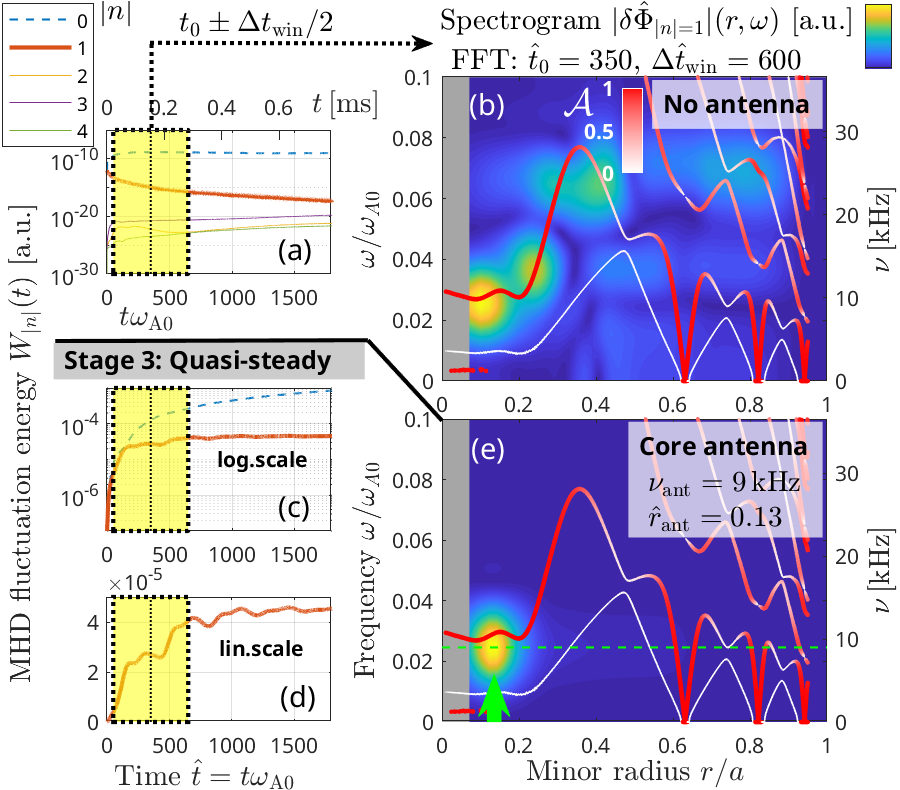}\vspace{-0.2cm}
	\caption{Long-time evolution (left) and spectrograms (right) for (a,b) the undriven case of Fig.~\protect\ref{fig:05_kstar-1.051_decay-ant_early}(a), and (c,d,e) the antenna-driven case of Fig.~\protect\ref{fig:05_kstar-1.051_decay-ant_early}(e). This is the final Stage 3 of the mode structure formation process described in the text: the attainment of a quasi-steady state. Panels (a) and (c) show the evolution of the (volume-integrated) MHD fluctuation energy $W_{|n|}(t)$ defined in Eq.~(\protect\ref{eq:dg_wn}) on a logarithmic scale and (d) on a linear scale. In the undriven case (a), we simulated the toroidal Fourier components $|n|=0,1,2,3,4$ after perturbing only $n=1$. In the antenna-driven case, we simulated only $|n|=0,1$, while filtering out $|n|>1$. The narrow antenna was applied in the central core at $\hat{r}_{\rm ant} = 0.13$ with frequency $\nu_{\rm ant} = 9\,{\rm kHz} \approx 0.25\,\omega_{\rm A0}/(2\pi)$ and $n_{\rm ant} = -1$. The yellow-shaded region in panels (a,c,d) indicates the time window $\hat{t}_0 \pm \Delta\hat{t}_{\rm win}/2$ that we used to compute the frequency-radius spectrograms $|\delta\hat{\Phi}_{|n|=1}|(r,\omega)$ that are shown in panels (b) and (e). Here, we used a window of width $\Delta\hat{t}_{\rm win} = 600$ centered around $\hat{t}_0 = 350$, Hanning weights and zero-padding were applied before performing the FFT to sharpen the image. The spectrograms are overlaid with the full MHD continua that were computed using the {\tt FALCON} code \protect\cite{Falessi20, Falessi19b}, and whose local Alfv\'{e}nicity $\A$ is indicated by shades of red (Alfv\'{e}nic) to white (acoustic).}\vspace{-0.5cm}
	\label{fig:06_kstar-1.051_decay-ant_spec}%
\end{figure}

For both cases in Fig.~\ref{fig:05_kstar-1.051_decay-ant_early}(a,e), the left column of Fig.~\ref{fig:06_kstar-1.051_decay-ant_spec} shows the evolution of the MHD fluctuation energy
\begin{equation}
	\hspace{-0.9cm} W_{|n|} \equiv \int\limits_{\rm plasma}{\rm d}^3 x\, \left(\frac{|\rho_{\rm mhd}^{{1/2}}\delta{\bm u}|_{|n|}^2}{2} + \frac{|\delta{\bm B}|^2_{|n|}}{2} + \frac{[\delta P_{\rm mhd}]_{|n|}}{\Gamma - 1}\right),
	\label{eq:dg_wn}
\end{equation}\vspace{-0.35cm}

\noindent integrated over the plasma volume, and the right column of Fig.~\ref{fig:06_kstar-1.051_decay-ant_spec} shows frequency-radius spectrograms $|\delta\hat{\Phi}_{|n|=1}|(r,\omega)$.

In Fig.~\ref{fig:06_kstar-1.051_decay-ant_spec}(a), the time trace of the energy $W_{n=0}$ contained in the axisymmetric ($n=0$) component saturates after a brief transient phase, showing that the undriven simulation approximately conserves energy. The decay of $W_{|n|=1}$, which contains the initially perturbed $n=1$ component, shows its stability in the present KSTAR Model 1. Nonlinearly excited Fourier harmonics with $|n|>1$ also remain negligibly small.

In the absence of an antenna, Figs.~\ref{fig:05_kstar-1.051_decay-ant_early}(a) and \ref{fig:06_kstar-1.051_decay-ant_spec}(b) show global broad-band fluctuations. In order to interpret frequency-radius spectrograms $|\delta\hat{\Phi}_{|n|=1}|(r|m)$ like that in Fig.~\ref{fig:06_kstar-1.051_decay-ant_spec}(b) we overlay them with continuous spectra computed using the {\tt FALCON} code \cite{Falessi20, Falessi19b}. This code also yields the continuum's local polarization, which is measured by the Alfv\'{e}nicity $\A \in [0,1]$ and plotted in shades of red ($\A \rightarrow 1$) to white ($\A \rightarrow 0$) to indicate dominant Alfv\'{e}nic or acoustic polarization, respectively. In Fig.~\ref{fig:06_kstar-1.051_decay-ant_spec}(b), one can see that the strongest (most slowly decaying) fluctuations tend to cluster near the continuum branches with dominant Alfv\'{e}nic polarization (red) in the plasma core ($\hat{r} \lesssim 0.5$).

Next, we activate the antenna with toroidal mode number $n_{\rm ant} = -1$ and frequency $\nu_{\rm ant} = 9\,{\rm kHz}$ in the central core at $(R_{\rm ant},z_{\rm ant}) = (1.98\,{\rm m},-0.03\,{\rm m})$, which corresponds to the minor radius $\hat{r}_{\rm ant} \approx 0.13$ on the outer midplane. The spectrogram in Fig.~\ref{fig:06_kstar-1.051_decay-ant_spec}(e) shows the antenna-driven plasma response near (slightly below) the central plateau of the low-frequency Alfv\'{e}nic continuum branch with poloidal and toroidal mode numbers $|m/n| = 1/1$, which has $\omega/\omega_{\rm A0} \approx 0.03$ ($\approx 10\,{\rm kHz}$) in the region $\hat{r} \lesssim 0.2$.

The evolution of $W_{|n|=1}$ in Fig.~\ref{fig:06_kstar-1.051_decay-ant_spec}(c,d) shows that it takes about $1000$ Alfv\'{e}n times (here $\approx 0.44\,{\rm ms}$) for the antenna-driven $n=-1$ fluctuations to reach a relatively steady state, which is typical for our simulations with low-frequency antennas. The continuous increase of $W_{n=0}$ in Fig.~\ref{fig:05_kstar-1.051_decay-ant_early}(c) can be attributed to the antenna effectively heating the plasma.\footnote{This heating effect vanishes if we artificially suppress pressure fluctuations by setting $\delta P_{\rm mhd} \rightarrow 0$ at each time step. See Fig.~\protect\ref{fig:16_kstar-1.051_dpre0_evol-spec}.}

While the amplitude of this antenna-driven quasi-mode takes a fairly long time to settle, its spatial structure is established within a few $100$ Alfv\'{e}n times. The snapshot of $\delta\hat{\Phi}_{|n|=1}(R,z)$ in Fig.~\ref{fig:05_kstar-1.051_decay-ant_early}(e) taken at $\hat{t} = 350$ is representative also for later times. On the way to this mode structure, the fluctuating field passes through several stages that are characterized by wave propagation on different time scales, and whose main features are shown in panel groups (b)--(d) of Fig.~\ref{fig:05_kstar-1.051_decay-ant_early} in terms of snapshots of the fluctuating component $\delta\hat{B}_{\rm tor}(R,z)$ of the toroidal magnetic field.

Altogether, we identify four stages in the mode structure formation process, which may be outlined as follows:
\begin{itemize}
	\item {\it Stage 0.} After the antenna is abruptly activated at $t=0$, fast magnetoacoustic waves with speed $v_{\rm A0}$ radiate in all directions and fill the plasma on the time scale $a/v_{\rm A} \sim 0.1\,\mu{\rm s}$. This is shown in panel set (b) of Fig.~\ref{fig:05_kstar-1.051_decay-ant_early}, where the fine structure of $\delta\hat{B}_{\rm tor}$ is brought forth via a suitably scaled and clipped color bar.
	\item {\it Stage 1.} Meanwhile, the actual peak of $\delta\hat{B}_{\rm tor}$ spreads poloidally via shear Alfv\'{e}n waves propagating along ${\bm B}$ with speed $v_{\rm A}$, and panel set (c) of Fig.~\ref{fig:05_kstar-1.051_decay-ant_early} shows how the local flux surface is filled by these waves on the poloidal Alfv\'{e}n time scale $\tau_\vartheta \sim qR/v_{\rm A} \sim 1\,\mu{\rm s}$
	\item {\it Stage 2.} The structure of the low-frequency quasi-mode gradually emerges as the wave field undergoes phase mixing and radial spreading in Fig.~\ref{fig:05_kstar-1.051_decay-ant_early}(d). In the present case with frequency $\omega = 2\pi\times 9\,{\rm kHz}$, this takes about $\tau_\omega \sim \pi/\omega \sim 50\,\mu{\rm s}$, in accordance with the universal uncertainty principle.\footnote{Only after observing half of a harmonic oscillation cycle (sacrificing time resolution), one has enough information (including the curve's symmetry) to anticipate a periodic waveform with a well-defined frequency.}
	\item {\it Stage 3.} Finally, there is the time scale $\tau_{\rm sat}$ of energy accumulation and saturation, which is completed when a balance between drive and damping has been established and a quasi-steady state is attained. For our core-localized low-frequency quasi-modes, this typically took about $\tau_{\rm sat} \sim 0.5\,{\rm ms}$, as can be inferred from the time traces of the MHD fluctuation energy $W_n(t)$ in Fig.~\ref{fig:06_kstar-1.051_decay-ant_spec}. Factors that can influence $\tau_{\rm sat}$ include the mode's frequency (Appendix~\ref{apdx:modeform_highfreq}) and the dissipation coefficients (Appendix~\ref{apdx:dissipation_sensitivity}).
\end{itemize}

\noindent A detailed discussion of this mode structure formation process along with a larger set of snapshots can be found in Appendix~\protect\ref{apdx:modeform} (for the same KSTAR Model 1, but with a remote antenna). Only a few remarks shall be made here.

A local change in the fields or currents caused by an antenna (or other disturbance) constitutes a violation of the previously existing force balance that determined the global equilibrium, which in our case is determined by the Grad-Shafranov equation. The numerical plasma then seeks to establish a new equilibrium by propagating information about the disturbance. In the MHD model, the fundamental speed of plasma response is the Alfv\'{e}n velocity $v_{\rm A}$, and this is indeed the speed at which we observe transient perturbation fronts propagating in our antenna-driven MHD simulations. When viewed in the poloidal $(R,z)$ plane of Fig.~\ref{fig:05_kstar-1.051_decay-ant_early}, the radially propagating Alfv\'{e}nic fronts are seen first, albeit at low amplitude (Stage 0), while those propagating along ${\bm B}$ become visible on a global scale at later times (Stage 1).

Although not shown here, it is also worth noting that the patterns seen during Stage 1 in Fig.~\ref{fig:05_kstar-1.051_decay-ant_early}(b) remain effectively unchanged when density and pressure fluctuations are suppressed (letting $\delta\rho_{\rm mhd} = \delta P_{\rm mhd} = 0$ at all times). This confirms our expectation from Section~\ref{sec:history} that, in a low-beta plasma like a tokamak, the sonic component is not essential for radial information transfer as such. Nevertheless, the sonic component does influence the local responsiveness of the plasma and, thus, the final form of the mode structure as we will see in Section~\ref{sec:discuss_compress} and Appendix~\ref{apdx:modeform_transients}.

\begin{figure}
	[tb]
	\centering\vspace{-0.4cm}
	\includegraphics[width=0.48\textwidth]{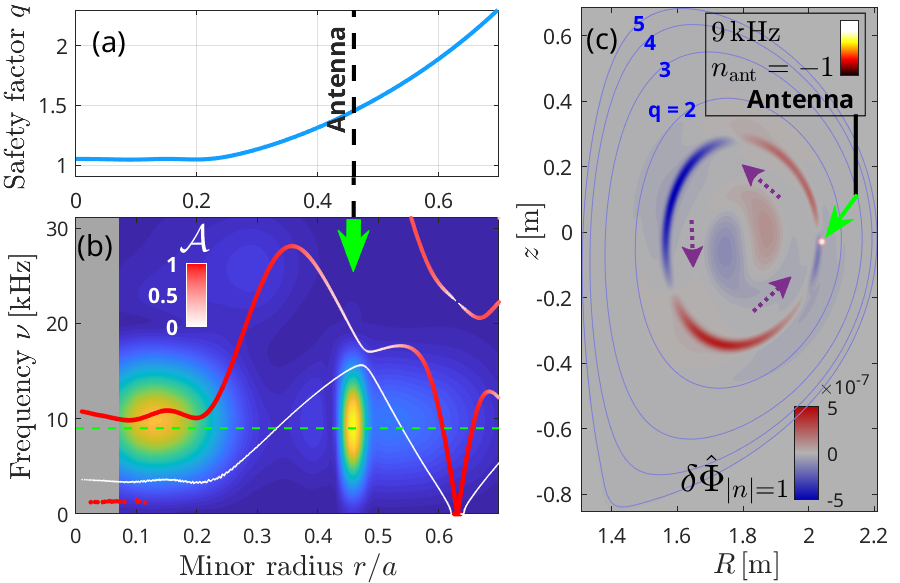}\vspace{-0.3cm}
	\caption{Demonstration of the possibility of exciting a coherent Alfv\'{e}nic quasi-mode near the central continuum plateau in KSTAR Model 1 via a resonant but remote antenna as defined in Eq.~(\protect\ref{eq:setup_ant}). The antenna has a narrow width of $w_{\rm ant} = 0.02\,{\rm m}$ and is centered around $(R_{\rm ant},z_{\rm ant}) = (2.04\,{\rm m}, -0.03\,{\rm m})$ on the outer midplane near mid-radius ($\hat{r}_{\rm ant} \approx 0.45$). It drives fluctuations with toroidal mode number $n_{\rm ant} = -1$ at frequency $\nu_{\rm ant} = 9\,{\rm kHz}$. Panel (a) shows the safety factor profile $q(r)$ in the relevant region. Panel (b) shows the spectrogram of the electrostatic potential $|\delta\hat{\Phi}_{|n|=1}|(r,\nu)$ computed as in Fig.~\protect\ref{fig:06_kstar-1.051_decay-ant_spec}. Panel (c) shows a snapshot of the potential's spatial structure $\delta\hat{\Phi}_{|n|=1}(R,z)$ in the poloidal plane at $\zeta=0$. The waves propagate counter-clockwise (arrows). Magnetic surfaces with $q = 2,3,4,5$ are drawn as blue contours and labeled.}\vspace{-0.6cm}
	\label{fig:07_kstar-1.051_late_spec-pol}%
\end{figure}

\begin{figure*}
	[tb]\vspace{-1.7cm}
	\centering
	\includegraphics[width=0.96\textwidth]{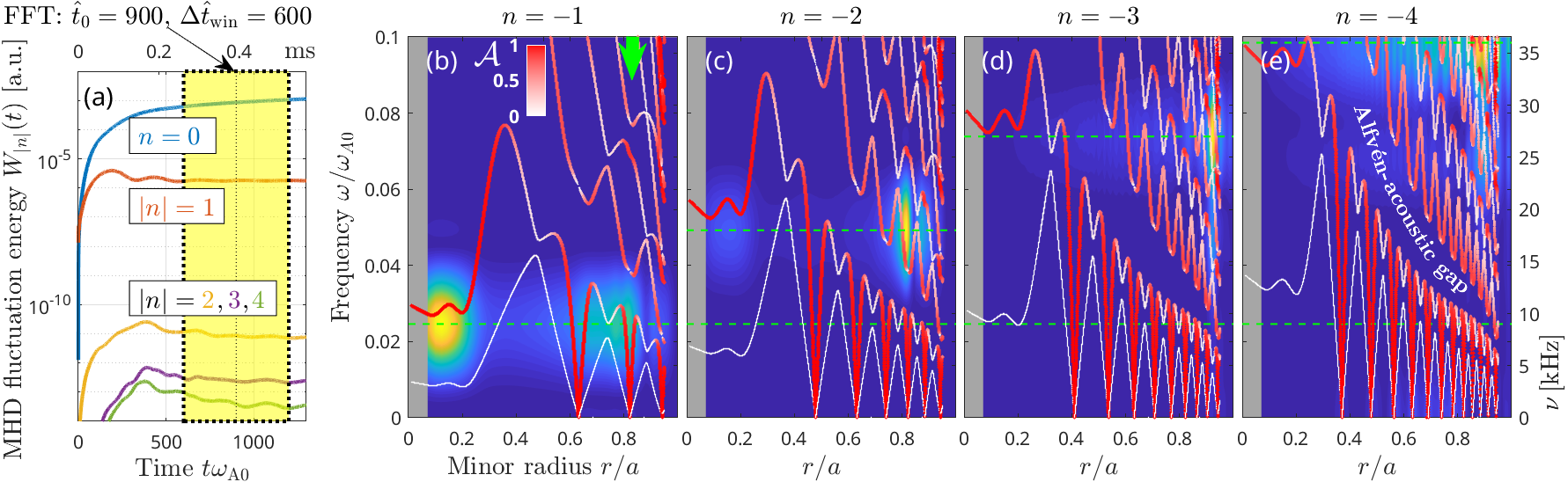}\vspace{-0.3cm}
	\caption{Fundamental component $|n|=1$ and higher toroidal harmonics $|n|>1$ of the MHD plasma response in KSTAR Model 1 driven by a narrow antenna with toroidal mode number $n_{\rm ant} = -1$, frequency $\nu_{\rm ant} = 9\,{\rm kHz}$ and width $w_{\rm ant} = 0.02\,{\rm m}$, here located near the plasma edge at $\hat{r}_{\rm ant} = 0.85$, as indicated by the green arrow at the top of panel (b). Panel (a) shows the temporal evolution of the MHD fluctuation energy $W_{|n|}(t)$ for $|n|=0,1,2,3,4$. Panels (b)--(e) show respective spectrograms $|\delta\hat{\Phi}_{|n|}|(r,\omega)$ for $-n=1,2,3,4$ in comparison with continua computed by {\tt FALCON}, whose color indicates strong (red) and weak (white) Alfv\'{e}nicity $\A$. The spectrograms have a Nyquist frequency of about $\omega_{\rm Nyq} \approx 0.6\,\omega_{\rm A0} \approx 2\pi\times 220\,{\rm kHz}$ and were computed by Fourier-analyzing the signals in the time window $\hat{t}_0 \pm \Delta\hat{t}_{\rm win}/2 = 900 \pm 600/2$, which is highlighted by a yellow vertical band in panel (a). This simulation was performed with enhanced spatial resolution in the toroidal angle: $N_R\times N_z\times N_\zeta = 400\times 400\times 128$.}\vspace{-0.2cm}
	\label{fig:08_kstar-1.051_n1234}%
\end{figure*}

\begin{figure*}
	[tbp]%\vspace{-0.2cm}
	\centering
	\includegraphics[width=0.96\textwidth]{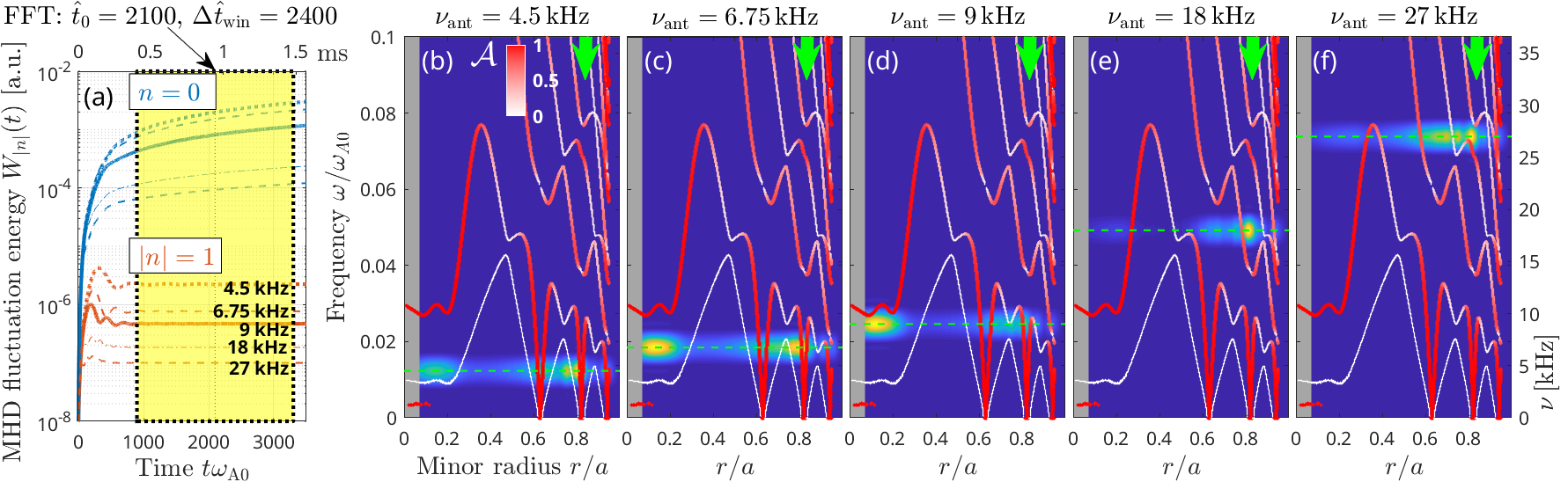}\vspace{-0.3cm}
	\caption{Scan of the antenna frequency $\nu_{\rm ant} = (4.5,\, 6.75,\, 9,\, 18,\, 27)\,{\rm kHz}$ in KSTAR Model 1. Other antenna parameters are fixed at their default values in Table~\protect\ref{tab:setup_ant}. Panel (a) shows the temporal evolution of the MHD fluctuation energy $W_{|n|}(t)$ for $n=0$ (blue) and $|n|=1$ (orange) for all values of $\nu_{\rm ant}$. Panels (b)--(f) show the respective spectrograms $|\delta\hat{\Phi}_{|n|=1}|(r,\omega)$ in comparison with continua computed by {\tt FALCON}, whose color indicates strong (red) and weak (white) Alfv\'{e}nicity $\A$. These spectrograms were computed by Fourier analyzing the signals in the time window $\hat{t}_0 \pm \Delta\hat{t}_{\rm win}/2 = 2100 \pm 2400/2$, which is highlighted by a yellow band in panel (a). Results of an extended antenna scan to higher frequencies $\nu \sim \O(100\,{\rm kHz})$ can be found in Fig.~\protect\ref{fig:b10_kstar-1.051_scan-hi-f_n1} of Appendix~\protect\ref{apdx:modeform_highfreq}.}
	\label{fig:09_kstar-1.051_scan-f_n1}\vspace{-0.5cm}
\end{figure*}

% -----------------------------------------------------------------------------
\subsection{``Action at a distance''}\vspace{-0.1cm}
\label{sec:results_distance}

Let us emphasize again that the simulation setup in Fig.~\ref{fig:03_kstar_profiles} is deliberately prepared in a way that facilitates the resonant excitation of a core-localized low-frequency mode around $9...10\,{\rm kHz}$ by shaping the $q$ profile in Fig.~\ref{fig:03_kstar_profiles}(b) such that the continuous spectrum of MHD waves has a flat piece of Alfv\'{e}nically polarized continuum with Fourier component $|m/n| = 1/1$ in the central region $\hat{r} \lesssim 0.2$. Moreover, instead of driving such plateaus directly as in our introductory examples in Figs.~\ref{fig:05_kstar-1.051_decay-ant_early} and \ref{fig:06_kstar-1.051_decay-ant_spec} above, the main purpose of this work is to study the nonlocal action of an antenna when it is placed some distance away from the responsive domain. An example is shown in Fig.~\ref{fig:07_kstar-1.051_late_spec-pol}, where a $9\,{\rm kHz}$ antenna is located on the outer midplane at radius $\hat{r}_{\rm ant} \approx 0.45$. The spectrogram in Fig.~\ref{fig:07_kstar-1.051_late_spec-pol}(b) still shows a clear response near the $|m/n| = 1/1$ Alfv\'{e}nic continuum plateau in the central core around $\hat{r} \lesssim 0.2$.

Fig.~\ref{fig:07_kstar-1.051_late_spec-pol} demonstrates that the MHD model allows an antenna to act from a distance. The present paper shall throw some light on the properties and parameter dependence of this phenomenon. The early transients and mode structure formation process are more or less the same as what we described in Section~\ref{sec:results_transients} above.\footnote{See Appendix~\protect\ref{apdx:modeform_transients} for a similar analysis using a remote antenna.}
We assume that, in the manner of Huygens' principle, those processes continue to operate in the background throughout the simulation, at each time step and every grid point, in microscopic amplitude increments. In the remainder of the present Section~\ref{sec:results}, we characterize the emerging quasi-steady state.

The spatial structure of $\delta\Phi_{|n|=1}(R,z)$ in the quasi-steady state is shown in Fig.~\ref{fig:07_kstar-1.051_late_spec-pol}(c). The flux surfaces near the antenna location exhibit a non-normal spatio-temporal structure dominated by a few poloidal harmonics satisfying $m \approx |n| q(r_{\rm ant}) \approx 1.5$, namely $m=1$ and $2$. The attribute ``non-normal'' refers to the fact that such a structure would not survive in the absence of the antenna. In contrast, the central component of $\delta\Phi_{|n|=1}$ in Fig.~\ref{fig:07_kstar-1.051_late_spec-pol}(c) is expected to be more robust as its frequency and spatial structure is supported by the $m \approx |n| q(0) \approx 1$ continuum plateau.

\begin{figure*}
	[tb]\vspace{-1.5cm}
	\centering
	\includegraphics[width=0.96\textwidth]{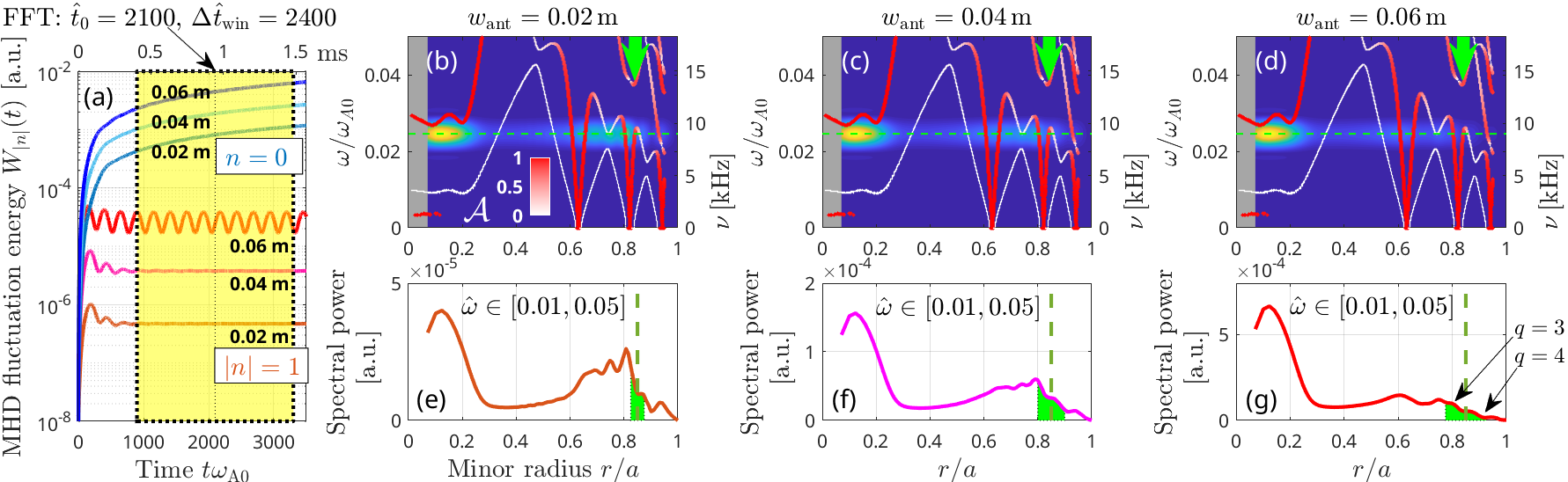}\vspace{-0.2cm}
	\caption{Scan of the antenna width $w_{\rm ant} = (0.02,\, 0.04,\, 0.06)\,{\rm m}$ in KSTAR Model 1. Other antenna parameters are fixed at their default values in Table~\protect\ref{tab:setup_ant}. Panel (a) shows the temporal evolution of the MHD fluctuation energy $W_{|n|}(t)$ for $n=0$ (shades of blue) and $|n|=1$ (shades of red) for all values of $w_{\rm ant}$. For $n=-1$, panels (b)--(d) show the respective spectrograms $|\delta\hat{\Phi}_{|n|=1}|(r,\omega)$ in comparison with {\tt FALCON} continua. The FFT window is highlighted yellow in panel (a). Panels (e)--(g) show the respective radial profiles of the fluctuation magnitude $\left<\smash{\delta\hat\Phi_{|n|=1}}\right>_\omega(r)$ defined in Eq.~(\protect\ref{eq:dg_spec_wband}), integrated over the frequency band $\hat{\omega} \equiv \omega/\omega_{\rm A0} \in [0.01, 0.05]$.}
	\label{fig:10_kstar-1.051_scan-w_n1}\vspace{-0.3cm}
\end{figure*}

Results obtained when the antenna was placed near $\hat{r}_{\rm ant} \approx 0.85$ (i.e., closer to the plasma edge), are shown in Fig.~\ref{fig:08_kstar-1.051_n1234} for toroidal mode numbers $|n|=1,2,3,4$. Since our antenna field contains only a single toroidal harmonic, $n_{\rm ant} = -1$, the higher harmonics $|n|>1$ in Fig.~\ref{fig:08_kstar-1.051_n1234} must arise from MHD nonlinearities, which are weak here due to the weak antenna field ($\hat{A}_{\rm ant} = 2\times 10^{-8}$). One can see in Fig.~\ref{fig:08_kstar-1.051_n1234} that the resulting frequencies satisfy $\nu_{|n|>1} = n\nu_{|n|=1}$. Moreover, one can see that, as $|n|$ increases, the plasma response near the central continuum plateau ($\hat{r} \lesssim 0.2$) becomes fainter and eventually invisible relative to the fluctuations near the antenna. In addition to possibly insufficient spatial resolution near the antenna (note the dense continua at large radii in Fig.~\ref{fig:08_kstar-1.051_n1234}(e)), we can think of several possible physical reasons, why edge-core coupling appears to become weaker with increasing $n$ in Fig.~\ref{fig:08_kstar-1.051_n1234}:
\begin{itemize}
	\item  One possible factor is that coherence is lost more rapidly with increasing distance from the source when the wavelength is short and magnetic shear is strong.
	\item  Another factor is the stronger field line bending associated with higher $|n|$, which consumes more energy near the antenna, leaving less for distant regions.
	\item  Finally, we note that only for $|n|=1$ the antenna's near-field at $\hat{r}_{\rm ant} = 0.85$ and its distant receiver at $\hat{r} \lesssim 0.2$ are both located below the Alfv\'{e}n-acoustic gap, where Alfv\'{e}nic and acoustic waves have comparable poloidal wavenumbers $k_\vartheta = m/r$ at a given frequency. In contrast, the antenna field's $|n| > 1$ harmonics lie above that gap, so we expect its polarization to become more Alfv\'{e}nic, while the core-localized plateau remains in the Alfv\'{e}n-acoustic band.
\end{itemize}

\noindent In any case, due to the weakness of $|n| > 1$ harmonics the following spectral analyses will focus on $|n|=1$ only.

In the following Sections~\ref{sec:results_scan-f}--\ref{sec:results_scan-r_focus}, we scan the antenna frequency $\nu_{\rm ant}$, its radial width $w_{\rm ant}$, and its location $r_{\rm ant}$ in order to gather more information about such antenna-plasma interactions with distant radial coupling. Section~\ref{sec:results_scan-r_focus} also contains a quantitative analysis of the volumetric focusing or dilution of wave energy propagating radially inward or outward. These results will be used later in Section~\ref{sec:discuss} to discuss the physics underlying such radial coupling in the visco-resistive full MHD model used here.

% -----------------------------------------------------------------------------
\subsection{Scan of the antenna frequency $\nu_{\rm ant}$}%\vspace{-0.2cm}
\label{sec:results_scan-f}

For KSTAR Model 1, which has one continuum plateau in the plasma center, Fig.~\ref{fig:09_kstar-1.051_scan-f_n1} shows results of simulations performed with narrow ($w_{\rm ant} = 0.02\,{\rm m}$) edge-localized ($\hat{r}_{\rm ant} = 0.85$) antennas oscillating at five different frequencies: $\nu_{\rm ant} = (4.5,\, 6.75,\, 9,\, 18,\, 27)\,{\rm kHz}$. 

The evolution of the axisymmetric MHD fluctuation energy $W_{n=0}(t)$, plotted as bluish lines in Fig.~\ref{fig:09_kstar-1.051_scan-f_n1}(a), shows how the antenna continuously heats the plasma. Meanwhile, the energy $W_{|n|=1}(t)$ of the $|n|=1$ component saturates after about $1000$ Alfv\'{e}n times ($\approx 0.44\,{\rm ms}$).\footnote{One may wonder why the final quasi-steady values of $W_{|n|=1}(t)$ in Fig.~\protect\ref{fig:09_kstar-1.051_scan-f_n1}(a) decrease with increasing antenna frequency $\nu_{\rm ant}$. Moreover, scans to higher frequencies in Fig.~\protect\ref{fig:b10_kstar-1.051_scan-hi-f_n1} of Appendix~\protect\ref{apdx:modeform_highfreq} show that this trend is reversed around $300\,{\rm kHz}$. We see two likely reasons for these anomalies. First, our antenna is usually nonresonant with the local plasma response, which means that it is likely to undo some of the waves it has previously excited, as can be seen in Fig.~\protect\ref{fig:07_kstar-1.051_late_spec-pol}(c). Second, our small antenna is poorly resolved by the spatial mesh in our simulation, which affects the way it couples with the plasma. We did not verify this hypothesis because it is computationally expensive and not really important here. We shall only note that our antenna model in Eq.~(\protect\ref{eq:setup_ant_dB}) differs from Faraday's law $\partial_t{\bm B} = -\nablab\times{\bm E}$ by a factor $\partial_t \rightarrow -i\omega$. We thus have $\delta B_{\rm ant} \propto \hat{A}_{\rm ant}/\omega_{\rm ant}$ with fixed antenna amplitude $\hat{A}_{\rm ant}$. This should cancel the usual $\omega^2$-dependence of electromagnetic field energy, so we may expect $W_{|n|=1}$ to be independent of the antenna frequency $\omega_{\rm ant}$ in a hypothetical simulation with perfect resolution, in the absence of nonlinearities and dissipation, and without the above-mentioned ``reabsorption''.}

After the $|n|=1$ fluctuations saturated, we computed their spectrograms $|\delta\hat{\Phi}_{|n|=1}|(r,\omega)$, whose parula-colored contours are shown in Fig.~\ref{fig:09_kstar-1.051_scan-f_n1}(b)--(f) together with continuous spectra computed by {\tt FALCON}. Although the antenna is located near the edge, Fig.~\ref{fig:09_kstar-1.051_scan-f_n1}(d) shows that it can cause a strong core response when its frequency $\nu_{\rm ant}$ is near the frequency of the Alfv\'{e}nic $|m/n| = 1/1$ continuum branch around $\hat{r} \lesssim 0.2$, $\omega \approx 0.025\, \omega_{\rm A0} \approx 2\pi\times 10\,{\rm kHz}$. A clearly visible response is seen even in Figs.~\ref{fig:09_kstar-1.051_scan-f_n1}(b,c) with lower antenna frequencies $\nu_{\rm ant} = 4.5\,{\rm kHz}$ and $6.75\,{\rm kHz}$ (we will return to this point in Section~\ref{sec:discuss_lowfreq}). In contrast, when the antenna frequency is increased to $18\,{\rm kHz}$ and $27\,{\rm kHz}$, Figs.~\ref{fig:09_kstar-1.051_scan-f_n1}(d,e) show only a weak response near the $1/1$ continuum branch, presumably due to its steep slope that implies strong radial phase mixing (``continuum damping'').

% -----------------------------------------------------------------------------
\subsection{Scan of the antenna width $w_{\rm ant}$}%\vspace{-0.2cm}
\label{sec:results_scan-w}

For KSTAR Model 1, which has one continuum plateau in the plasma center, Fig.~\ref{fig:10_kstar-1.051_scan-w_n1} shows results of simulations performed with an edge-localized ($\hat{r}_{\rm ant} = 0.85$) antenna with frequency $\nu_{\rm ant} = 9\,{\rm kHz}$ and three different spatial widths: $w_{\rm ant} = (0.02,\, 0.04,\, 0.06)\,{\rm m}$.

The evolution of the axisymmetric MHD fluctuation energy $W_{n=0}(t)$, plotted as bluish lines in Fig.~\ref{fig:10_kstar-1.051_scan-w_n1}(a), shows again how the antenna heats the plasma, and the energy $W_{|n|=1}(t)$ of the antenna-driven $|n|=1$ component saturates within $1000$ Alfv\'{e}n times ($\approx 0.44\,{\rm ms}$). As expected, the deposited energy increases with increasing width $w_{\rm ant}$.

\begin{figure*}
	[tb]
	\centering\vspace{-1.5cm}
	\includegraphics[width=0.96\textwidth]{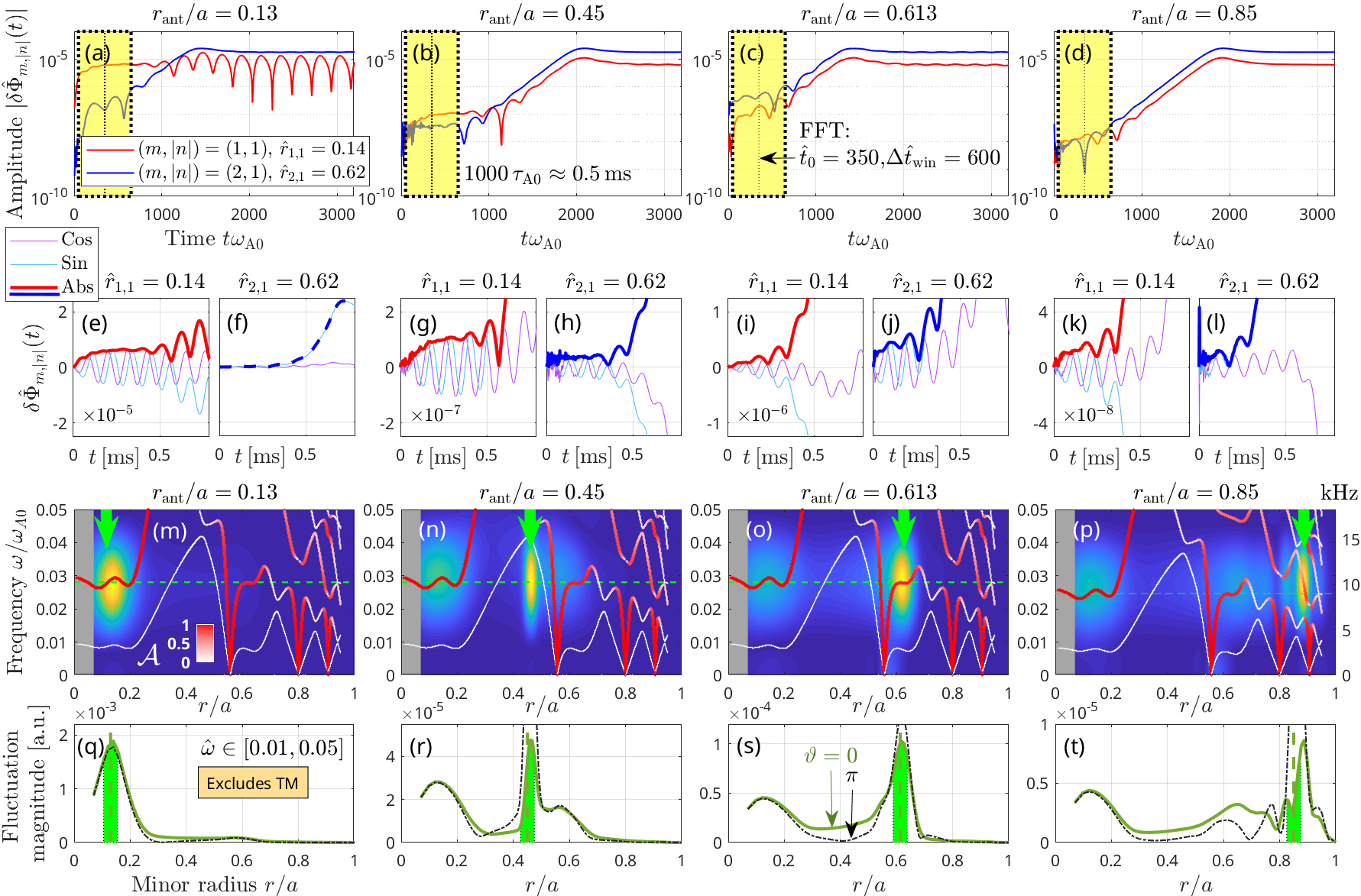}\vspace{-0.2cm}
	\caption{Scan of the antenna location $\hat{r}_{\rm ant} = 0.13,\, 0.45,\, 0.613,\, 0.85$ in KSTAR Model 2 with two continuum plateaus: one near the center ($\hat{r} \lesssim 0.2$) and one near mid-radius ($0.6 \lesssim \hat{r} \lesssim 0.65$). Other antenna parameters are fixed at their default values in Table~\protect\ref{tab:setup_ant}. Panels (a)--(d) in the first row show the temporal evolution of the magnitude $|\delta\hat{\Phi}_{m,|n|}|(t)$ of the electrostatic potential's Fourier components $|m/n|=1/1$ (red) and $2/1$ (blue), measured at $\hat{r}_{1,1} = 0.14$ and $\hat{r}_{2,1} = 0.62$, respectively. The same data are plotted again in the second row, panels (e)--(l), together with the complete waveforms of the cosine (magenta) and sine (cyan) components, $\delta\hat{\Phi}_{m,|n|}^{\rm (c)}(t)$ and $\delta\hat{\Phi}_{m,|n|}^{\rm (s)}(t)$. For each case, panels (m)--(p) in the third row show the respective spectrograms $|\delta\hat{\Phi}_{|n|=1}|(r,\omega)$ evaluated at the outer midplane ($\vartheta = 0$) in comparison with {\tt FALCON} continua. Panels (q)--(t) in the fourth row show the respective radial profiles of the magnitude $\left<\smash{\delta\hat\Phi_{|n|=1}}\right>_\omega(r)$ defined in Eq.~(\protect\ref{eq:dg_spec_wband}), integrated over the frequency band $\hat{\omega} \equiv \omega/\omega_{\rm A0} \in [0.01, 0.05]$. As indicated by the arrows in panel (s), the solid green curves are obtained from the spectrograms at the outer midplane ($\vartheta = 0$), and the dash-dotted black lines are measured at the inner midplane ($\vartheta = \pi$). The antenna frequency $\nu_{\rm ant} = 9\,{\rm kHz}$ and its locations $r_{\rm ant}$ are indicated by green arrows and dashed lines in panels (m)--(t) of the third and fourth row. Note that the present KSTAR Model 2 is subject to a slowly growing tearing mode (TM) at the $q = 2$ surface ($\hat{r} \approx 0.56$), which begins to dominate after about $600$ Alfv\'{e}n times ($0.3\,{\rm ms}$), forcing us to use a relatively short FFT window $\hat{t}_0 \pm \Delta\hat{t}_{\rm win}/2 = 350 \pm 600/2$, which is highlighted yellow in the top panels (a)--(d). This tearing mode, details of which can be found in Appendix~\protect\ref{apdx:dissipation_tm}, also affects measurements of $\delta\hat{\Phi}_{2,1}^{\rm (c,s)}(t)$ in the upper two rows, especially in panels (j) and (l), but it is largely excluded in panels (q)--(t) by via the band-pass filter with $\hat{\omega} \in [0.01, 0.05]$.}\vspace{-0.3cm}
	\label{fig:11_kstar-q1.051-flat-q2.15_rant-scan}%
\end{figure*}

After the $|n|=1$ fluctuations saturated, we computed their spectrograms $|\delta\hat{\Phi}_{|n|=1}|(r,\omega)$, whose parula-colored contours are shown in Fig.~\ref{fig:10_kstar-1.051_scan-w_n1}(b)--(d) together with continuous spectra computed by {\tt FALCON}. The result for the narrow antenna $w_{\rm ant} = 0.02\,{\rm m}$ from Fig.~\ref{fig:09_kstar-1.051_scan-f_n1}(d) is plotted again in Fig.~\ref{fig:10_kstar-1.051_scan-w_n1}(b). Panels (c) and (d) show that, independently of the antenna width $w_{\rm ant}$, a very similar coherent response occurs near the central plateau of the Alfv\'{e}nic $|m/n| = 1/1$ continuum branch around $\hat{r} \lesssim 0.2$ with frequency $\omega \approx 0.025\, \omega_{\rm A0} \approx 2\pi\times 10\,{\rm kHz} \approx \nu_{\rm ant} = 9\,{\rm kHz}$.

As may be expected, the fluctuations near the antenna become smoother and the relative amplitude in that region decreases with increasing width $w_{\rm ant}$. This is shown more clearly in Fig.~\ref{fig:10_kstar-1.051_scan-w_n1}(e)--(g), where we plotted the radial profiles of the fluctuation magnitude integrated over the frequency band $\omega \in [0.01,0.05]\,\omega_{\rm A0} \approx 2\pi\times[4,18]\,{\rm kHz}$ as\vspace{-0.3cm}
\begin{equation}
	\left<\smash{\delta\hat\Phi_n}\right>_\omega(r) \equiv \left[\int_{\omega_1}^{\omega_2}{\rm d}\omega\,|\delta\hat{\Phi}_n|^2(r,\omega)\right]^{1/2}.
	\label{eq:dg_spec_wband}
\end{equation}

The results in Fig.~\ref{fig:10_kstar-1.051_scan-w_n1} tell us that the phenomena we are studying in this paper are qualitatively independent of the antenna width and that we may continue to work with the narrow antenna $w_{\rm ant} = 0.02\,{\rm m}$, even though it is barely resolved by our computational mesh. The advantage of the narrow antenna is that it directly perturbs only a small range of magnetic flux surfaces with little variation in magnetic pitch. In contrast, a wider antenna is more likely to drive multiple continuum branches with different $m$ numbers. We believe this to be the reason for the strong pulsations (beating) of $W_{|n|=1}(t)$ in Fig.~\ref{fig:10_kstar-1.051_scan-w_n1}(a) for the wider antenna with $w_{\rm ant} = 0.06\,{\rm m}$ at $\hat{r}_{\rm ant} = 0.85$, which overlaps with the $q=3$ and $q=4$ radii as indicated in Fig.~\ref{fig:10_kstar-1.051_scan-w_n1}(g). More information about beating in the antenna's near-field can be found in Appendix~\ref{apdx:modeform_pulse}.

% -----------------------------------------------------------------------------
\subsection{Scan of the antenna's radial location $r_{\rm ant}$}%\vspace{-0.2cm}
\label{sec:results_scan-r_focus}

For KSTAR Model 2, which has two continuum plateaus --- one in the center, $\hat{r} \lesssim 0.2$, and one near mid-radius, $0.6 \lesssim \hat{r} \lesssim 0.65$ --- Fig.~\ref{fig:11_kstar-q1.051-flat-q2.15_rant-scan} shows results of simulations performed with a narrow ($w_{\rm ant} = 0.02\,{\rm m}$) antenna with frequency $\nu_{\rm ant} = 9\,{\rm kHz}$ and four different radial locations: $\hat{r}_{\rm ant} = 0.13,\, 0.45,\, 0.613,\, 0.85$. Our KSTAR Model 2 is subject to an unstable $|m/n| = 2/1$ tearing mode at the $q=2$ surface ($\hat{r}_{q=2} \approx 0.55$), just inside of the outer plateau (see Figs.~\ref{fig:c02_kstar-q1.051-flat-q2.15_tearing} and \ref{fig:c03_kstar-1.051-flat-q2.15_dB} in Appendix~\ref{apdx:dissipation_tm} for details). Fortunately, there is a window of several 100 Alfv\'{e}n times during which the antenna-driven fluctuations dominate and can thus be analyzed. This time window is highlighted by yellow boxes in panels (a)--(d) in the upper row of Fig.~\ref{fig:11_kstar-q1.051-flat-q2.15_rant-scan}

The first column of Fig.~\ref{fig:11_kstar-q1.051-flat-q2.15_rant-scan} shows results obtained with $\hat{r}_{\rm ant} = 0.13$ located within the central continuum plateau. Panels (a), (e) and (f) show the evolution of
\begin{equation}
	\delta\hat{\Phi}_{m,1}(t) \equiv \delta\hat{\Phi}_{|n|=1}(t|\hat{r}_{m,1},m)
	\label{eq:philoc}
\end{equation}

\noindent for $n=-1$, which denotes the amplitude of the $m$-component of the electrostatic potential fluctuations at a certain radius $\hat{r}_{m,|n|}$. Here, the radii $\hat{r}_{1,1}=0.14$ and $\hat{r}_{2,1}=0.62$ are chosen to lie in the inner $(m=1)$ and outer ($m=2$) continuum plateau, respectively. At the inner plateau, we measured the $m=1$ component $\delta\hat{\Phi}_{1,1}(t)$, and at the outer plateau and we measured the $m=2$ component $\delta\hat{\Phi}_{2,1}(t)$. During the first $0.5\,{\rm ms}$ ($\approx 1000$ Alfv\'{e}n times), $\delta\hat{\Phi}_{1,1}$ in Fig.~\ref{fig:11_kstar-q1.051-flat-q2.15_rant-scan}(e) dominates and exhibits harmonic oscillations. Meanwhile, Fig.~\ref{fig:11_kstar-q1.051-flat-q2.15_rant-scan}(f) shows that $\delta\hat{\Phi}_{2,1}$ at the outer plateau remains negligibly small until the tearing mode emerges.

The second column of Fig.~\ref{fig:11_kstar-q1.051-flat-q2.15_rant-scan} shows results obtained with $\hat{r}_{\rm ant} = 0.45$, closer to but not quite in the outer continuum plateau. The time traces in panels (b), (g) and (h) as well as the spectrogram in panel (n) and its band-pass-filtered radial profile in panel (r) all show that the fluctuations $\delta\hat{\Phi}_{1,1}$ on the inner plateau are stronger by about a factor 2 than $\delta\hat{\Phi}_{2,1}$ on the outer plateau, in spite of the fact that the inner plateau is further away from the antenna.

The fluctuations $\delta\hat{\Phi}_{2,1}$ at the outer plateau are seen to dominate here only in the third column of Fig.~\ref{fig:11_kstar-q1.051-flat-q2.15_rant-scan}, where the antenna directly drives that domain. It seems that the antenna also slightly enhances or accelerates the emergence of the tearing mode, traces of which can be seen in the spectrogram in panel (o) near $\omega = 0$. In order to exclude the tearing mode's contribution --- which also affects the magnitudes of $\delta\hat{\Phi}_{m,|n|}(t)$ in panels (i) and (j) --- we have plotted in the bottom row the profiles $\left<\smash{|\delta\hat\Phi_{|n|=1}|}\right>_\omega(r)$ defined in Eq.~(\ref{eq:dg_spec_wband}), here integrated over the frequency band $\omega \in [0.01,0.05]\,\omega_{\rm A0} \approx 2\pi\times[4,18]\,{\rm kHz}$. The result in panel (s) shows that $\left<\smash{|\delta\hat\Phi_{|n|=1}|}\right>_\omega(\hat{r}_{1,1}=0.13)$ reaches nearly half the magnitude of $\left<\smash{|\delta\hat\Phi_{|n|=1}|}\right>_\omega(\hat{r}_{2,1}=0.62)$, in spite of the fact that the latter is directly driven by the antenna while the former responds from a distance. This stands in stark contrast to the reverse situation in Fig.~\ref{fig:11_kstar-q1.051-flat-q2.15_rant-scan}(q).

Finally, we moved the antenna further towards the plasma boundary and applied the drive at $\hat{r}_{\rm ant} = 0.85$. The results for this case are summarized in the right column of Fig.~\ref{fig:11_kstar-q1.051-flat-q2.15_rant-scan}. One can see that the remote antenna is able to drive fluctuations at both the inner and outer continuum plateaus. The tearing mode is again enhanced and, thus, biases the measurements of $\delta\hat{\Phi}_{1,1}$ and $\delta\hat{\Phi}_{2,1}$ in panels (k) and (l). The band-pass filtered profile $\left<\smash{|\delta\hat\Phi_{|n|=1}|}\right>_\omega(r)$ in Fig.~\ref{fig:11_kstar-q1.051-flat-q2.15_rant-scan}(t) is more suitable for a quantitative comparison and shows that the magnitude of the plasma response at the core-localized plateau is about $1.4$ to $2.2$ times stronger than the response at the outer plateau. These ratios differ significantly on the outer ($\vartheta = 0$) and inner midplanes ($\vartheta = \pi$) as one can see by comparing the solid green and dash-dotted black curves in the bottom row of Fig.~\ref{fig:11_kstar-q1.051-flat-q2.15_rant-scan}. Such poloidal asymmetries in the mode structures are typical in our simulations and indicate that the radial channeling and volumetric focusing paradigms introduced in Section~\ref{sec:history} manifest themselves in nontrivial ways in the nontrivial geometry of our system.

\begin{figure}
	[tb]
	\centering\vspace{-0.7cm}
	\includegraphics[width=0.48\textwidth]{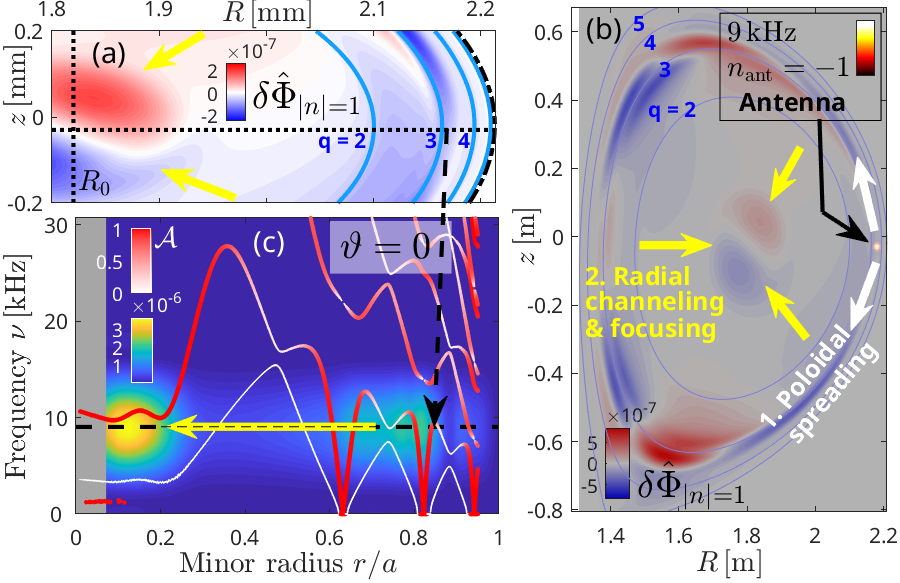}\vspace{-0.3cm}
	\caption{Illustration of volumetric focusing of antenna-driven MHD waves in our KSTAR case Model 1. This is an adaptation of Fig.~\protect\ref{fig:07_kstar-1.051_late_spec-pol}, here with the antenna located at $(R,z)_{\rm ant} = (2.18\,{\rm m}, -0.03\,{\rm m})$ ($\hat{r}_{\rm ant} \approx 0.85$) on the outer midplane close to the plasma edge. This is the same case as in Figs.~\protect\ref{fig:09_kstar-1.051_scan-f_n1}(d) and \protect\ref{fig:10_kstar-1.051_scan-w_n1}(b,e). Panel (a) shows an enlargement of the $\delta\hat{\Phi}_{|n|=1}(R,z)$ field from panel (b) near the outer midplane ($\vartheta=0$), where the spectrogram in panel (c) was computed. The area shown in panel (a) is similar to that covered by the ECEI diagnostics in Fig.~\protect\ref{fig:01_kstar_ecei}(b,d), and the horizontal $R$-axis in (a) is roughly aligned to scale with the horizontal $r$-axis in (c), so that the magnetic axis $R=R_0$ (dotted) matches $r=0$, and the plasma boundary (dash-dotted) matches $r/a=1$.As discussed in the text of Section~\protect\ref{sec:discuss_focus}, the poloidal localization of our antenna suggests that the drive is realized here in two stages: 1.\ Poloidal spreading of shear Alfv\'{e}n waves on flux surfaces intersecting the antenna (white arrows). 2.\ Radial channeling and focusing of energy from the established annular shear Alfv\'{e}n wave field (yellow arrows).}
	\label{fig:12_focus_kstar}\vspace{-0.5cm}
\end{figure}

% =============================================================================
\section{Discussion}%\vspace{-0.3cm}
\label{sec:discuss}

% -----------------------------------------------------------------------------
\subsection{Volumetric focusing}%\vspace{-0.2cm}
\label{sec:discuss_focus}

The trend observed in Fig.~\ref{fig:11_kstar-q1.051-flat-q2.15_rant-scan} and discussed in Section~\ref{sec:results_scan-r_focus} above is consistent with the notion of volumetric focusing for inward drive and volumetric dilution for outward drive. The concept was schematically illustrated in Fig.~\ref{fig:02_focus_schematic} of the introductory Section~\ref{sec:history} from a generic point of view, and its application to one of our simulations for the KSTAR tokamak is illustrated in Fig.~\ref{fig:12_focus_kstar}.

It seems safe to conclude that outward drive is generally less efficient because the power density is necessarily reduced as energy is transferred from smaller to larger values of $r$ and, thus, spread over a larger volume. Drawing conclusions about the efficiency of inward drive is less trivial and the schematic yellow arrows in Fig.~\ref{fig:12_focus_kstar} pointing straight inward require some discussion. These arrows are meant to illustrate the trend, but may be an inaccurate depiction of the underlying coupling and energy transfer processes. For instance, we observe that the ratio of the fluctuation amplitudes at the inner and outer continuum plateaus in Fig.~\ref{fig:11_kstar-q1.051-flat-q2.15_rant-scan} do not follow the $1/r$ scaling of the inverse volume $1/\Delta V(r) \propto 1/(R_0 r \Delta r)$ of toroidal shells with a given minor radial width $\Delta r$ in configuration space.\footnote{Of course, the $1/r$ scaling cannot be expected to hold exactly because the effective volume in magnetically confined plasmas is more adequately measured in terms of magnetic flux ($\Delta\Psi$) rather than spatial distance ($\Delta r$). Nevertheless, given the fact that here (and generally in typical tokamaks) the normalized toroidal flux and volume-averaged minor radius satisfy $\Delta\hat{\Psi} \approx 2\hat{r}\Delta\hat{r}$, the $1/r$ scaling should be a reasonable approximation, so the significant deviation from this scaling in Fig.~\protect\ref{fig:11_kstar-q1.051-flat-q2.15_rant-scan} points towards the existence of other contributing factors.}
One should thus also consider other factors, besides mere volumetric focusing and dilution, that influence the rate and coherence of spatial accumulation of wave energy that ultimately determines the fluctuation intensity at certain radii.

First, although we certainly have a {\it geometric} $1/r$ singularity, there is no guarantee that the wave vectors are perfectly aligned with $-\hat{\bm e}_r$, so there is no guarantee for an on-axis cusp to form in the wave field. That is why in catastrophe theory, such singularities are classified as non-robust (or ``unstable''): imperfect alignment will smear the focus. For instance, radial energy transfer rates may vary with $r$ and $\vartheta$ due to factors such as local magnetic shear, thermal pressure and plasma compressibility. It is even conceivable that energy transmissivity is radially asymmetric in the sense of favoring either inward or outward energy transfer (e.g., due to magnetic surface curvature). In short, we expect radial focusing to be partially compromised by ``optical aberrations'' caused by the plasma's internal nonuniformity and noncircular poloidal shape.

Second, the extreme poloidal localization of our numerical antenna (and, to a lesser degree, real antennas in real experiments) implies that volumetric focusing in our setup must be preceded by a relatively rapid poloidal spreading of antenna-driven fluctuations on nearby magnetic flux surfaces. The process is thus thought to involve Stages 1 and 2 that were introduced in Section~\ref{sec:results_transients} above and occur simultaneously but on different time scales:
\begin{enumerate}
	\item $\tau_\vartheta$: First, our poloidally and radially localized antenna deposits energy on the magnetic flux surfaces that it intersects. As indicated by the white arrows in Fig.~\ref{fig:12_focus_kstar}(b), the fluctuations spread from the antenna over the entire plasma circumference. This happens on the parallel Alfv\'{e}n time scale $\tau_\vartheta \sim qR/v_{\rm A} \sim 1\,\mu{\rm s}$, as can be observed in Fig.~\ref{fig:05_kstar-1.051_decay-ant_early}(c) of Section~\ref{sec:results_transients} and in Fig.~\ref{fig:b01_transient-early}(a,d,e) of Appendix~\ref{apdx:modeform_transients}.
	
	\item $\tau_\omega$: Second, a global mode structure forms in accordance with the given magnetic configuration and plasma profiles. This can be expected to occur on the time scale $\tau_\omega \gtrsim 0.5/\nu_{\rm ant}$ dictated by the universal uncertainty principle. Consistently with this threshold, in Fig.~\ref{fig:05_kstar-1.051_decay-ant_early}(d) of Section~\ref{sec:results_transients} and Fig.~\ref{fig:b02_transient-late}(a,b) of Appendix~\ref{apdx:modeform_transients} shows the appearance of the core response at $\nu_{\rm ant} \sim 10\,{\rm kHz}$ after about $50\,\mu{\rm s}$.
\end{enumerate}

\noindent Since $\tau_\vartheta \ll \tau_\omega$ for the low-frequency drive considered in this work ($\tau_\omega^{-1} = \nu_{\rm ant} \sim 10\,{\rm kHz}$), we can expect that the effective drive should not be seen as coming from the local antenna directly but from the entire fluctuating flux surfaces that intersect the local antenna. These fluctuating flux surfaces act like an effective annular antenna, images of which can be clearly seen in Figs.~\ref{fig:07_kstar-1.051_late_spec-pol}(c) and \ref{fig:12_focus_kstar}(b).

Indeed, we confirmed that the global mode structure --- whose details are shown in Fig.~\ref{fig:13_kstar-1.051_mstruc} and discussed in Section~\ref{sec:discuss_coupling} below --- is largely insensitive to the poloidal location of the antenna. Similar results (not shown) were obtained with an antenna placed on the high-field-side midplane ($\vartheta = \pi$). We consider this ability of the plasma to transform our poloidally localized drive into an effective ``surface antenna'' to be one of the factors that facilitate volumetric focusing in our present setup, partly justifying our drawing of radially-inward-pointing yellow arrows from several directions in Fig.~\ref{fig:12_focus_kstar}(a,b).

Conversely, we expect that volumetric focusing becomes less efficient at antenna frequencies exceeding $1/(10\tau_\vartheta) \sim \O(100\,{\rm kHz})$, because then $\tau_\omega < 10\tau_\vartheta$ and the poloidal localization of the antenna becomes increasingly noticeable on the wave oscillation time scale.\footnote{Figs.~\protect\ref{fig:b10_kstar-1.051_scan-hi-f_n1}, \protect\ref{fig:b11_transient_416kHz} of Appendix~\protect\ref{apdx:modeform_highfreq} show an example, where a remote antenna drives a core-localized $|m/n| = 0/1$ continuum plateau at $420\,{\rm kHz}$.}

% -----------------------------------------------------------------------------
\subsection{Core-edge coupling physics}
\label{sec:discuss_coupling}

In this section, we inspect some prominent structural features of our antenna-driven quasi-modes and discuss implications that seem relevant for our ongoing study of core-edge coupling physics that may contribute to the double-peaked fishbone phenomenon in Fig.~\ref{fig:01_kstar_ecei}.

\begin{figure}
	[tb]\vspace{-1.4cm}
	\centering
	\includegraphics[width=0.48\textwidth]{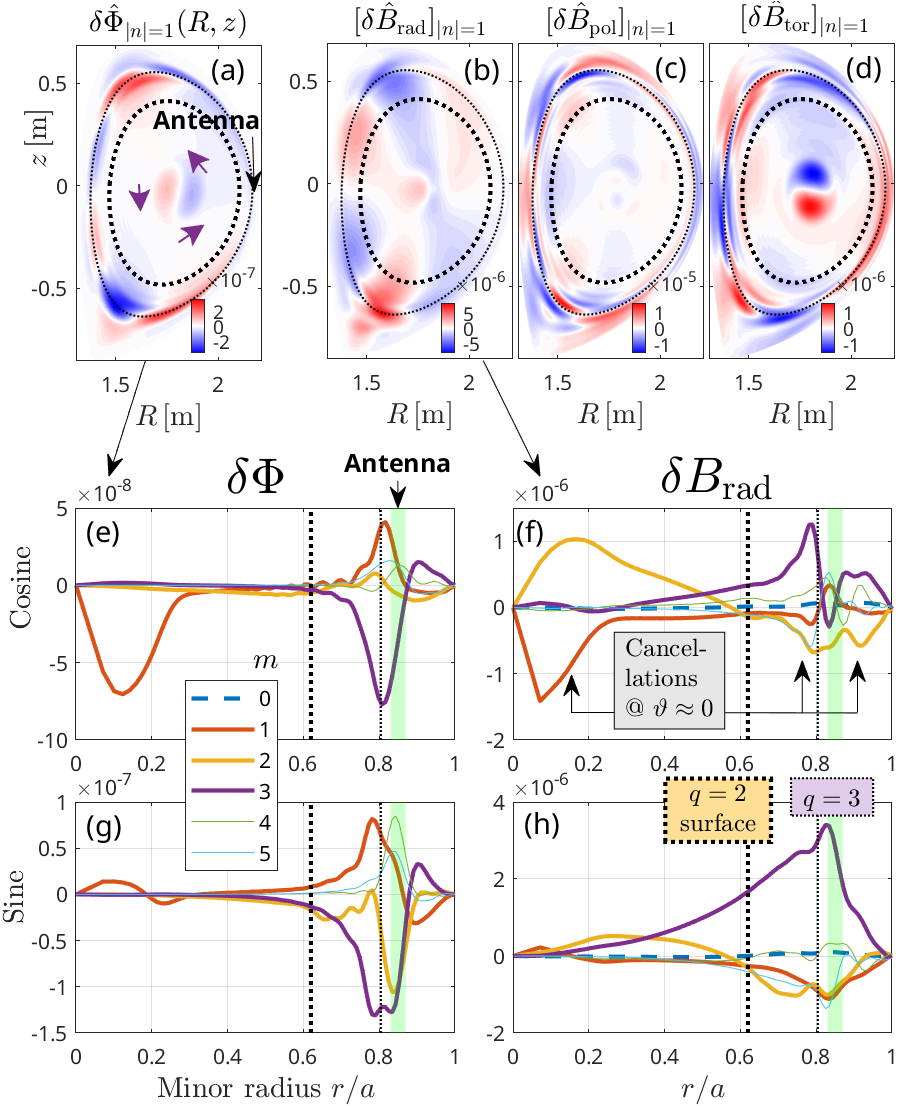}\vspace{-0.2cm}
	\caption{Mode structures of the electrostatic potential $\delta\hat{\Phi}_{|n|=1}$ and magnetic field perturbation $\delta\hat{\bm B}_{|n|=1}$ in KSTAR Model 1 driven by a narrow $9\,{\rm kHz}$ antenna near the edge ($\hat{r}_{\rm ant} = 0.85$). This is the same case as in Fig.~\protect\ref{fig:12_focus_kstar}, with the same snapshot time $t = 2100\,\omega_{\rm A0}^{-1} \approx 1\,{\rm ms}$ as in Fig.~\protect\ref{fig:09_kstar-1.051_scan-f_n1}(d). Panels (a)--(d) show the mode structures in the poloidal $(R,z)$ plane. It should be noted that these plots were obtained by synthesizing the $m/n=(0...12)/(-1)$ Fourier components. At the low frequencies considered here ($\sim 10\,{\rm kHz}$) this yields an accurate representation of the raw data, snapshots of which can be seen in Fig.~\protect\ref{fig:b02_transient-late} of Appendix~\protect\ref{apdx:modeform_transients}. Panels (e)--(h) show the respective cosine and sine components of individual poloidal Fourier components, $\delta\hat{\Phi}_{|n|=1}(r|m)$ and $[\delta\hat{B}_{\rm rad}]_{|n|=1}(r|m)$, for $m = 0...5$. Dotted lines indicate surfaces where $q=2,3$. The green-shaded area indicates the antenna location and width. The ``cancellations @ $\vartheta = 0$'' indicated in (f) are the Fourier-way of producing the inboard-ouboard asymmetry of $\delta\hat{B}_{\rm rad}$ in panel (b), which we attribute to the Shafranov shift of the magnetic axis: We think that $\delta\hat{B}_{\rm rad}$ is weaker on the low- and stronger on the high-field side because flux surfaces are respectively denser and sparser.}\vspace{-0.3cm}
	\label{fig:13_kstar-1.051_mstruc}\vspace{-0.3cm}
\end{figure}

We consider the case of Fig.~\ref{fig:12_focus_kstar}, where a $9\,{\rm kHz}$ antenna is located at $\hat{r}_{\rm ant} = 0.85$ near the plasma edge and resonates with the central $|m/n| = 1/1$ Alfv\'{e}nic low-frequency continuum plateau. The upper row of Fig.~\ref{fig:13_kstar-1.051_mstruc} shows the spatial structure of the electrostatic potential $\delta\Phi$ and the three components of the fluctuating magnetic field, $\delta B_{\rm rad}$, $\delta B_{\rm pol}$, and $\delta B_{\rm tor}$, in the poloidal $(R,z)$ plane. In addition, Figs.~\ref{fig:13_kstar-1.051_mstruc}(e)--(h) show the profiles of the main poloidal Fourier components $m=0,...,5$ of $\delta\hat{\Phi}_{|n|=1}$ and $[\delta\hat{B}_{\rm rad}]_{|n|=1}$. Here, the antenna is located on the outer midplane ($\vartheta = 0$), but similar results (not shown) were obtained when the antenna was placed on the inner midplane crossing ($\vartheta = \pi)$ of same magnetic flux surface.

Although we do not currently have complete explanations for all the features discussed below, we can say that they are consistently seen in all our simulations.

\subsubsection{Radial mode structure and parity}
\label{sec:discuss_coupling_radparity}

The $m=1$ component of $\delta\hat{\Phi}_{|n|=1}$ in Figs.~\ref{fig:13_kstar-1.051_mstruc}(e,g) has two distinct peaks, one near the antenna and one in the central core, the latter being the main subject of interest in this work. The $m>1$ components of $\delta\hat{\Phi}_{|n|=1}$ are localized near the antenna, with relatively short tails extending only barely inside the $q=2$ surface.

In contrast, $[\delta\hat{B}_{\rm rad}]_{|n|=1}$ in Figs.~\ref{fig:13_kstar-1.051_mstruc}(f,h) has both $m=1$ and $m=2$ peaking in the central core, with the $m=2$ component extending all the way to the $q=2$ surface (bold dotted line in Figs.~\ref{fig:13_kstar-1.051_mstruc}), slightly inside of which it crosses zero and changes sign. When viewed in isolation, this looks like a signature of (Doppler-shifted) tearing parity. However, in the present case, the $m=3$ component, which peaks at the antenna radius, has a long tail reaching deep into the plasma core across the $q=2$ surface. Due to this, the total mode structure in Fig.~\ref{fig:13_kstar-1.051_mstruc}(b) shows no evidence for $|m/n|=2/1$ tearing.\footnote{For comparison, an actual $2/1$ tearing mode as found in our KSTAR Model 2 is shown in Fig.~\protect\ref{fig:c02_kstar-q1.051-flat-q2.15_tearing} of Appendix~\protect\ref{apdx:dissipation_tm}.}
Meanwhile, the structure of $[\delta\hat{B}_{\rm rad}]_{|n|=1}$ bears resemblance to the radial displacement $\delta\xi_{\rm rad}$ predicted for GAEs as in Fig.~4(a) of Ref.~\cite{Appert82}, whose $m$-components are geometrically coupled here.

GAEs and tearing modes both constitute kink-like perturbations that can facilitate core-edge (or edge-core) radial coupling, especially in toroidal geometry. Although our antenna-driven fluctuations do not seem to classify as any particular pure eigenmode, which is not surprising for a stable plasma configuration, we believe that the plasma's mere capacity to support marginally stable kink-like perturbations facilitates the remote excitation of the quasi-modes seen in our simulations. We will return to this discussion in Section~\ref{sec:discuss_lowfreq} below.

\subsubsection{Phasing \& causation}
\label{sec:discuss_coupling_phasing}

Fig.~\ref{fig:13_kstar-1.051_mstruc} shows that the core- and edge-localized components of both $\delta\hat{\Phi}_{|n|=1}$ and $[\delta\hat{B}_{\rm tor}]_{|n|=1}$ exhibit a significant phase shift on the low-field side of the plasma, where the edge component seems to be leading and the core component seems to be trailing.

\begin{figure}
	[tb]\vspace{-1.8cm}
	\centering
	\includegraphics[width=0.48\textwidth]{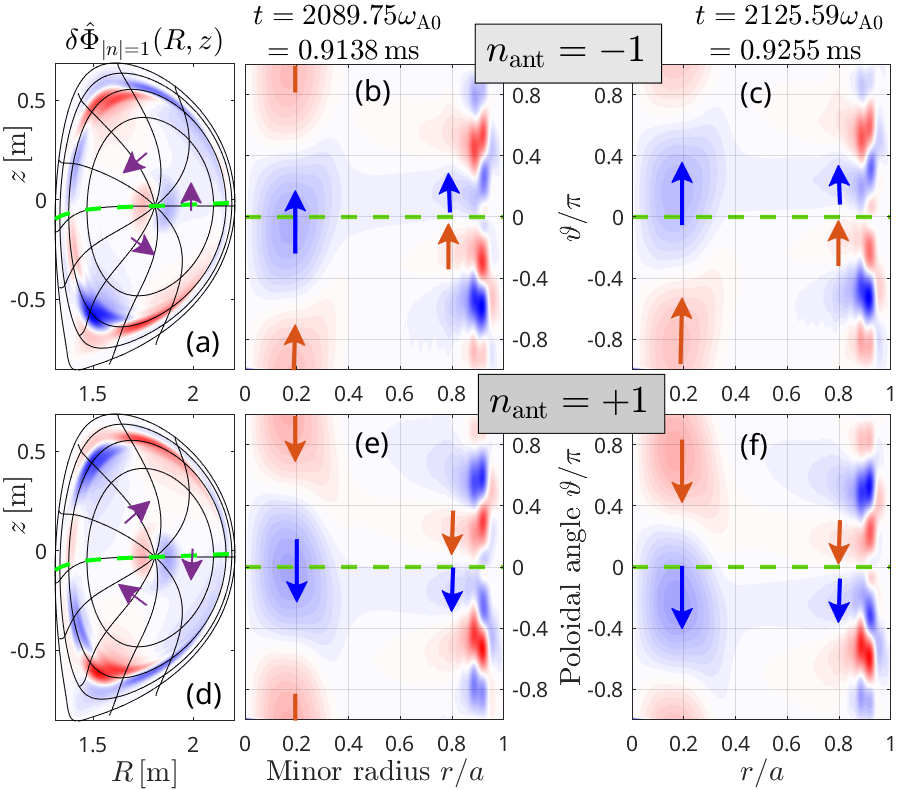}\vspace{-0.2cm}
	\caption{Detailed view of the phase relation between edge-localized waves near the $9\,{\rm kHz}$ antenna ($\hat{r}=0.85$) and the core-localized quasi-mode driven by it. Panel (a) shows the same snapshot $\delta\hat{\Phi}_{|n|=1}(R,z)$ of the electrostatic potential as Fig.~\protect\ref{fig:13_kstar-1.051_mstruc}(a), but here overlaid with the $\hat{r}={\rm const}$.\ and $\vartheta={\rm const}$.\ lines of the toroidal flux coordinates that were used to project the mode structure into the rectangular $(\hat{r},\vartheta)$-plane in panel (b). Panel (c) shows a snapshot taken about $12\,\mu{\rm s}$ ($\sim 10\%$ of the oscillation period) later. (More snapshots for this case can be found in Fig.~\ref{fig:b02_transient-late} of Appendix~\ref{apdx:modeform_transients}.) The dashed green line is the magnetic midplane, which coincides with $\vartheta = 0,\pi$ for the straight-field-line coordinates chosen here (commonly known as ``PEST coordinates''). The arrows indicate the propagation direction of the ${\bm E}\times{\bm B}$ vortices. Our default antenna has $n_{\rm ant} = -1$, implying propagation in the $+\vartheta$ (counter-clockwise) direction. Panels (d)--(f) show results obtained with $n_{\rm ant} = +1$, giving propagation in the $-\vartheta$ (clockwise) direction, as was the case for KSTAR's double-peaked fishbones in Fig.~\protect\ref{fig:01_kstar_ecei}(b,d).}\vspace{-0.55cm}
	\label{fig:14_kstar-1.051_phasing}%
\end{figure}

A clearer view of this is provided in Fig.~\ref{fig:14_kstar-1.051_phasing}, where the mode structure of the electrostatic potential $\delta\hat{\Phi}_{|n|=1}(R,z)$ (reproduced in the left column) was mapped into the rectangular plane spanned by the minor radius $\hat{r}$ and poloidal angle $\vartheta$. The upper panels (a) and (b) show the same data as in Fig.~\ref{fig:13_kstar-1.051_mstruc}(a), which were obtained with $n_{\rm ant}=-1$. Fig~\ref{fig:14_kstar-1.051_phasing}(c) shows another snapshot taken about $12\,\mu{\rm s}$ ($\sim 10\%$ of the oscillation period) later. The lower panels (d)--(f) show corresponding results for the case with $n_{\rm ant} = +1$, where waves travel in the positive-$\zeta$ (out-of-plane) and negative-$\vartheta$ direction (clockwise in the $(R,z)$-plane), as observed in Fig.~\ref{fig:01_kstar_ecei}(b,d) for KSTAR's double-peaked fishbones.

While it is difficult to infer the relative phasing via a visual inspection of ECEI data like those in Fig.~\ref{fig:01_kstar_ecei}(b,d), phase analyses of the experimental ECE radiometer signals in Ref.~\cite{Lee26} also revealed a leading edge component near the low-field-side midplane, at least in cases with strong and moderate fishbones.

Our interpretation of spatial channeling theory \cite{Kolesnichenko20} that we discussed in Section~\ref{sec:history} implies that there should be a phase shift between the driver and receiver, because it takes time for the information to propagate from one peak to another. While a quantitative prediction of energy fluxes between two distant peaks requires more theoretical work, it is tempting to draw a connection between these relative phase shifts and the causal relationship between the inner or the outer components of the fluctuating fields. In the case of our simulations (Figs.~\ref{fig:13_kstar-1.051_mstruc}, \ref{fig:14_kstar-1.051_phasing}), we know that the edge-localized antenna is driving the core-localized quasi-mode. One may then conjecture that the leading phase of the outer component of the experimentally observed double-peaked fishbone (Fig.~\ref{fig:01_kstar_ecei}) is also driving the trailing core component. However, this logic has some pitfalls.

\begin{figure}
	[tb]\vspace{-1.8cm}
	\centering
	\includegraphics[width=0.48\textwidth]{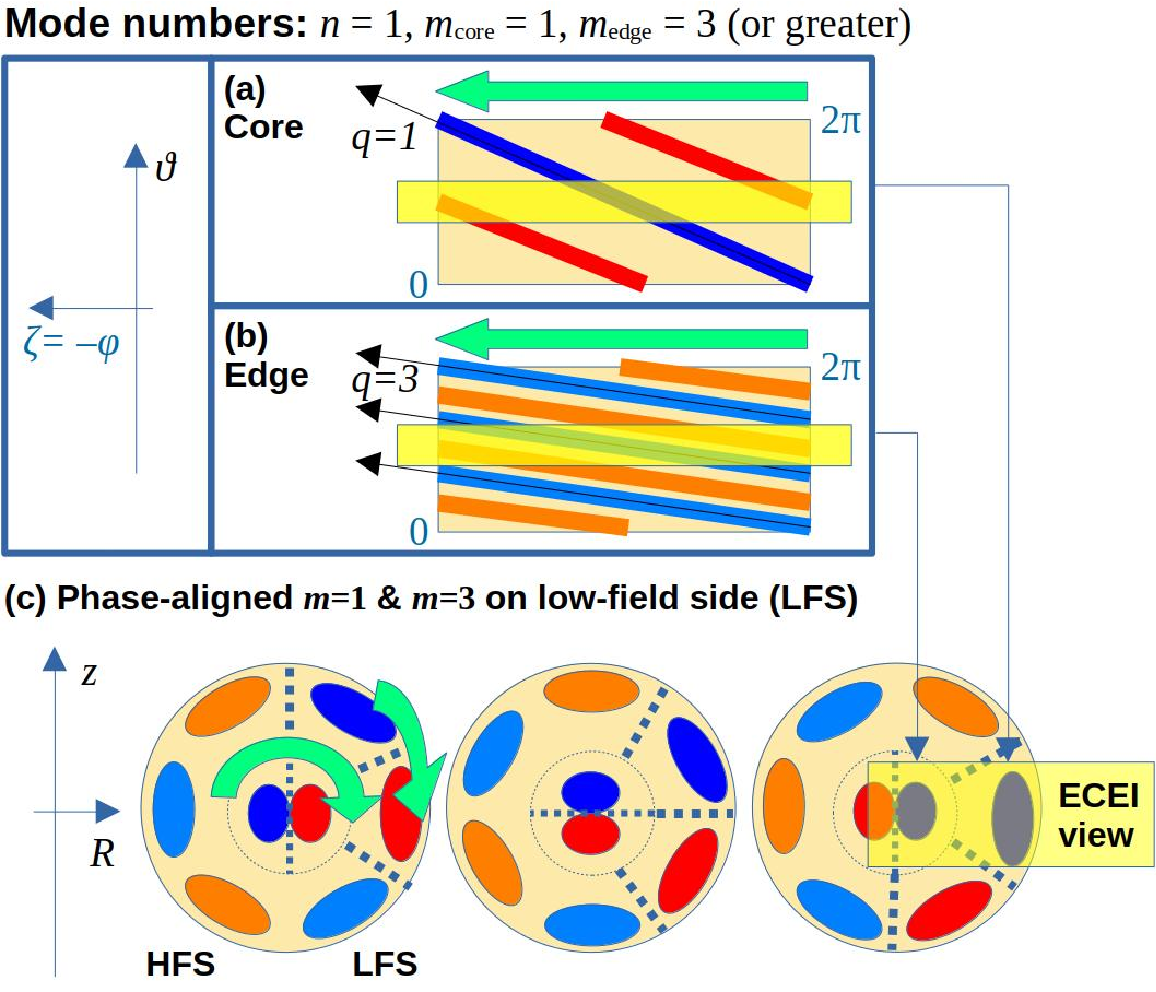}\vspace{-0.2cm}
	\caption{Schematic illustration of phase relations between a core-localized $m/n=1/1$ Fourier component and an edge-localized $3/1$ Fourier component. For simplicity, this sketch depicts a field-aligned ($k_\parallel = 0$) mode in a plasma subject to rigid toroidal rotation. Panels (a) and (b) show the respective structure in the $(\vartheta,\zeta)$-plane of flux surfaces with $q=1$ and $q=3$. The green arrows indicate the toroidal phase velocity in the $\zeta = -\varphi$ direction. Row (c) shows three snapshots of the global structure in the $(R,z)$ plane for half an oscillation period (left to right). The curved green arrows indicate the poloidal phase velocity, which coincides with the KSTAR observations. (Most of our simulations were set up the other way around.)}\vspace{-0.4cm}
	\label{fig:15_mode_rotation}%
\end{figure}

Notice that --- at least in our simulations, and presumably also in the experiments --- the inner and outer mode components have different dominant poloidal mode numbers. This means that their relative phasing varies along the poloidal angle $\vartheta$. To illustrate this in a simplified manner, we show in Fig.~\ref{fig:15_mode_rotation} a sketch of a core-localized $m/n=1/1$ component and an edge-localized $m/n=3/1$ component that are initially phase-aligned on both the low- and high-field side ($\vartheta = 0,\pi$), and out-of-phase at the top and bottom ($\vartheta = \pm\pi/2$).\footnote{This is representative for any odd-odd or even-even combination of $m$-numbers. In contrast, for odd-even pairs that are in-phase on one side, the other side will be out-of-phase.}
As the waves propagate along the green arrows, the phase alignment is assumed to be maintained on the low-field side ($\vartheta = 0$). At other angles, on can then observe different combinations of one component leading and the other trailing. Here, the rate of phase slippage is maximal near the high-field side midplane ($\vartheta = \pi$).

Clearly, there is no way of telling from such phasing data alone, which component is primary and which is secondary. In order to perform such a feat, one first has to postulate that such a distinction actually exists due to a poloidal asymmetry in the radial coupling mechanism.\footnote{This assumption of poloidally asymmetric coupling is also linked to our discussion of imperfect volumetric focusing in Section~\protect\ref{sec:discuss_focus} above.}
Moreover, one would need to have sufficiently detailed understanding of the underlying coupling mechanism to know (i) at which angle the coupling is strongest, and (ii) whether the primary component should really be leading and the secondary trailing in the poloidal plane, or the other way around, because this is not immediately evident in nontrivial geometries with potentially nontrivial wave propagation paths. Only when equipped with all that knowledge, one may conclusively determine the primary and secondary components from their relative phasing in the ``poloidal region of interest''.

This is a tall order, beyond our current capabilities. Moreover, the scare quotes around ``poloidal region of interest'' are meant to attract attention to a conceptual ambiguity, rooted in the fact that the metric of the poloidal angle coordinate $\vartheta$ is arbitrary, so we do not even know how to best measure phase profiles along lines of ``$\vartheta = {\rm const}$.''

What we can do is making educated guesses based on geometric considerations and aided by numerical experiments based on physical models. First, while the choice of poloidal angle coordinates seems arbitrary, it must be noted that the magnetic midplane has physical meaning and can therefore be defined uniquely, even in up-down asymmetric toroidal plasmas, at least in domains where the plasma can be treated as being topologically confined (see Section 3.4 of Ref.~\protect\cite{Bierwage25} for a related recent discussion). The magnetic midplane provides a preferred choice for $\vartheta = 0,\pi$, which is automatically satisfied by the magnetic straight-field-line coordinates (also known as ``PEST coordinates'') that we used for our analysis in Fig.~\ref{fig:14_kstar-1.051_phasing}.

Second, it is reasonable to assume that the coupling is generally strongest near the magnetic midplane. This is empirically supported by the fact that normal modes in a toroidal plasma peak either on the low-field (``ballooning'') or on the high-field side (``anti-ballooning'') but rarely in upper or lower half-plane (see Refs.~\cite{Bierwage22c,Bierwage22d} for related discussions).\footnote{The antenna's near-field in Figs.~\protect\ref{fig:13_kstar-1.051_mstruc} and \protect\ref{fig:14_kstar-1.051_phasing} clearly violates this rule, showing its non-normal character.}
In our simulations, the inner and outer mode components always  appear to be phase-aligned near the low-field-side midplane, whereas phase slippage occurs on the high-field side, regardless of the antenna's location. (The schematic drawing in Fig.~\ref{fig:15_mode_rotation} is based on that observation.) It is thus reasonable to conjecture that the coupling is strongest somewhere near the low-field side midplane. However the exact location (above, at, or below the outer midplane) remains uncertain, especially since the plasmas under consideration are up-down asymmetric (and subject to differential rotation in the experiments). Such factors are likely to contribute to the observed phase shift between the inner and outer mode components in ways that remain to be untangled.

While uncertainties will perhaps always remain, we may at least work towards conditional claims of the following form: {\it If} the core-edge coupling mechanism responsible for the experimentally observed double-peaked fishbone modes is the same as the mechanism that couples the edge antenna to the core-localized quasi-mode in our simulations, {\it then} our simulations provide weak evidence for the possibility that the fishbone's edge-localized peak is the primary component (e.g., directly excited by beam ions) and the core-localized peak is the secondary (parasitic) component. At present, this conjecture can only be made under the premise that the neglected differential plasma rotation and kinetic effects have no substantial impact on the phasing, which remains to be ascertained in future simulations.

\subsubsection{Polarization \& compressibility}
\label{sec:discuss_coupling_polcomp}

We observe that $[\delta\hat{B}_{\rm pol}]_{|n|=1}$ in Fig.~\ref{fig:13_kstar-1.051_mstruc}(c) is dominated by waves on magnetic surfaces near the antenna location, and is relatively weak in the center of the plasma. In contrast, $[\delta\hat{B}_{\rm tor}]_{n=1}$ in Fig.~\ref{fig:13_kstar-1.051_mstruc}(d) consists of a peak near the antenna and a comparable or even larger peak near the magnetic axis in the region of the central $|m/n|=1/1$ continuum plateau. The structure of $[\delta\hat{B}_{\rm tor}]_{|n|=1}$ closely resembles that of $\delta\hat{P}_{\rm mhd}$ in Fig.~\ref{fig:b02_transient-late} of Appendix~\ref{apdx:modeform_transients}, and both are about 90 degrees phase-shifted with respect to $\delta\hat{\Phi}_{|n|=1}$ in Fig.~\ref{fig:13_kstar-1.051_mstruc}(a). This indicates the presence of a compressional component, both near the antenna and in the central plasma, where our quasi-mode is located. The role of plasma compressibility deserves more discussion, and the following Section~\ref{sec:discuss_compress} is dedicated to this topic.

% -----------------------------------------------------------------------------
\subsection{Role of plasma compressibility}%\vspace{-0.2cm}
\label{sec:discuss_compress}

It has recently been demonstrated experimentally that fishbone-like low-frequency modes can have a significant `acoustic polarization' \cite{Du24,Heidbrink25d}. The observation in Section~\ref{sec:discuss_coupling_polcomp} above show that the core-localized quasi-mode's structure in our simulations also contains a significant acoustic component --- that is, sound waves associated with plasmas compressibility (not purely magnetic field compression) --- raises questions like the following:
\begin{enumerate}
	\item[Q1.]  Is plasma compressibility qualitatively essential for efficient core-edge coupling?
	\item[Q2.]  Even if not, how large is its quantitative effect?
	\item[Q3.]  To what extent can we rely on the MHD model; when are (gyro)kinetic models required; and to what order of accuracy should they include the effect of parallel magnetic fluctuations $\delta B_\parallel$? ($\approx \delta B_{\rm tor}$ near the center)
\end{enumerate}

\noindent While conclusive answers will require more work that goes beyond the scope of the present paper (especially question Q3, which requires a benchmark study), the following discussion is meant to pave the way in that direction.

\begin{figure}
	[tb]\vspace{-0.6cm}
	\centering
	\includegraphics[width=0.48\textwidth]{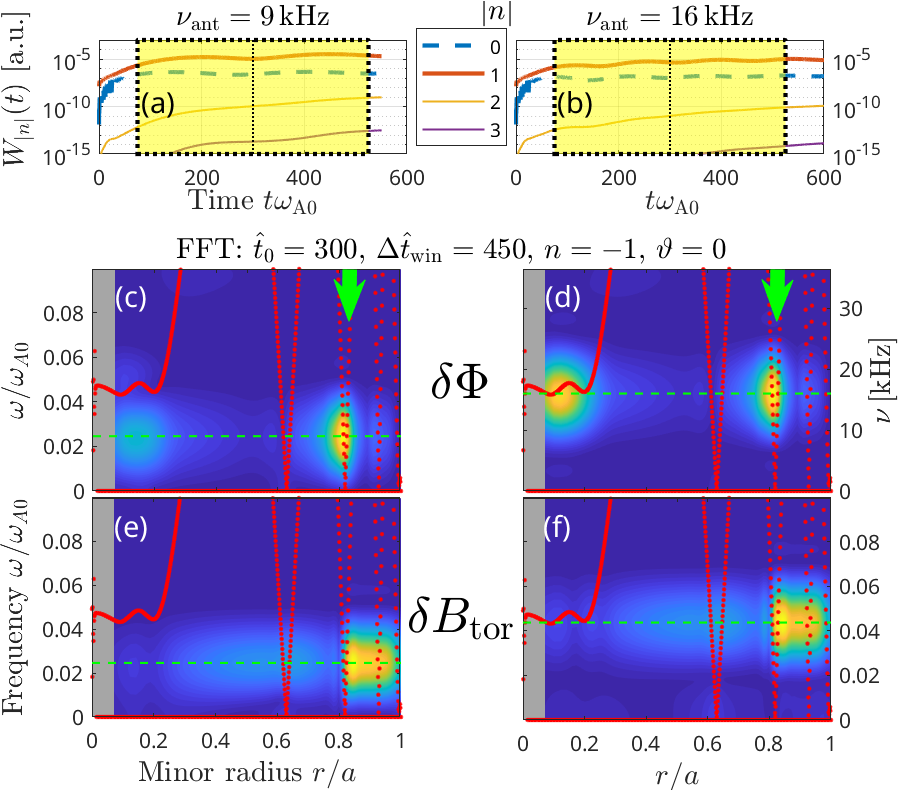}\vspace{-0.2cm}
	\caption{Edge-antenna-driven plasma response without compressibility. The left column shows results for the same case as in Fig.~\protect\ref{fig:12_focus_kstar}, except that here the MHD density and pressure perturbation were artificially suppressed at each time step ($\delta\rho_{\rm mhd},\delta P_{\rm mhd} \rightarrow 0$). This eliminates the acoustic branch of MHD waves and merges the central plateaus of the low- and high-frequency Alfv\'{e}nic branches (located around $10\,{\rm kHz}$ and $100\,{\rm kHz}$, respectively) into a single Alfv\'{e}nic $|m/n|=1/1$ continuum plateau near $16\,{\rm kHz}$. In the right column, the antenna frequency has been raised accordingly to $\nu_{\rm ant} = 16\,{\rm kHz}$. Panels (a) and (b) show the temporal evolution of the MHD fluctuation energy $W_{|n|}(t)$ for $|n|=0,1,2,3$ in the two simulations. Panels (c)--(f) show the spatio-temporal structure of the electrostatic potential $\delta\hat{\Phi}_{|n|=1}(r,\omega)$ (middle row) and toroidal magnetic fluctuation $[\delta\hat{B}_{\rm tor}]_{|n|=1}(r,\omega)$ (bottom row). The spectrograms are overlaid with MHD wave continua (red dots) that were computed by solving the equations in Appendix~A of Ref.~\protect\cite{Deng12} with $\delta P_{\rm mhd} = 0$.}\vspace{-0.5cm}
	\label{fig:16_kstar-1.051_dpre0_evol-spec}%
\end{figure}

\begin{figure*}
	[tbp]
	\centering\vspace{-1.6cm}
	\includegraphics[width=0.96\textwidth]{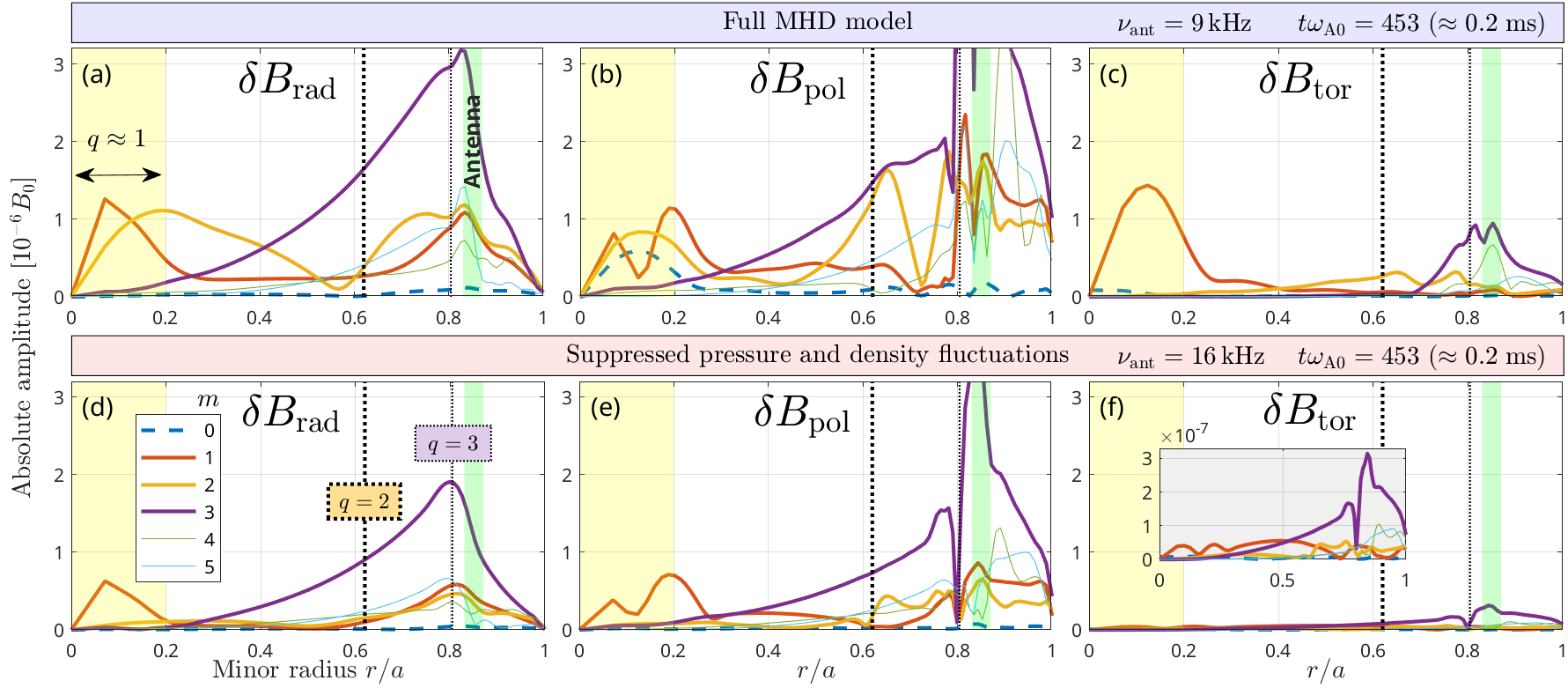}\vspace{-0.2cm}
	\caption{Quantitative and qualitative effect of the acoustic component of the plasmas response on the structure of the magnetic fluctuations $\delta\hat{\bm B}_{|n|=1}$. Shown here are the dominant poloidal Fourier components $m=0...5$ of $|\delta\hat{B}_{\rm rad}|_{|n|=1}$ (left), $|\delta\hat{B}_{\rm pol}|_{|n|=1}$ (center), and $|\delta\hat{B}_{\rm tor}|_{|n|=1}$ (right). The top row shows results from the same self-consistent simulation as in Fig.~\protect\ref{fig:13_kstar-1.051_mstruc}, with a $9\,{\rm kHz}$ antennna at $\hat{r}_{\rm ant} = 0.85$. The bottom row shows results for the same simulation as Fig.~\protect\ref{fig:16_kstar-1.051_dpre0_evol-spec}(b,d,f) with suppressed $\delta\rho_{\rm mhd},\delta P_{\rm mhd} \rightarrow 0$, and with increased antenna frequency $\nu_{\rm ant} = 16\,{\rm kHz}$ matching the shifted $|m/n|=1/1$ continuum plateau. The $q\approx 1$ region is shaded yellow. The antenna location and its radial width are shaded green. Vertical black dotted lines indicate the $q=2$ (bold) and $q=3$ (thin) surfaces.}\vspace{-0.3cm}
	\label{fig:17_kstar-1.051_dpre0_mstruc}%
\end{figure*}

Concerning question Q1, we found that, within the realm of the full MHD model, the acoustic component is not essential for the core-edge coupling effects that we studied in this paper. We obtained qualitatively similar results in modified {\tt MEGA} simulations where the MHD density and pressure fluctuations, $\delta\rho_{\rm mhd}$ and $\delta P_{\rm mhd}$, were artificially suppressed at each time step. The results of such numerical experiments are summarized in Fig.~\ref{fig:16_kstar-1.051_dpre0_evol-spec}, where we show data obtained with two different antenna frequencies: our usual $9\,{\rm kHz}$ (left) and a higher value of $16\,{\rm kHz}$ (right). The artificial suppression of $\delta\rho_{\rm mhd}$ and $\delta P_{\rm mhd}$ eliminates the acoustic wave branches and with them the Alfv\'{e}n-acoustic continuum gap in panels (c)--(f). Moreover, the low-frequency Alfv\'{e}n continuum plateau at $\omega \approx 0.027\times\omega_{\rm A0} \approx 10\,{\rm kHz}$ merges with the high-frequency Alfv\'{e}n continuum plateau (around $100\,{\rm kHz}$) into a single Alfv\'{e}nic $|m/n|=1/1$ continuum plateau at $0.044\times\omega_{\rm A0} \approx 16\,{\rm kHz}$, which is why we increased the antenna frequency $\nu_{\rm ant}$ to that value in the right column of Fig.~\ref{fig:16_kstar-1.051_dpre0_evol-spec}. One can clearly see that the central plasma responds well to the remote antenna located at $\hat{r}_{\rm ant} = 0.85$: The electrostatic potential $\delta\hat{\Phi}_{|n|=1}$ in Fig.~\ref{fig:16_kstar-1.051_dpre0_evol-spec}(d) has a distinct peak at the central continuum plateau in the region $\hat{r} \lesssim 0.2$, just as in the self-consistent simulation in Figs.~\ref{fig:09_kstar-1.051_scan-f_n1}--\ref{fig:13_kstar-1.051_mstruc}.

There are two notable qualitative differences compared to the self-consistent simulations in Fig.~\ref{fig:09_kstar-1.051_scan-f_n1}. First, in spite of continuous antenna drive, the axisymmetric fluctuation energy $W_{n=0}(t)$ in Fig.~\ref{fig:16_kstar-1.051_dpre0_evol-spec}(a,b) saturates. This is to be expected because the artificial suppression of $\delta P_{\rm mhd}$ eliminates the plasma heating that had caused $W_{n=0}(t)$ to rise continuously in the self-consistent simulation in Fig.~\ref{fig:09_kstar-1.051_scan-f_n1}(a). Second, $[\delta B_{\rm tor}]_{|n|=1}$ in the bottom row of Fig.~\ref{fig:16_kstar-1.051_dpre0_evol-spec} does not show a core-localized response. This, of course, is consistent with the acoustic waves being suppressed.

These qualitative differences do not however affect our answer to question Q1: Acoustic waves are not essential for the core-edge coupling studied here.

Let us now proceed to question Q2 concerning the quantitative importance of plasma compressibility. For this purpose, Fig.~\ref{fig:17_kstar-1.051_dpre0_mstruc} shows the radial structures of the dominant poloidal Fourier components $m=0...5$ for the three components of the fluctuating magnetic field $\delta\hat{\bm B}_{n=1}$ in the self-consistent simulation (top) and in the case with suppressed $\delta P_{\rm mhd},\delta\rho_{\rm mhd}\rightarrow 0$ (bottom). For the radial and poloidal components, $[\delta\hat{B}_{\rm rad}]_{|n|=1}$ and $[\delta\hat{B}_{\rm pol}]_{|n|=1}$, we observe an overall reduction in the amplitudes by a factor of about $2$. In part, this may be due to the different antenna frequencies ($9\,{\rm kHz}$ and $16\,{\rm kHz}$) that were chosen to match the respective $|m/n|=1/1$ continuum plateau in Figs.~\ref{fig:12_focus_kstar}(c) and \ref{fig:16_kstar-1.051_dpre0_evol-spec}(d,f), but we suspect that plasma compressibility also affects the strength of the coupling between plasma and antenna. While most $m$-components are subject to moderate attenuation, the $m=2$ component, which is relatively strong in panels (a) and (b) of Fig.~\ref{fig:17_kstar-1.051_dpre0_mstruc}, is nearly absent in panels (d) and (e). This may be explained by the excitation of the $|m/n|=2/1$ acoustic continuum branches near $\hat{r} \approx 0.3$ and $0.5$ in Fig.~\ref{fig:12_focus_kstar}(c), which are absent in Fig.~\ref{fig:16_kstar-1.051_dpre0_evol-spec}(d,f). Hence, the suppression of the $m=2$ component may be taken to be a qualitative difference rather than a quantitative one (like the strong attenuation of $[\delta\hat{B}_{\rm tor}]_{|n|=1}$ discussed above).

In summary, our preliminary answers to questions Q1 and Q2 are that the acoustic component of the plasma response can have a notable qualitative and quantitative influence, but is not essential for the core-edge coupling phenomena studied here, at least in the realm of the MHD model.\footnote{When we suppress pressure and density fluctuations, we obtain the fast magnetoacoustic waves of the zero-beta limit whose dispersion relation was shown in Eq.~(\protect\ref{eq:imhd_disp_fw}). Indeed, those are effectively the waves we see in Figs.~\protect\ref{fig:05_kstar-1.051_decay-ant_early}(b) and \protect\ref{fig:b01_transient-early}(b,c), because the plasma beta in conventional tokamaks is so low (few $\%$), that the pressure remains passive in a fast wave. In Appendix~\protect\ref{apdx:modeform_highfreq}, we confirm that the remote antenna drive works also at frequencies in the range of the toroidicity- and ellipticity-induced gaps ($\gtrsim 100\,{\rm kHz}$), where plasma compressibility effects are also weak. Only at low frequencies $\O(10\,{\rm kHz})$ do we find a notable contribution.}
To address question Q3, it remains to be verified how reliable the MHD results are for predicting the actual low-frequency plasma response in experiments. For this purpose, comparisons with kinetic models are underway and will be reported in a separate paper of this series.

In preparation for such a kinetic benchmark study, let us add a few remarks concerning the role of $\delta B_\parallel$. The electric current that flows through the toroidal volume enclosed by a magnetic surface decreases with decreasing distance to the magnetic axis. In our case, we have $q \approx r B_{\rm tor}/(R B_{\rm pol}) \approx 1$ near that axis, which together with $B_{\rm tor} \approx B_0$ implies that the poloidal magnetic field strength scales approximately as $B_{\rm pol} \propto r$ near the plasma center. This, in turn, implies that the fluctuating component $\delta B_{\rm pol}(r\rightarrow 0)$ must also vanish towards the magnetic axis, so that an Alfv\'{e}nic plasma response in that region can be realized only via (i) a transverse displacement of the magnetic axis itself, for instance in the form of a torsional (nonplanar) Alfv\'{e}n wave, or (ii) a modulation of the surrounding poloidal currents that induce $\delta B_\parallel$ ($\approx \delta B_{\rm tor}$ near $r \approx 0$). It is thus worth examining the role of near-axis contributions of $\delta B_\parallel$ beyond MHD force balance.\footnote{Although $\delta B_\parallel$ does not appear explicitly in reduced MHD and widely-used self-consistent gyrokinetic models, its leading-order contribution is implicitly accounted for via cancellations that are interpreted as a manifestation of MHD force balance \cite{Tang80, Scott24_arxiv} (see Section~\protect\ref{sec:history}). Failure to do so can have a large unphysical impact on the simulation results \protect\cite{Joiner07_09}. The inclusion of additional correction terms has also been suggested \protect\cite{Graves19}.}
This point shall be elaborated in the forthcoming kinetic benchmark study.

% -----------------------------------------------------------------------------
\subsection{Responsiveness at lower frequencies}
\label{sec:discuss_lowfreq}

In the original KSTAR reference case, whose $q$ profile and MHD continua are plotted gray in Figs.~\ref{fig:03_kstar_profiles} and \ref{fig:04_kstar_continua}(a), the core-localized low-frequency Alfv\'{e}nic continuum branch with $|m/n|=1/1$ covers the frequency band from about $20\,{\rm kHz}$ at the axis down to $\nu=0$ at the $q=1$ surface. Although continuum damping (radial phase mixing) can be expected to be significantly stronger than in our modified KSTAR Models 1 and 2 with flat continuum plateaus, the $q$ profile of the original reference case would provide a path for the fishbone frequency to chirp down along the $|m/n|=1/1$ continuum towards the $q=1$ surfaces in the fashion of a typical energetic particle mode (EPM) \cite{Chen94}.\footnote{An illustration for the BAE-TAE frequency range can be found in Fig.~1 of Ref.~\protect\cite{Bierwage16b}, and we assume that the same is possible along the low-frequency (sub-BAAE) Alfv\'{e}nic continuum when the $1/1$ kink (Appendix~\ref{apdx:kink}) is stabilized, for instance by mirror-trapped fast ions \protect\cite{Porcelli91}.}

Interestingly, it turns out that even in our KSTAR Model 1, which does not posses a $q=1$ surface, the central core plasma responds rather well at frequencies {\it below} the low-frequency Alfv\'{e}nic $|m/n|=1/1$ continuum plateau; that is, without a matching continuum. For the self-consistent simulation, this was shown in Fig.~\ref{fig:09_kstar-1.051_scan-f_n1}(b,c). The same behavior is observed in the ``incompressible'' case in Fig.~\ref{fig:16_kstar-1.051_dpre0_evol-spec}(c). If this prediction of the MHD model can be trusted, this observation has the important implication that the experimentally observed dynamic frequency chirping of a fishbone mode is possible even in the absence of a $q=1$ surface, as long as $q$ is above but close to unity ($q \gtrsim 1$). Fishbones that are subject to a fast-ion-induced frequency shift as described by the nonperturbative theory by Chen~{\it et al}.~\cite{Chen84} will presumably benefit from such a responsive background plasma.

Currently, we do not have a conclusive explanation for the fairly good responsiveness at sub-resonant frequencies, but as we already noted in Section~\ref{sec:discuss_coupling_radparity} above, some features of the observed radial mode structure indicate that nearby marginally stable kink-like modes could play a facilitating role, both from above and from below the frequency band of interest:
\begin{itemize}
	\item At $\nu \sim 10\,{\rm kHz}$, the height of the central $1/1$ continuum plateau, there would be the kink-like GAE \cite{Appert82}. As we noted near the end of Section~\ref{sec:discuss_compress}, we may assume $rB_{\rm pol} \approx {\rm const.}$ near the magnetic axis ($r \approx 0$), so that Eq.~(7) of Ref.~\cite{Appert82} reduces to\vspace{-0.2cm}
	\begin{equation}
		\omega^2 \approx \omega_{\rm A}^2\left[1 - \left(\frac{2 n}{q R k_{\rm rad}}\right)^2\right].
		\label{eq:gae}
	\end{equation}\vspace{-0.4cm}
	
	A core-localized $|n|=1$ mode with radial width $k_{\rm rad}^{-1} \sim \Delta r \approx 0.2a$ in a plasma with inverse aspect ratio $a/R \approx 1/3$ is thus predicted to exist $1 - \sqrt{1 - (2\times 0.2/3)^2} \approx 1\%$ below the $1/1$ continuum plateau frequency, so effectively at the plateau itself.
	
	\item At $\nu \sim 0$, we anticipate the contribution of marginally stable kink-tearing modes. Indeed, we found that tearing modes at $q=2,3$ become unstable in KSTAR Model 1 when the resistivity is increased and the viscosity is reduced by a factor 10, respectively (details for this case will be reported elsewhere).
\end{itemize}

\noindent With the plasma supporting kink-like perturbations around $0$ and $10\,{\rm kHz}$, our observation that kink-like quasi-modes can be excited at intermediate frequencies as in Figs.~\ref{fig:09_kstar-1.051_scan-f_n1}(b) and \ref{fig:16_kstar-1.051_dpre0_evol-spec}(c) does not seem unreasonable, but a rigorous theoretical explanation would certainly be desirable.

% =============================================================================
\section{Summary \& outlook}%\vspace{-0.4cm}
\label{sec:summary}

Motivated by the observation of coherent low-frequency fluctuations in the central core and edge regions of KSTAR H-mode and (a few) L-mode plasmas \cite{Lee23, Lee26} --- a phenomenon we call double-peaked fishbone mode --- we explored one out of several possible explanations that were outlined in the introductory Section~\ref{sec:intro}; namely, Option 3: {\it The resonant coupling and synchronization of two independent Alfv\'{e}nic mode components, one in the central core and one near the edge, into a coherently oscillating entity that is held together by MHD waves.}

In this work, we studied the radial coupling process at a fundamental level, using suitably prepared numerical simulations of visco-resistive single-fluid full MHD equations, where an artificial internal antenna was used as a driver (acting as a {\it primary mode}) and radially distant flat plateaus of the low-frequency Alfv\'{e}nic continuum were used as receivers (acting as parasitic {\it secondary modes}).

In Section~\ref{sec:results}, we elucidated the mode structure formation process for our antenna-driven quasi-modes, and confirmed the possibility and effectiveness of ``action at a distance'' in a KSTAR-shaped MHD plasma. We presented evidence for inward drive (edge-to-core) being more efficient than outward drive (core-to-edge), although the latter is still a viable option for double-peaked fishbones since the outer peak has a significantly smaller amplitude, while occupying a potentially larger volume. The appendices of this paper contain additional results on various aspects of our simulations, such as the role of visco-resistive dissipation.

In Section~\ref{sec:discuss}, the main results were discussed in more detail, contemplating some of the underlying mechanisms and processes based on our current level of understanding.

We began with a discussion of volumetric focusing (Fig.~\ref{fig:11_kstar-q1.051-flat-q2.15_rant-scan}), and how we think it is physically realized in our simulation setup with a highly localized antenna. Then we discussed core-edge coupling physics, considering the form of the radial mode structure, poloidal phase relations between core and edge fluctuations, and wave polarization.

Concerning the relative poloidal phase between the inner and outer peaks of the wave field, our analysis (Fig.~\ref{fig:14_kstar-1.051_phasing}) indicated qualitative agreement with the findings of the comprehensive survey of experimental data in Ref.~\cite{Lee26}: namely, that the outer component seems to be leading and the inner component seems to be trailing around the outer midplane. This led us to the question of causation. Since we know that the core-localized quasi-mode in our simulations was driven by the edge-localized antenna, the agreement with experimental data constitutes weak evidence for the possibility that the double-peaked fishbone's outer component may be the primary and the inner component the secondary. However, we also explained that this hinges on several assumptions that are currently difficult to verify, so the evidence we have is only indicative, not conclusive.

Concerning polarization effects, we found that the acoustic component is not strictly essential for core-edge coupling but significantly enhances its efficiency (Fig.~\ref{fig:17_kstar-1.051_dpre0_mstruc}).

Finally, we found that a core-localized quasi-mode can form also sub-resonantly; that is, at frequencies $0 \lesssim \nu \lesssim 10\,{\rm kHz}$ below the central Alfv\'{e}nic continuum plateau. Since this behavior was seen both with and without the acoustic component (Figs.~\ref{fig:09_kstar-1.051_scan-f_n1} and \ref{fig:16_kstar-1.051_dpre0_evol-spec}), we conjecture that our antenna-driven quasi-modes are primarily kink-like perturbations that do not rely on plasma compressibility (Section~\ref{sec:discuss_lowfreq}). On the physics side, we see this broad-band responsiveness as being complementary to and facilitating for fast-ion-induced frequency shifts of kink-type modes \cite{Chen84}. On the modeling side, this finding gives reason for hope that the present results of the single-fluid MHD model can be verified by future benchmark studies against kinetic models, even though we are operating in a frequency band that is usually assumed to require a kinetic treatment of thermal ions. This is important because we expect that the observed sub-resonant responsiveness of the plasma can facilitate dynamic frequency chirping, even without a matching Alfv\'{e}nic continuum. Frequency chirping is a characteristic feature of fishbone-like modes. Being able to justify the simulation of such dynamics with an MHD model would allow us to gain more insight through numerical experiments with relatively low physical complexity and relatively little computational effort.

Indeed, much work remains to be done on the theory, modeling and simulation side. For instance, one should examine whether and how volumetric focusing as well as the coupling and phasing between the antenna and quasi-mode in our simulations are altered by differential plasma rotation. One should also look into other core-edge coupling mechanism, such as Options 2 (free-boundary eigenmodes) and 4 (beam ions on wide orbits) that were proposed in Section~\ref{sec:intro} as candidates contributing to the formation of KSTAR's double-peaked fishbones. On the driving side, the antenna would have to be replaced by a realistic population of fast ions. On the receiving side, there could be continuum plateaus supporting quasi-modes (as assumed in the present work), marginally stable internal or external kink modes, or drift-Alfv\'{e}n modes:\vspace{-0.2cm}
\begin{itemize}
	\item  In the KSTAR plasmas of interest, a continuum plateau supporting a quasi-mode may exist in the central core if one assumes that a flat safety factor near unity ($q\sim 1$) is left behind by a sawtooth crash \cite{Kadomtsev75} and maintained by quasi-interchange or ``flux pumping'' processes \cite{Wesson86, Krebs17, Jardin20}.
	\item  Other candidates for the inner mode component include the internal kink and drift-Alfv\'{e}n modes, which are thought to underlie classical fishbones in resonant \cite{Chen84,Chen94}, diamagnetic \cite{Coppi86} or mixed \cite{Zhang89} flavors.
	\item  Candidates for the outer mode component include kink-tearing modes \cite{Du21,Liu22} and external kinks that are thought to underlie energetic particle-driven wall-stabilized modes (EWM) or, more generally, off-axis fishbone modes (OAFB) \cite{Matsunaga09, Okabayashi09, Matsunaga10, Okabayashi11, Heidbrink11}. In H-mode plasmas, the $q$ profile often exhibits a flat portion near the edge due to bootstrap currents, which may support an edge-localized quasi-mode. One may even consider (standing) Alfv\'{e}nic waves in the scrape-off layer \cite{Stepanenko23}.
\end{itemize}

Some of the tasks outlined above are planned to be tackled in future papers of this series. Currently underway is a numerical simulation study of wave-particle resonances, with particular attention to the possibility of core-edge coupling via barely trapped orbits that traverse both the inner and outer plasma as suggested in Ref.~\cite{Lee23} (Option 4 in Section~\ref{sec:intro}). Preparations are also underway to search for double-peaked global eigenmodes subject to differential rotation (Option 2 in Section~\ref{sec:intro}), using a free-boundary MHD model. It is also necessary to test the applicability of the MHD model in the low-frequency domain populated by fishbones ($\sim 10\,{\rm kHz}$) by comparisons with kinetic models, so we commenced a validation study. Last but not least, the experimentally observed influence of external magnetic perturbations on the form and strength of KSTAR fishbones \cite{Lee26} remains to be explained. Since double-peaked fishbones were seen with and without external magnetic perturbations, we believe that their effect is mostly indirect via their influence on the magnetic topology and plasma profiles.

On the experimental side, an extensive survey and statistical analysis of presently available experimental data was reported in Part I of this paper series \cite{Lee26}. Further measurements, also on other tokamaks, would help to constrain theoretical analyses and numerical modeling. For instance, it would be useful to obtain information about the mode structure away from the outer midplane. More accurate measurements of plasma rotation would also be helpful.

One reason for our interest in double-peaked fishbone modes is that their existence constitutes evidence for robust long-range coupling, a thorough understanding of which could be useful for tackling practical problems. For instance, one may envision the possibility of fluctuations in the central core regulating or mitigating deleterious edge-localized modes (ELM). In the KSTAR plasmas that we analyzed here, where the $q \sim 1$ region encompassed no more than $20\%$ of the minor radius, a few \% of the plasma volume. In contrast, models of steady-state scenarios considered for JT-60SA and ITER plasmas have $q \sim 1$ radii around $40\%$, corresponding to $10$-$20\%$ of the plasma volume. If fishbone oscillations can be excited under such conditions, their coupling to the plasma edge (as seen in KSTAR \cite{Lee23, Lee26}) may then have interesting effects that should be explored.

% =============================================================================
\section*{Acknowledgments}

A.B.\ would like to thank Yasushi Todo (NIFS) for providing the {\tt MEGA} code \cite{Todo98,Todo05} and acknowledges helpful discussions with Valentin Igochine (IPP Garching), Kimin Kim (KFE), Kouji Shinohara (U Tokyo), Timur Zh.\ Esirkepov and Shuhei Sumida (QST). The hybrid simulations were carried out using the supercomputers JFRS-1, Plasma Simulator, and HPE SGI8600 in the National Institutes for Quantum Science and Technology (QST).\vspace{-0.3cm}

%\clearpage
\section*{Appendices}%\vspace{-0.3cm}

\appendix
% =============================================================================
\section{Stability of the $n=1$ internal kink}%\vspace{-0.2cm}
\label{apdx:kink}

Fishbone-like modes tend to occur in plasmas with a marginally stable $n=1$ internal kink mode, and it is also known that this mode can have two stability boundaries in a tokamak plasma \cite{Kerner80, Tokuda82, Manickam84}. Since the central value $q_0$ of the safety factor $q(\psi_{\rm P})$ is notoriously difficult to determine experimentally, and since fishbone-like modes have also been found in KSTAR plasmas where the central safety factor was estimated to have a fairly low value of $q_0 \approx 0.7$ \cite{Jhang23}, we have performed an ideal MHD stability analysis for a range of $q$ profiles using the code {\tt MINERVA} \cite{Aiba09}.

\begin{figure}
	[tb]\vspace{-0.7cm}
	\centering
	\includegraphics[width=0.48\textwidth]{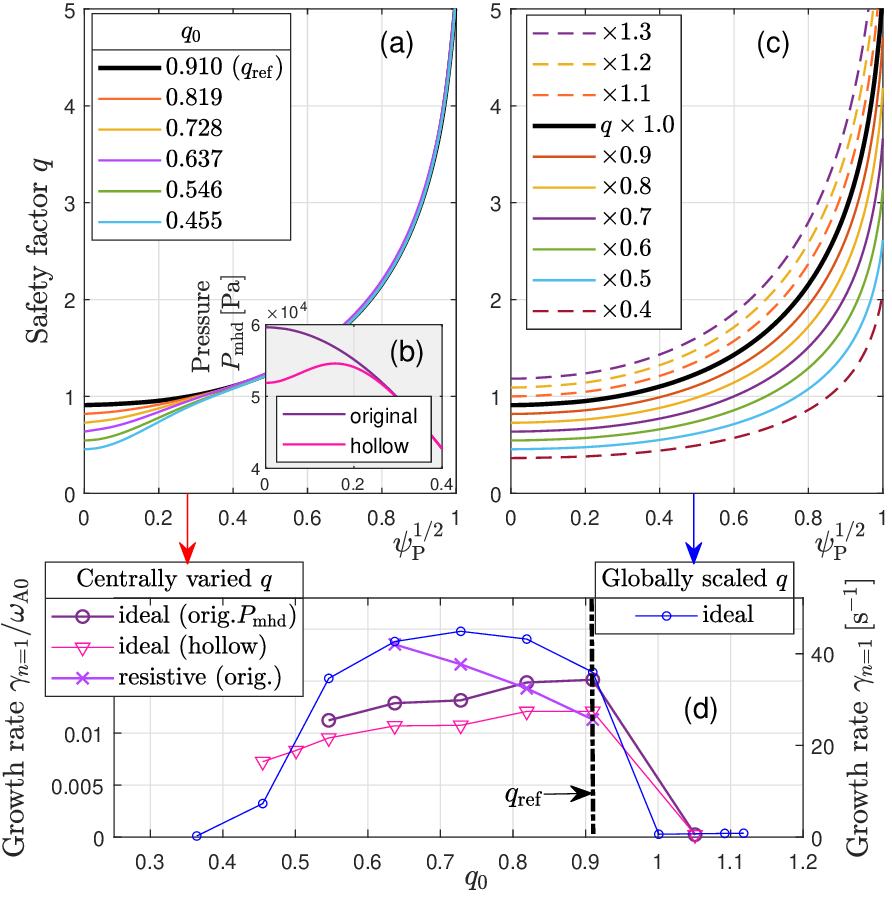}\vspace{-0.3cm}
	\caption{$n=1$ internal kink stability analysis using the ideal MHD solver {\tt MINERVA} \protect\cite{Aiba09} (circles \& triangles) and the resistive MHD solver of {\tt MEGA} \protect\cite{Todo98, Todo05} (crosses). Panels (a)--(c) show the sets of profiles used for the safety factor $q$ and MHD pressure $P_{\rm mhd}$. All other parameters and the plasma shape are based on our KSTAR reference shot \#18567 in Fig.~\protect\ref{fig:03_kstar_profiles}. The equilibria were constructed using the code {\tt CHEASE} \protect\cite{Luetjens96}. Panel (d) summarizes the computed linear growth rates $\gamma_{n=1}$ (left normalized, right in SI units) as functions of $q_0 \equiv q(\psi_{\rm P}=0)$ at the magnetic axis.}\vspace{-0.3cm}
	\label{fig:a01_kink_stability}%
\end{figure}

First, we varied only the central part of the $q$ profile as shown in Fig.~\ref{fig:a01_kink_stability}(a), because the experimental uncertainty for the outer part of the $q$ profile is expected to be small. The results are shown as violet circles in Fig.~\ref{fig:a01_kink_stability}(d), where one can see that the $n=1$ internal kink mode becomes unstable when $q_0$ crosses unity from above and its growth rate decreases gradually in an almost linear manner until at least $q_0 \approx 0.55$. Below this value, we were not able to obtain convergence with the code {\tt CHEASE} \protect\cite{Luetjens96} that we used here to compute these MHD equilibria. Replacing the centrally peaked pressure profile by a slightly hollow one as shown in Fig.~\ref{fig:a01_kink_stability}(b), we were able to scan a little further down to $q_0 \approx 0.46$, before encountering the same problem. The growth rates plotted as magenta triangles in Fig.~\ref{fig:a01_kink_stability}(d) show the same trend as with the peaked pressure profile. Results from resistive MHD simulations using {\tt MEGA} \cite{Todo98, Todo05} with the same numerical parameters as in Section~\ref{sec:setup} yield increasing growth rates down to at least $q_0 \approx 0.64$ as can be seen from the violet crosses in Fig.~\ref{fig:a01_kink_stability}(d).

These results indicate that it is unlikely that the second stability domain of the $n=1$ internal kink mode can be reached in practice when the outer portion of the $q$ profile is fixed as in Fig.~\ref{fig:a01_kink_stability}(a). We note that previous studies that showed the existence of a second stable domain have used different sets of $q$ profiles, similar to those shown in Fig.~\ref{fig:a01_kink_stability}(c), where the entire $q$ profile is simply rescaled, so that the $q=1$ surfaces moves towards the edge as $q_0$ is reduced. In this case, we can indeed find a growth rate that peaks around $q_0 \approx 0.7$ and essentially vanishes for $q_0 \lesssim 0.45$ as shown by the blue circles in Fig.~\ref{fig:a01_kink_stability}(d). However, the large $q=1$ radii associated with such profiles seem to be incompatible with our experimental observations in KSTAR such as Fig.~\ref{fig:01_kstar_ecei}.

Based on these results, we suspect that $q_0$ is near and possibly above unity in the KSTAR plasmas of interest here and have chosen the $q$ profiles in Fig.~\ref{fig:03_kstar_profiles} accordingly.

\begin{figure*}
	[tbp]
	\centering
	\includegraphics[width=0.96\textwidth]{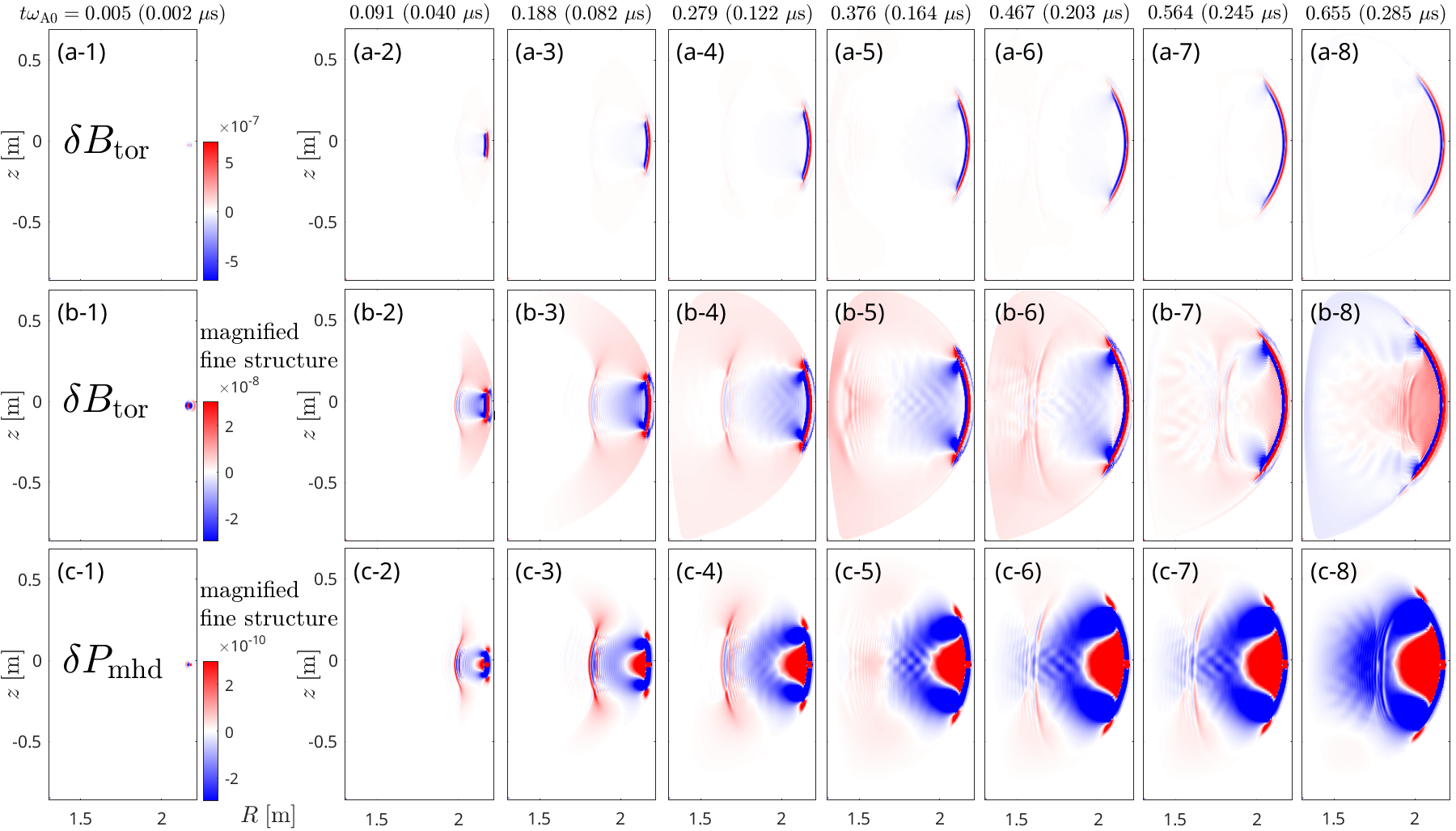} \\
	\includegraphics[width=0.96\textwidth]{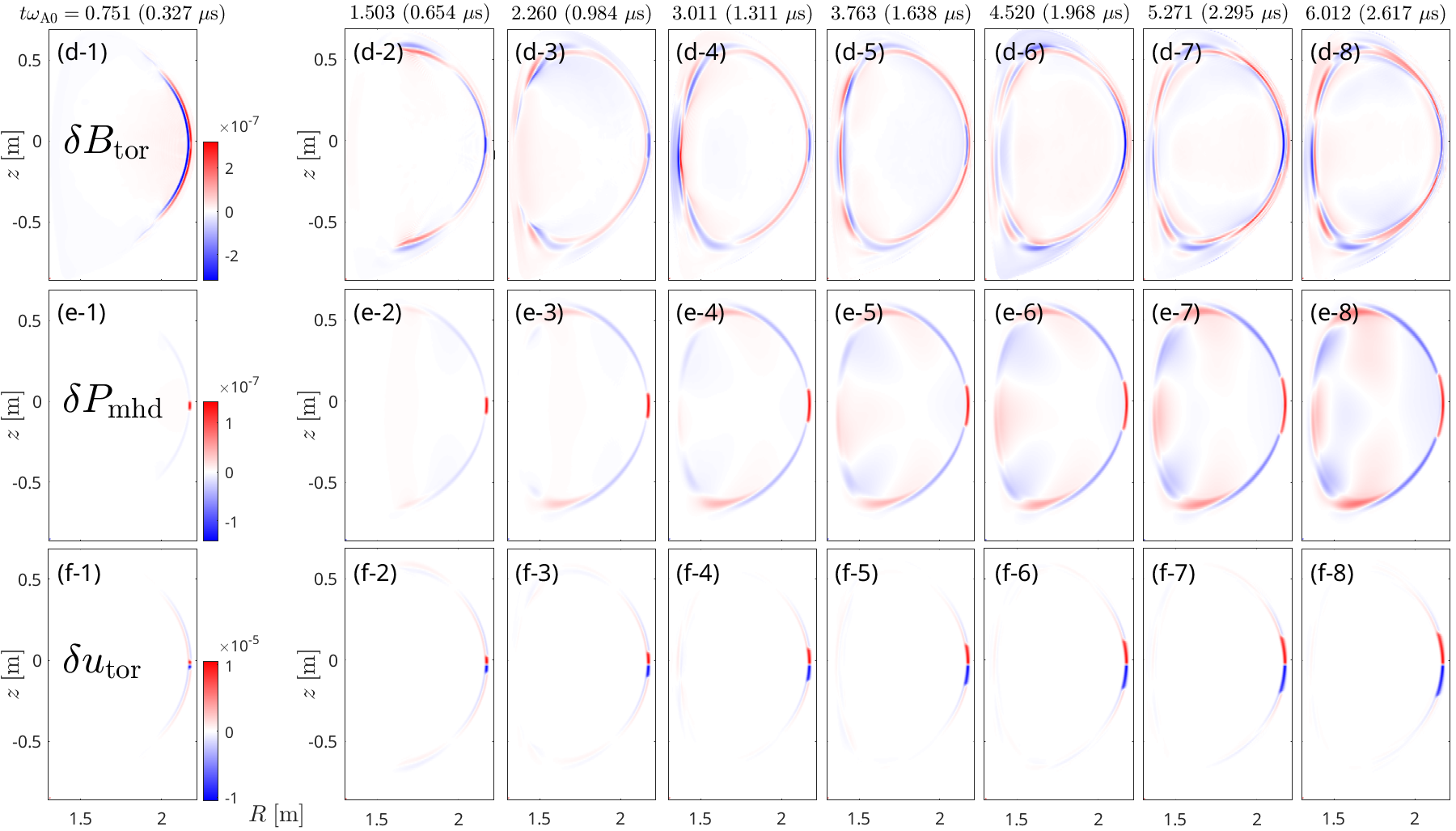}
	\caption{Early MHD transients arising after the abrupt activation of the radially and poloidally localized antenna, which is located on the low-field-side midplane at $(R_{\rm ant},z_0) = (2.18\,{\rm m},-0.03\,{\rm m})$. This is the same case as in Fig.~\protect\ref{fig:12_focus_kstar}. For the first $0.3\,\mu{\rm s}$ of the simulation, rows (a) and (b) show the early evolution of $\delta B_{\rm tor}$ and $\delta P_{\rm mhd}$. While (a) shows largely the poloidal spreading of the perturbations via shear Alfv\'{e}n waves propagating along and counter to ${\bm B}$, the color scales in (b) and (c) are chosen to reveal weaker perturbations that propagate radially at what appears to be local Alfv\'{e}n speed $v_{\rm A}$. The poloidal mode structure largely forms during the first few $\,\mu{\rm s}$ that are covered by the series of snapshots in rows (d)--(f), where we plotted $\delta B_{\rm tor}$, $\delta P_{\rm mhd}$ and $\delta u_{\rm tor}$, respectively. Note that this figure shows the raw data of simulations that were run with a low-pass filter that eliminates $|n|>1$ components. The contour plots shown here and in Figs.~\protect\ref{fig:b02_transient-late}--\protect\ref{fig:b04_transient_dpre0} and \protect\ref{fig:b11_transient_416kHz} below hence contain $n=0,\pm 1$ components as well as radial and poloidal fine structures down to the resolution limit of our $(R,z)$ grid. Therefore, the present mode structures contain more detail than the contour plots of Fourier-synthesized fields such as $[\delta B_{\rm tor}]_{|n|=1}(R,z)$ in Fig.~\protect\ref{fig:13_kstar-1.051_mstruc}, which are composed of the $m=0...12$ components of $n=-1$.}%\vspace{-0.1cm}
	\label{fig:b01_transient-early}%
\end{figure*}

\begin{figure*}
	[tb]%\vspace{-0.5cm}
	\centering
	\includegraphics[width=0.96\textwidth]{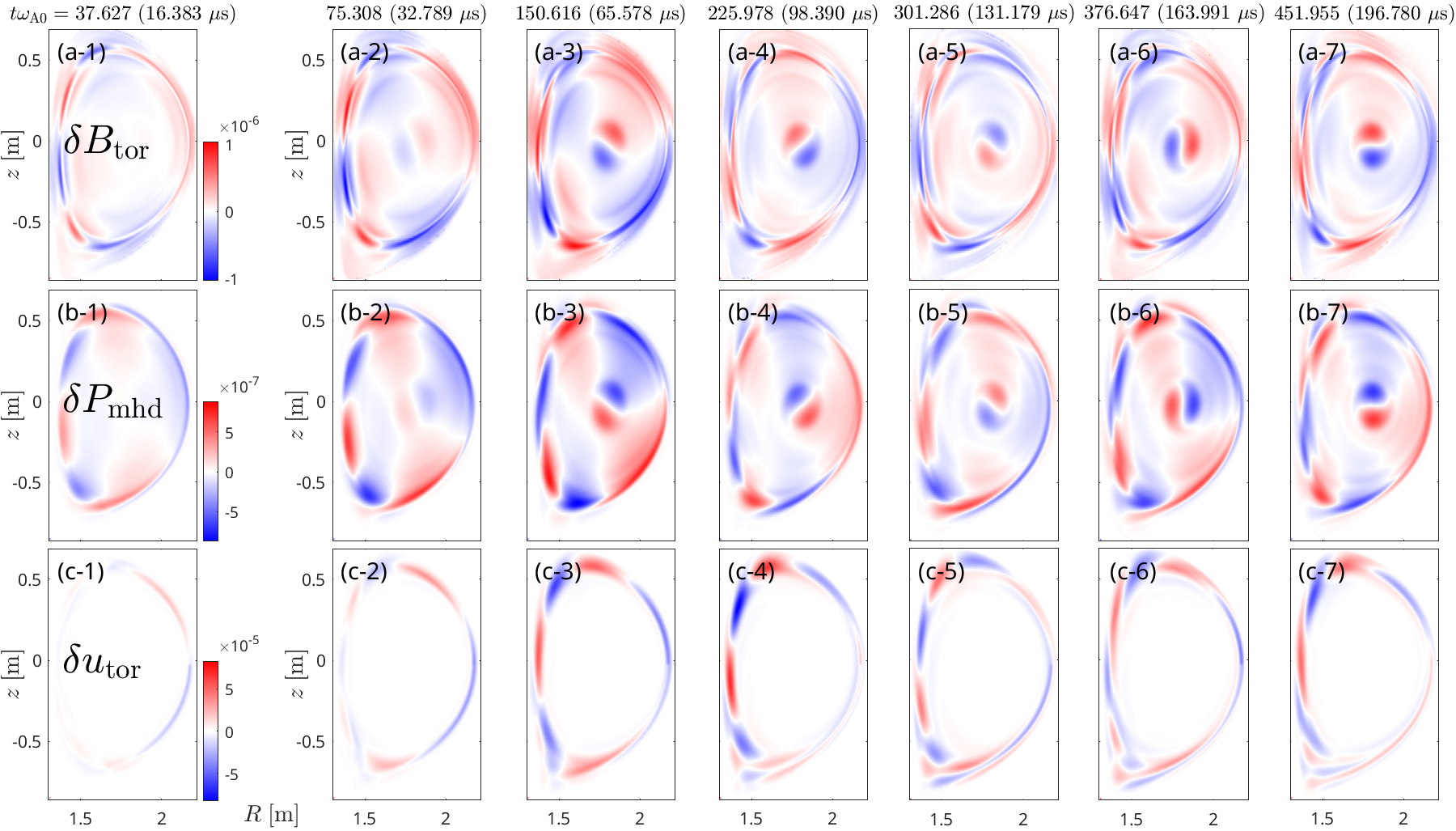}
	\caption{Continuation of Figs.~\protect\ref{fig:b01_transient-early}(d,e,f), showing the completion of the radial mode structure formation on the time scale $\nu_{\rm ant}^{-1} \approx 100\,\mu{\rm s}$.}\vspace{-0.05cm}
	\label{fig:b02_transient-late}%
\end{figure*}

\begin{figure*}
	[tb]
	\centering
	\includegraphics[width=0.96\textwidth]{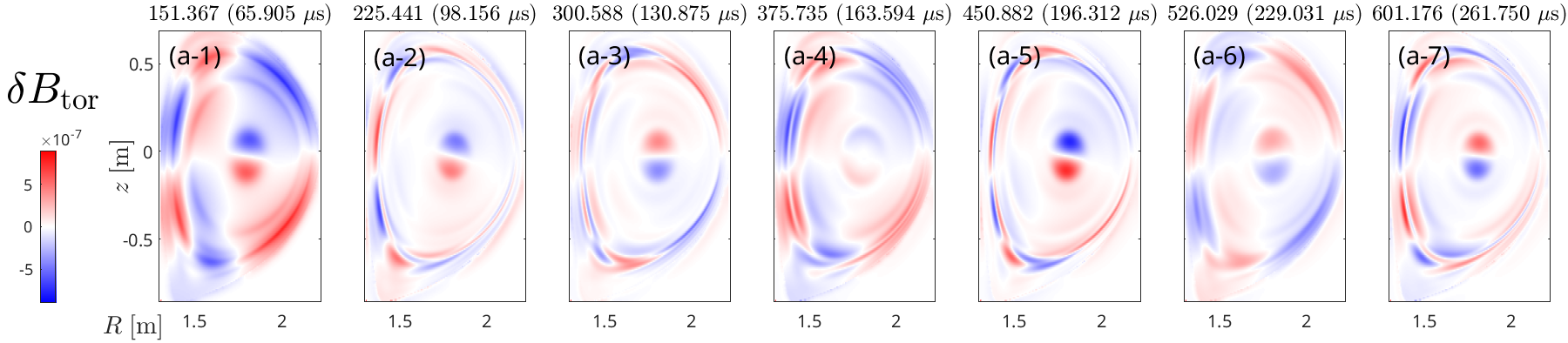}
	\caption{Similar to Fig.~\protect\ref{fig:b02_transient-late}(a), but with a standing wave antenna ($n_{\rm ant} = \pm 1$). The reason for the undulation of the mode phase is thought to be the plasma's up-down asymmetry, which implies different path length' for co- and counter propagating shear Alfv\'{e}n waves.}%\vspace{-0.1cm}
	\label{fig:b03_transient-standing}%
\end{figure*}

% =============================================================================
\section{Details of edge-driven core response}
\label{apdx:modeform}

In Section~\ref{sec:results_transients}, we described the early transients that were seen when driving a core-localized quasi-mode with a core-localized antenna directly. In Sections~\ref{sec:results_distance}--\ref{sec:results_scan-r_focus}, we then demonstrated that the same quasi-mode can also be driven by a remote antenna located near the plasma edge and examined its dependence on antenna parameters.

This appendix contains supplementary material concerning the edge-driven core plasma response. The early transients and mode structure formation process in such a case are described and discussed in Appendix~\ref{apdx:modeform_transients}. Additional information concerning the stronger residual beating seen with a wider antenna are provided in Appendix~\ref{apdx:modeform_pulse}. Finally, Appendix~\ref{apdx:modeform_highfreq} contains results for edge-localized antenna drive applied at higher frequencies in the range of the toroidicity- and ellipticity-induced Alfv\'{e}nic gaps.

\begin{figure*}
	[tb]\vspace{-0.5cm}
	\centering
	\includegraphics[width=0.96\textwidth]{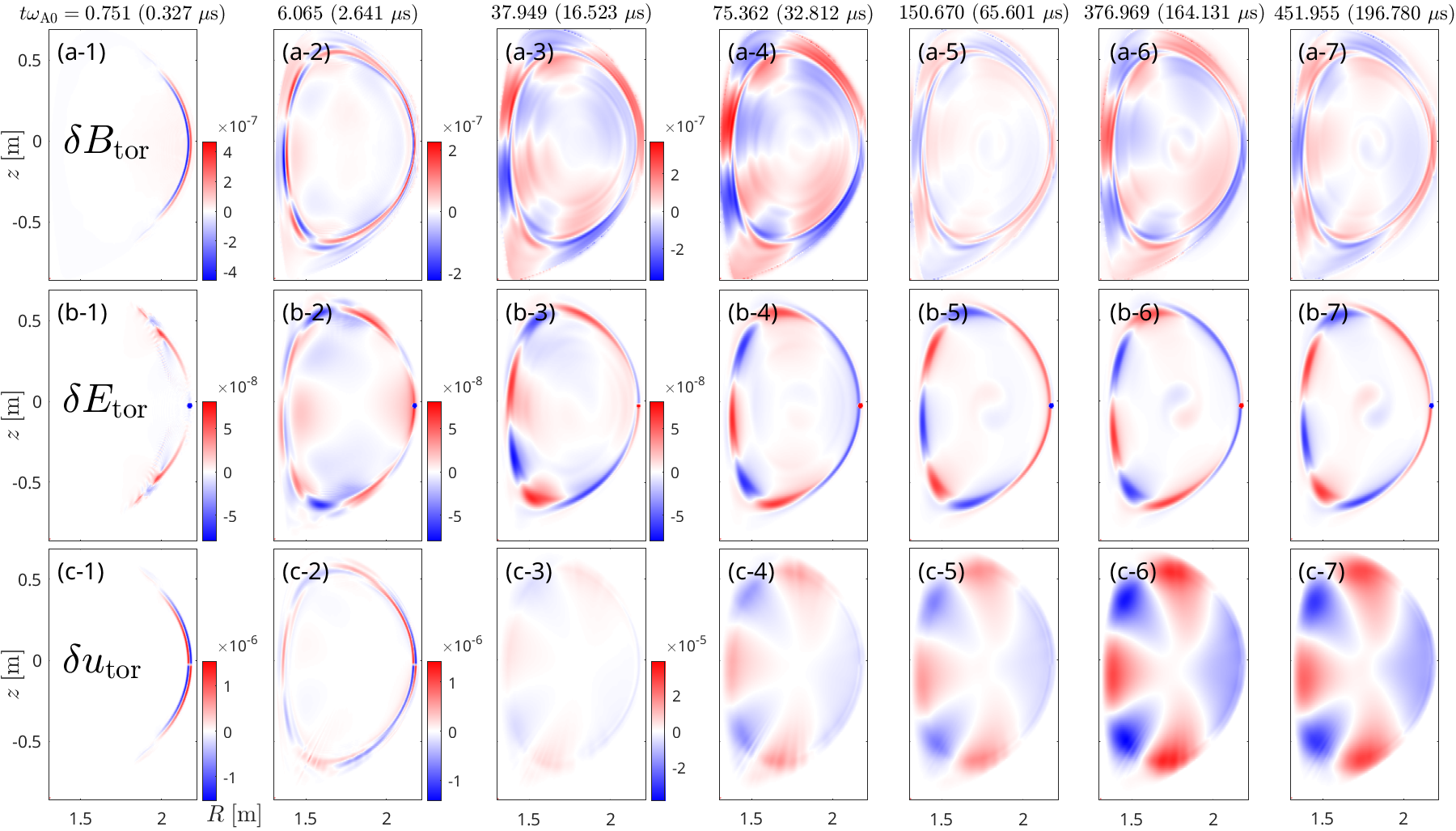}%\vspace{-0.15cm}
	\caption{Demonstration of the role of plasma compressibility for mode structure formation by artificially suppressing pressure and density fluctuations ($\delta P_{\rm mhd},\delta\rho_{\rm mhd} \rightarrow 0$). The antenna frequency is raised from $9\,{\rm kHz}$ to $16\,{\rm kHz}$ in order to match the central $|m/n|=1/1$ continuum plateau as in Figs.~\protect\ref{fig:16_kstar-1.051_dpre0_evol-spec}(d,f). The snapshot times cover some of the early and advanced stages from Figs.~\protect\ref{fig:b01_transient-early} and \protect\ref{fig:b02_transient-late}. Arranged as Fig.~\protect\ref{fig:b02_transient-late}, except that row (b) shows $\delta E_{\rm tor}$ (approximately resembling $\delta\Phi$) instead of $\delta P_{\rm mhd}$ since the latter is zero.}%\vspace{-0.2cm}
	\label{fig:b04_transient_dpre0}%
\end{figure*}

% -----------------------------------------------------------------------------
\subsection{Early transients \& quasi-mode formation}
\label{apdx:modeform_transients}

Our numerical antenna is located on the outer (low-field-side) midplane and covers the entire toroidal circumference of the tokamak. When this antenna is abruptly activated at the beginning of a simulation in the manner of a Heaviside step function, it launches shear Alfv\'{e}n waves in both directions, along and counter to ${\bm B}$. This process can be observed in rows (a) and (d) of Fig.~\ref{fig:b01_transient-early}, where we plotted a series of snapshots of $\delta B_{\rm tor}(R,z)$ at the (arbitrary) toroidal angle $\zeta = 0$. The antenna is located near the plasma boundary at $\hat{r}_{\rm ant} \approx 0.85$. With the local safety factor and mean Alfv\'{e}n speed being respectively $q\approx 3$ and $\left<v_A\right>_{r_{\rm ant}} \approx v_{\rm A0}/\sqrt{N_{\rm i}(r_{\rm ant})/N_{\rm i0}} \approx v_{\rm A0}/\sqrt{0.13} \approx 12.6\times 10^6\,{\rm m/s}$, the two wave fronts should meet on the inner (high-field-side) midplane after about $0.5\times 2\pi q R_0/\left<v_A\right>_{r_{\rm ant}} \approx 0.5\times 6\pi\times 1.81\,{\rm m}/(12.6\times 10^6\,{\rm m/s}) \approx 1.4\,\mu{\rm s}$. This estimate agrees with our observations around snapshot (d-4) in Fig.~\ref{fig:b01_transient-early}.

The data from row (a) are shown again in row (b) of Fig.~\ref{fig:b01_transient-early}, but with a different color scale in order to see $\delta B_{\rm tor}$ fluctuations at lower amplitudes. This reveals another wave front that propagates radially away from the antenna in all directions. Along the horizontal axis, this front crosses the major radial width $\Delta R \approx 0.8\,{\rm m}$ of our KSTAR plasma in about $0.16\,\mu{\rm s}$; that is, with a mean velocity $5\times 10^6\,{\rm m/s}$ comparable to the on-axis Alfv\'{e}n speed $v_{\rm A0} = 4.2\times 10^6\,{\rm m/s}$. This radially propagating wave field develops caustic structures (cusps and folds), presumably due to the smooth nonunformity of $v_{\rm A}$ and apparently also influenced by the curved magnetic surfaces. The waves are reflected off our fixed numerical plasma boundary, which acts like a concave mirror and causes some (imperfect) focusing of the wave front. Within the course of a microsecond, these low-amplitude fluctuations largely phase-mix.

Row (d) of Fig.~\ref{fig:b01_transient-early} is a continuation of row (a) and shows how the poloidal mode structure forms on the time scale of a few microseconds. The associated pressure perturbations $\delta P_{\rm mhd}(R,z)$ are shown in rows (c) and (e). However, we note that the dynamics of $\delta B_{\rm tor}$ seen in Fig.~\ref{fig:b01_transient-early} remain essentially the same when MHD density and pressure fluctuations are artificially suppressed. This implies that $\delta P_{\rm mhd}$ behaves merely as a {\it passive} scalar field during this initial transient relaxation up to the microsecond time scale, where the plasma on flux surfaces near the antenna adapts to the antenna's presence.

Row (f) of Fig.~\ref{fig:b01_transient-early} shows the evolution of the toroidal MHD velocity component $\delta u_{\rm tor}(R,z)$. In addition to the above-mentioned poloidal spreading of shear Alfv\'{e}n waves, which appear only faintly in $\delta u_{\rm tor}$, Fig.~\ref{fig:b01_transient-early}(f) shows yet another wave front that slowly advances poloidally away from the antenna at what appears to be the local speed of sound, $\left<v_{\rm S}\right>_{r_{\rm ant}} = \sqrt{\Gamma \beta(r_{\rm ant}) /2}\left<v_{\rm A}\right>_{r_{\rm ant}} \approx \sqrt{5/3\times 0.0016/2}\times 12.6\times 10^6\,{\rm m/s} \approx 0.5\times 10^6\,{\rm m/s}$, where $\beta(r_{\rm ant}) \approx \beta_0/30 \approx 0.16\%$ for $\beta_0 \approx 4.7\%$.

According to charge exchange spectroscopy (CES) measurements, the KSTAR plasma on which the present simulations are based had an on-axis ion temperature of about $T_{\rm i0}^{\rm CES} \approx 3\,{\rm keV}$, and about $1\,{\rm keV}$ in the region where our antenna is located. The local thermal speed of deuterons in that region would then be about $v_{\rm th,i} \approx 0.3\times 10^6\,{\rm m/s}$ for deuterons, so that an abrupt density perturbation would be expected to fill the flux surface within $0.5\times 2\pi q R_0/v_{\rm ti} \approx 60\,\mu{\rm s}$. The value estimated for $\left<v_{\rm S}\right>_{r_{\rm ant}}$ in the previous paragraph implies that our $\delta u_{\rm tor}$ fluctuations in Fig.~\ref{fig:b01_transient-early}(f) do so in about half of that time, so it lies in the right ballpark. This suggests that the early MHD response in Fig.~\ref{fig:b01_transient-early} should be quite realistic in the sense that a simulation with a kinetic ion model should yield similar results on this short time scale. Notable differences may be expected to arise only later on.

Since Fig.~\ref{fig:b01_transient-early} shows only the early dynamics on the time scale of a few microseconds, we expect to see established patterns (if any) only for modal components with relatively high characteristic frequencies $\nu \gtrsim 50\,{\rm kHz}$. In contrast, the antenna's oscillation at $\nu_{\rm ant} = 10\,{\rm kHz}$ is far from complete, so it cannot be know to the plasma yet.

The low-frequency response of the central core plasma becomes visible after about half of the antenna's oscillation cycle, $1/(2\nu_{\rm ant}) = 50\,\mu{\rm s}$, as one can see in Fig.~\ref{fig:b02_transient-late}, which shows the continuation of Fig.~\ref{fig:b01_transient-early}(d)--(f).

For completeness, Fig.~\ref{fig:b03_transient-standing} shows the situation in the case of a standing wave antenna, which drives a nearly standing waveform in the plasma center. The core mode's phase shows weak undulations, which are presumably due to the plasma's up-down asymmetry, which implies different path length' for co- and counter propagating shear Alfv\'{e}n waves above and below the midplane.

Finally, let us make a few more remarks concerning the role of plasma compressibility. In Figs.~\ref{fig:16_kstar-1.051_dpre0_evol-spec} and \ref{fig:17_kstar-1.051_dpre0_mstruc} of Section~\ref{sec:discuss_compress} of the main text, we found that suppressing the MHD pressure and density fluctuations ($\delta\rho_{\rm mhd},\delta P_{\rm mhd} \rightarrow 0 \; \forall t$) attenuates the core plasma response driven by a remote antenna but does not eliminate it. Indeed, effectively indistinguishable results were obtained during the first few dozen Alfv\'{e}n times as can be seen by comparing Figs.~\ref{fig:b04_transient_dpre0}(a-1)--(a-3) with corresponding snapshots in Figs.~\ref{fig:b01_transient-early}(a,d) and \ref{fig:b02_transient-late}(a-1). (The magnified fine structure in Fig.~\ref{fig:b01_transient-early}(b,c) at earlier times was so similar that we did not plot it again.) At later times, the main differences compared to the self-consistent simulation are
\begin{itemize}
	\item  the near-absence of a $\delta B_{\rm tor}$ response in the central core, where $\delta B_{\rm tor} \approx \delta B_\parallel$ (Fig.~\ref{fig:b04_transient_dpre0}(a) {\it vs}.\ Fig~\ref{fig:17_kstar-1.051_dpre0_mstruc});
	\item  a somewhat weaker central core response in $\delta E_{\rm tor}$ (approximately resembling $\delta\Phi$) shown in Fig.~\ref{fig:b04_transient_dpre0}(b);
	\item  the lack of dynamic evolution in the toroidal MHD velocity component $\delta u_{\rm tor}$ in Fig.~\ref{fig:b04_transient_dpre0}(c) due to the complete absence of acoustic waves.
\end{itemize}

\noindent These results confirm once more the expectations we expressed in the introductory Section~\ref{sec:history} that the acoustic component is not essential for the core-edge coupling studied in this paper, although it does have a significant quantitative impact on the low-frequency MHD plasma response.

\begin{figure}
	[tb]\vspace{-1.5cm}
	\centering
	\includegraphics[width=0.48\textwidth]{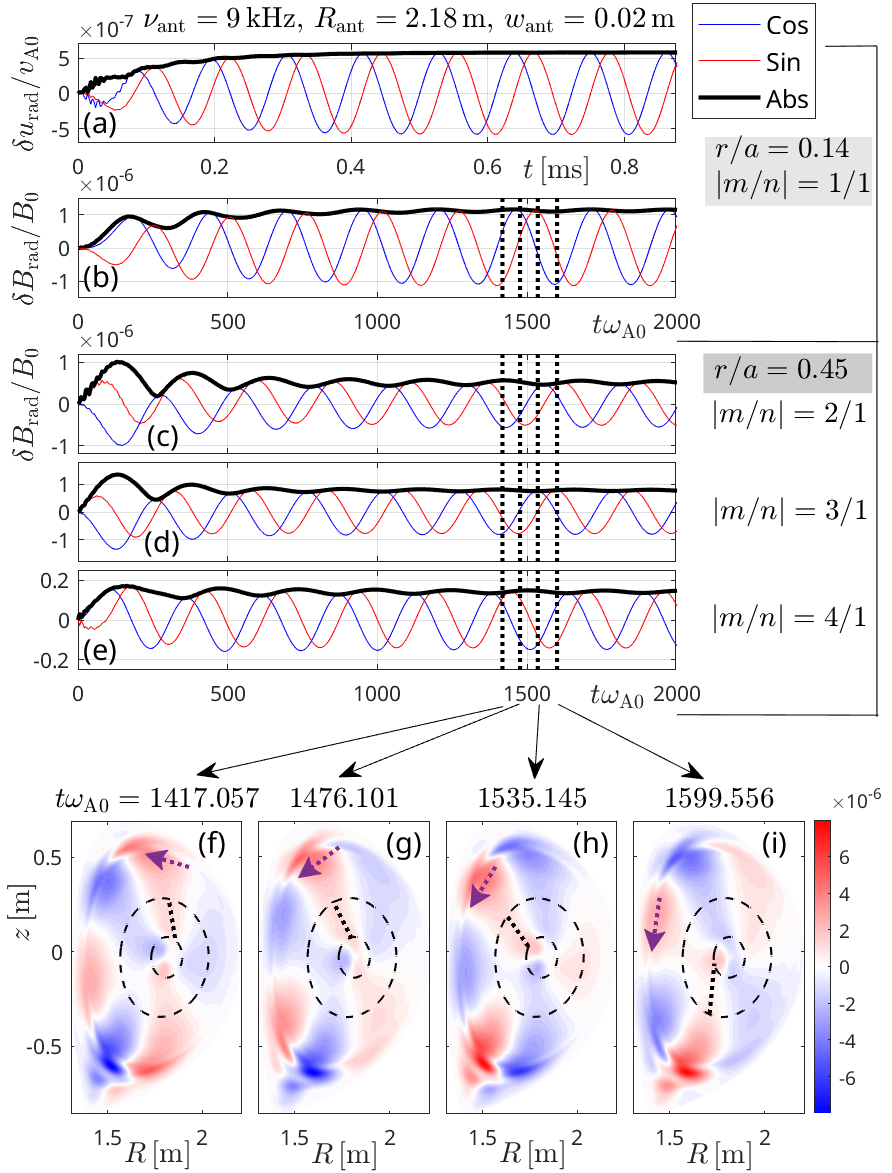}\vspace{-0.2cm}
	\caption{Mode pulsations in the presence of an edge-localized $9\,{\rm kHz}$ antenna with narrow width $w_{\rm ant} = 0.02\,{\rm m}$ and toroidal mode number $n_{\rm ant}=-1$. This is the same case as in Fig.~\protect\ref{fig:10_kstar-1.051_scan-w_n1}(b,e). Panels (a) and (b) show the time traces of the $m=1$ component of $[\delta\hat{u}_{\rm tor}]_{|n|=1}(t)$ and $[\delta\hat{B}_{\rm tor}]_{|n|=1}(t)$ measured at $\hat{r}=0.14$ on the outer midplane ($\vartheta = 0$). Panels (c)--(e) show the time traces of the $m=2,3,4$ components of $[\delta B_{\rm tor}]_{n=1}(t)$ at a larger radius $\hat{r} = 0.45$. For one oscillation period around $\hat{t} \approx 1500$ ($0.7\,{\rm ms}$), panels (f)--(i) show four snapshots of $[\delta\hat{B}_{\rm tor}]_{|n|=1}(R,z)$. The dashed ellipses in (f)--(i) indicate the flux surface contours at the above radii $\hat{r} = 0.14$ and $0.45$, which approximately delimit the toroidal shells dominated by poloidal Fourier components $m=1$, $m=2$, and $m\geq 3$. The phase of the $m=3$ component in the outer shell is highlighted by a violet arrow going counter-clockwise. The black dotted line shows how the $m=2$ component's phase is aligned with the $m=3$ component in (f)--(h) and then slips forward on the high-field side between snapshots (h) and (i).}\vspace{-0.5cm}
	\label{fig:b05_pulse_w0.02}%
\end{figure}

\begin{figure}
	[tb]\vspace{-1.5cm}
	\centering
	\includegraphics[width=0.48\textwidth]{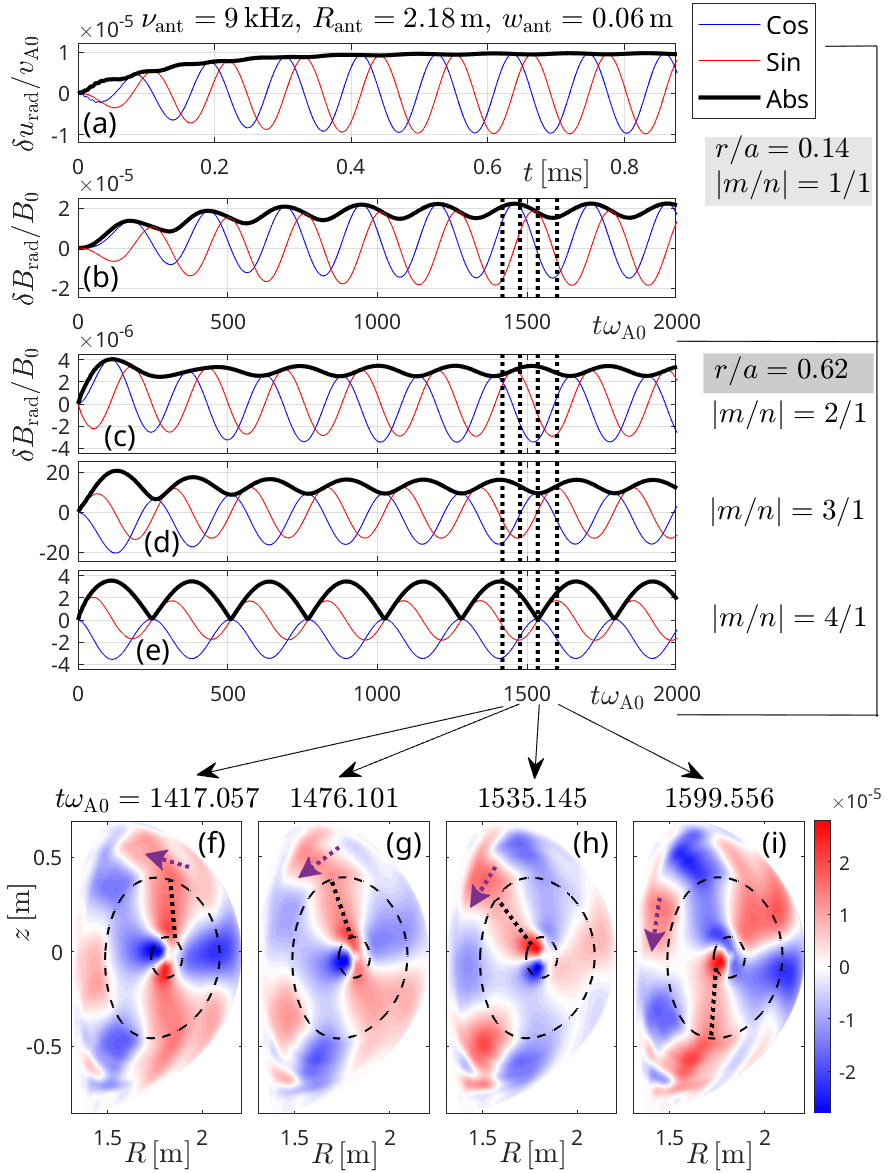}\vspace{-0.2cm}
	\caption{Mode pulsations in the presence of an edge-localized $9\,{\rm kHz}$ antenna with broader width $w_{\rm ant} = 0.06\,{\rm m}$. This is the same case as in Fig.~\protect\ref{fig:10_kstar-1.051_scan-w_n1}(d,g). Arranged as Fig.~\protect\ref{fig:b05_pulse_w0.02}.}\vspace{-0.35cm}
	\label{fig:b06_pulse_w0.06}%
\end{figure}

% -----------------------------------------------------------------------------
\subsection{Residual beating \& role of antenna width}
\label{apdx:modeform_pulse}

During the course of the two antenna oscillation cycles covered by Figs.~\ref{fig:b01_transient-early} and \ref{fig:b02_transient-late}, the radial mode structure formation process is largely completed. Details of the remaining variations in the mode structure and amplitude are shown in Figs.~\ref{fig:b05_pulse_w0.02} and \ref{fig:b06_pulse_w0.06} for the narrow ($w_{\rm ant} = 0.02\,{\rm m}$) and broader ($w_{\rm ant} = 0.06\,{\rm m}$) edge-localized $9\,{\rm kHz}$ antenna, respectively. Panels (a) and (b) of both figures show the time traces of the $m=1$ Fourier component of the toroidal flow $[\delta\hat{u}_{\rm tor}]_{|n|=1}$ and radial magnetic fluctuations $[\delta\hat{B}_{\rm rad}]_{|n|=1}$ at the point $(\hat{r},\vartheta) = (0.14,0)$ near the center. Panels (c)--(e) show the $m=2,3,4$ components of $[\delta\hat{B}_{\rm rad}]_{n=1}$ near mid-radius: $\hat{r} = 0.45$ in Fig.~\ref{fig:b05_pulse_w0.02} and $\hat{r} = 0.62$ in Fig.~\ref{fig:b06_pulse_w0.06}. These two radii are chosen to coincide roughly with the flux surfaces where the poloidal structure of $[\hat{B}_{\rm rad}]_{|n|=1}(R,z)$ in panels (f)--(i) changes from dominant $m=2$ to dominant $m=3$ in the respective figure.

\begin{figure}
	[tb]%\vspace{-0.3cm}
	\centering
	\includegraphics[width=0.48\textwidth]{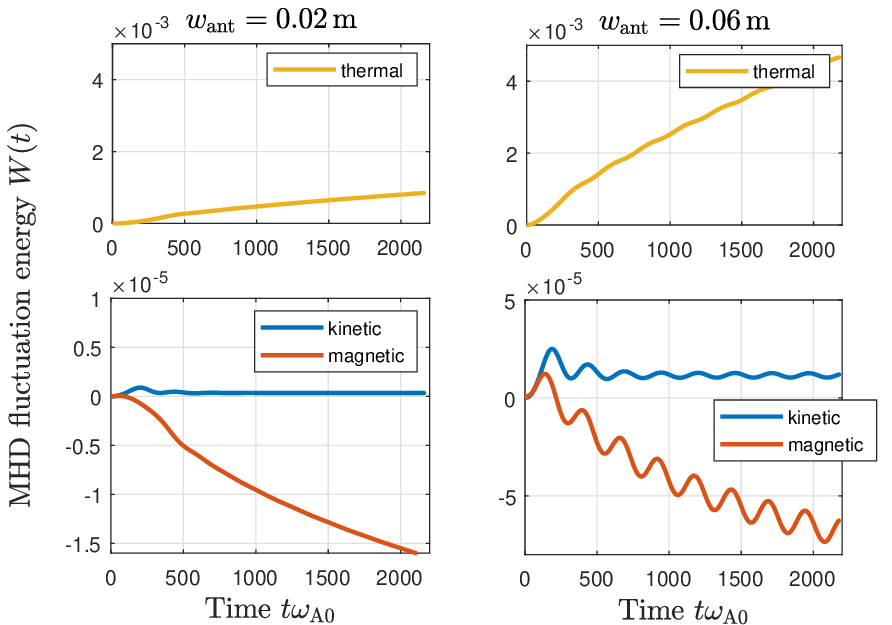}\vspace{-0.1cm}
	\caption{Time traces of (a,b) thermal and (c,d) kinetic and magnetic component of the MHD fluctuation energy (\protect\ref{eq:dg_wn}) in the cases with narrower (left) and broader antenna (right), whose details are shown in Figs.~\protect\ref{fig:b05_pulse_w0.02} and \protect\ref{fig:b06_pulse_w0.06}, respectively.}\vspace{-0.3cm}
	\label{fig:b07_pulse_wphys_w0.02-0.06}%
\end{figure}

In panel (a) of both figures, one can see that $\delta u_{\rm tor}$ takes a few $100\,\mu{\rm s}$ to saturate, with small residual pulsations. These pulsations are significantly larger in the time traces of $\delta B_{\rm rad}$ in panels (b)--(e). In the case with narrow antenna in Fig.~\ref{fig:b05_pulse_w0.02}, the pulsation amplitude is large initially but decays to a small fraction after a few oscillation cycles. In the case with the broader antenna in Fig.~\ref{fig:b06_pulse_w0.06}, the pulsation amplitude remains close to its initial level. The most extreme case is the $m=4$ component in Fig.~\ref{fig:b06_pulse_w0.06}(e) at $\hat{r} = 0.62$, where the pulsation magnitude equals the field's total oscillation amplitude. In this case, the pulsations are clearly visible even in the volume-integrated energy, whose time traces are plotted in Fig.~\ref{fig:b07_pulse_wphys_w0.02-0.06}.

All time traces in Figs.~\ref{fig:b05_pulse_w0.02} and \ref{fig:b06_pulse_w0.06} are perfectly sinusiodal and one can see that, mathematically, the pulsations are the result of a DC offset in the cosine component that is plotted as a blue curve in panels (a)--(f) in both Figs.~\ref{fig:b05_pulse_w0.02} and \ref{fig:b06_pulse_w0.06}. In the limit where the DC component is as large as the AC component, namely in Fig.~\ref{fig:b06_pulse_w0.06}(e), the absolute value (black curve) has the form: $[(1 + \cos(\omega t))^2 + \sin^2(\omega t)]^{1/2} = |2 + 2\cos(\omega t)|^{1/2} = 2\left|\cos\left(\tfrac{\omega}{2}t\right)\right| = \text{Env}\{\cos(\Omega t) + \cos(\Omega t+\omega t)\}$, which one can recognize to be the envelope of a perfect beat wave with arbitrary carrier frequency $\Omega/2 \gg |\omega|$.

Information hinting at the physical cause of the DC offset and associated pulsations can be gleaned from the relative phases of different $m$-components. Figs.~\ref{fig:b05_pulse_w0.02}(c,d) show that the $m=2$ and $m=3$ components of $\delta B_{\rm rad}$ are not quite but nearly phase-aligned on the low-field-side midplane, where our measurements are taken. This implies that their phases must slip on the high-field side, which is indeed observed in the snapshots shown in panels (f)--(i) of both figures. Meanwhile, $m=1$ in panel (b) is approximately out-of-phase with $m=2$ in (c), and $m=4$ in panel (e) is approximately out-of-phase with $m=3$ in (d). This means that $m=3$ and $m=4$ are aligned on the high-field side, which is consistent with the observed trend of the antenna-driven $\delta B_{\rm rad}$ being weaker on the low-field side and stronger on the high-field side.\footnote{A possible explanation was provided in the caption of Fig.~\protect\ref{fig:13_kstar-1.051_mstruc}.}
In contrast, the $m=4$ component is forced to oscillate in-phase with $m=3$ in Fig.~\ref{fig:b06_pulse_w0.06}, which is consistent with the fact that the wider antenna in that case extends over both the $q=3$ and $q=4$ surfaces. Apparently, the antenna-enforced phase-locking between $m=3$ and $m=4$ on the low-field side is (for currently unknown reasons) ``unnatural'' for this configuration and thus leads to pulsations.

Finally, Fig.~\ref{fig:b07_pulse_wphys_w0.02-0.06} shows that the pulsations are most pronounced in the magnetic (red) and less so in the kinetic (blue) component of the volume-integrated MHD fluctuation energy $W$. Moreover, while the kinetic part saturates after a few $100$ Alfv\'{e}n times, it is interesting to see that the magnetic energy continues to increase towards negative values. Apparently, our weak antenna ($\hat{A}_{\rm ant} = 2\times 10^{-8}$) is slowly transforming the magnetic equilibrium in its vicinity, taking it gradually to a state of lower total magnetic energy. One likely reason is that the antenna alters the local pressure profile due to its thermal heating effect (yellow), which is larger by two orders of magnitude than the change in magnetic energy perturbation and, thus, is the antenna's dominant effect on the system's total energy.

\begin{figure*}
	[tb]\vspace{-1.6cm}
	\centering
	\includegraphics[width=0.96\textwidth]{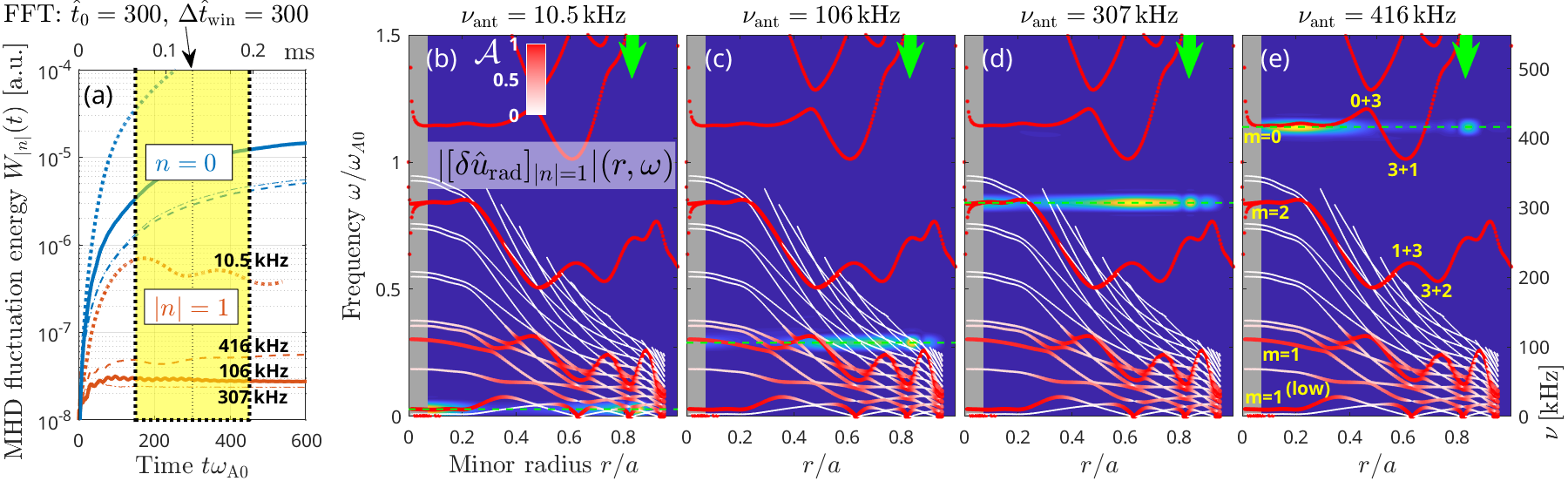}\vspace{-0.3cm}
	\caption{Antenna frequency scan into the range of toroidicity- and ellipticity-induced gaps in our KSTAR Model 1. Arranged as Fig.~\protect\ref{fig:09_kstar-1.051_scan-f_n1}, except that the FFT window is reduced to $\Delta\hat{t}_{\rm win} = 300$, and {\tt FALCON} \protect\cite{Falessi20, Falessi19b} continua cover only the lower part of the spectrograms (up to where the white lines end). Shear Alfv\'{e}n continua above $200\,{\rm kHz}$ were computed by solving the equations in Appendix~A of Ref.~\protect\cite{Deng12} in the slow-sound approximation. The antenna of width $w_{\rm ant} = 0.02\,{\rm m}$ is again located at $\hat{r}_{\rm ant} = 0.85$ (green arrow).}\vspace{-0.1cm}
	\label{fig:b10_kstar-1.051_scan-hi-f_n1}%
\end{figure*}

\begin{figure*}
	[tb]
	\centering
	\includegraphics[width=0.96\textwidth]{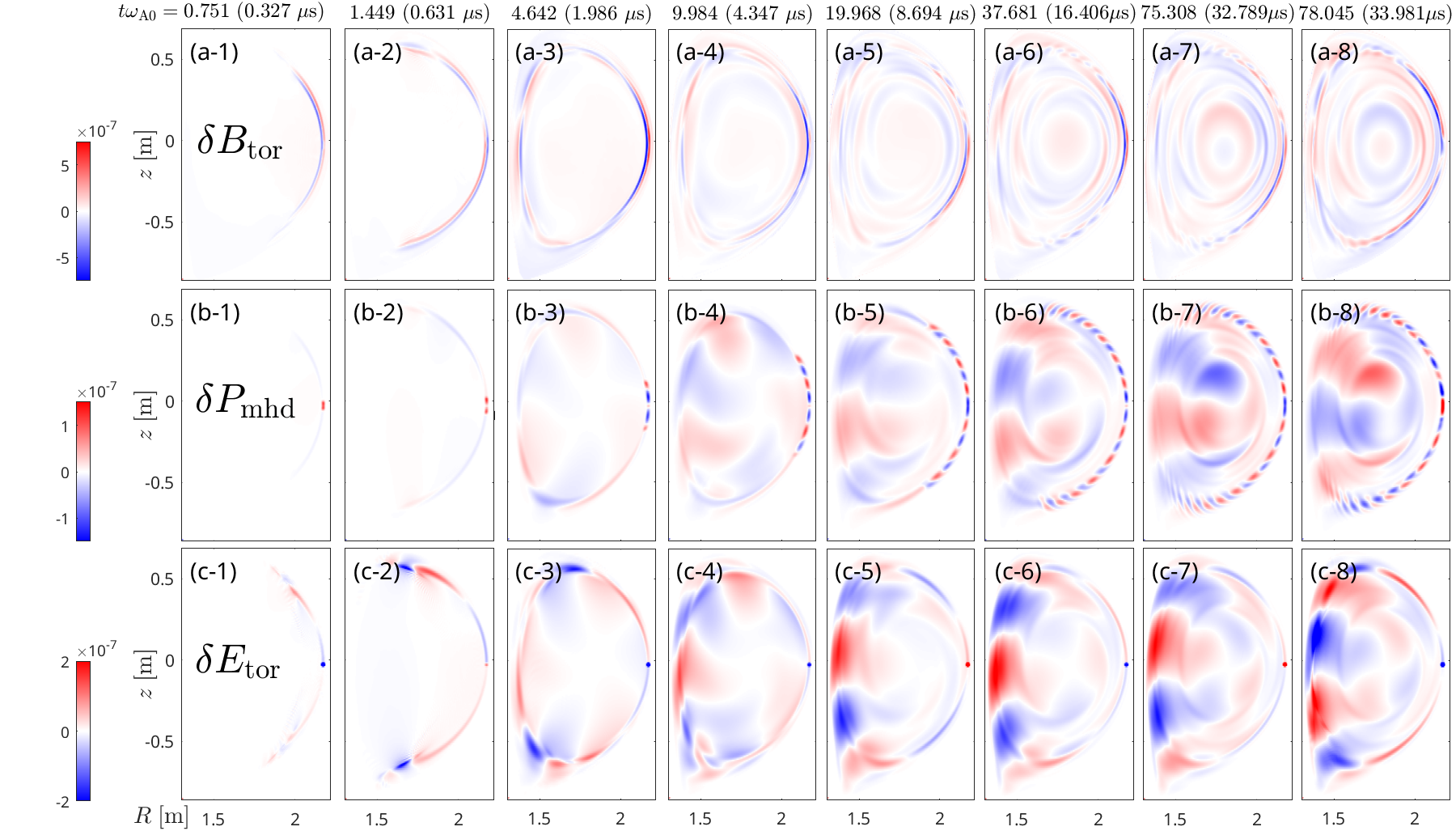}\vspace{-0.3cm}
	\caption{Early transients and mode structure formation at high antenna frequency $\nu_{\rm ant} = 416\,{\rm kHz}$, matching the central $|m/n|=0/1$ continuum plateau in our KSTAR Model 1 as in Fig.~\protect\ref{fig:b10_kstar-1.051_scan-hi-f_n1}(e). Arranged as Figs.~\protect\ref{fig:b01_transient-early} and \protect\ref{fig:b02_transient-late}, except that row (c) shows the toroidal component $\delta E_{\rm tor}$ of the electric field. ($\delta u_{\rm tor}$ is not shown since it is similar to the high-$m$ component of $\delta P_{\rm mhd}$ near the antenna.)}\vspace{-0.5cm}
	\label{fig:b11_transient_416kHz}%
\end{figure*}

% -----------------------------------------------------------------------------
\subsection{Response at higher frequencies}
\label{apdx:modeform_highfreq}

The present paper focused on the low-frequency response ($\sim 10\,{\rm kHz}$) of an MHD plasma to an internal antenna. We demonstrated that a core-localized Alfv\'{e}nic continuum plateau can respond with coherent oscillations to a distant but resonant antenna located near the plasma edge. We also found that the antenna-driven low-frequency MHD response had a significant compressional component, which is perhaps not surprising within the beta-induced gap, which in the KSTAR plasma studied here occupies the frequency range $\nu \lesssim 50\,{\rm kHz}$.

Out of curiosity and for completeness, we have also tested whether such edge-to-core coupling works in the range of toroidicity-induced and ellipticity-induced  gaps and eigenmodes, which occupy the frequency range $100\,{\rm kHz} \lesssim \nu \lesssim 400\,{\rm kHz}$, where compressional effects are expected to be weaker \cite{Du24,Heidbrink25d}. Multi-peaked compound modes in this frequency range were described in Ref.~\cite{Gorelenkov05, Mishchenko14}.

Similarly to the frequency scan in Fig.~\ref{fig:09_kstar-1.051_scan-f_n1} of the main text, Fig.~\ref{fig:b10_kstar-1.051_scan-hi-f_n1} shows (a) the time traces of the MHD fluctuation energy $W_{|n|=0,1}(t)$ and (b)--(e) spectrograms of the plasma response with narrow antennas located near the plasma edge and oscillating at $\nu_{\rm ant} = 10.5$, $106$, $307$ and $416\,{\rm kHz}$. Here, the spectrograms where computed from $\delta \hat{u}_{|n|=1}$ instead of $\delta\hat{\Phi}_{|n|=1}$ from Eq.~(\ref{eq:epot}) because we are also going to look at the $m=0$ component.

The $10.5\,{\rm kHz}$ case in Fig.~\ref{fig:b10_kstar-1.051_scan-hi-f_n1}(b) is similar to that Fig.~\ref{fig:09_kstar-1.051_scan-f_n1}(d), where we excited the central plateau of the lower Alfv\'{e}nic $|m/n|=1/1$ continuum branch. The $106\,{\rm kHz}$ antenna in Fig.~\ref{fig:b10_kstar-1.051_scan-hi-f_n1}(c) was meant to excite the central plateau ($\hat{r}\lesssim 0.2$) of the upper Alfv\'{e}nic $|m/n| = 1/1$ continuum branch, but the signal is dominated by a peak in the range $0.4 \lesssim \hat{r} \lesssim 0.8$, which may be a toroidal Alfv\'{e}n eigenmode (TAE) \cite{Cheng85}. The $307\,{\rm kHz}$ antenna in Fig.~\ref{fig:b10_kstar-1.051_scan-hi-f_n1}(d) was meant to excite the central plateau of the $|m/n|=2/1$ Alfv\'{e}n continuum branch. There seems to be some response, but the signal is dominated by a peak in the range $0.3 \lesssim \hat{r} \lesssim 0.7$, which may be an ellipticity-induced Alfv\'{e}n eigenmode (EAE) \cite{Betty91}. The $416\,{\rm kHz}$ antenna in Fig.~\ref{fig:b10_kstar-1.051_scan-hi-f_n1}(e) yields a clear response on the wide central plateau ($\hat{r} \lesssim 0.4$) of the $|m/n|=0/1$ Alfv\'{e}n continuum branch.

The formation process of the quasi-mode at $416\,{\rm kHz}$ can be observed in Fig.~\ref{fig:b11_transient_416kHz}(a), where a core-localized $m=0$ structure forms in snapshots of $\delta B_{\rm tor}$ after about $30\,\mu{\rm s}$, which is comparable to the saturation time $\tau_{\rm sat} \sim 50\,\mu{\rm s}$ that can be inferred from Fig.~\ref{fig:b10_kstar-1.051_scan-hi-f_n1}(a). The mode structure formation takes significantly longer than $\tau_\omega \approx 0.5/\nu_{\rm ant} \sim 1\,\mu{\rm s}$ (introduced in Section~\ref{sec:results_transients}), which may be due to the fact that $\tau_\omega$ is short and comparable to $\tau_\vartheta \sim 1\,\mu{\rm s}$, so that volumetric focusing may be expected to be less efficient here, as discussed in Section~\ref{sec:discuss_focus} of the main text.

One may wonder why $\delta P_{\rm mhd}$ in row (b) and $\delta E_{\rm tor}$ in row (c) of Fig.~\ref{fig:b11_transient_416kHz} do not show the $m=0$ structure in the central core, but something closer to $m=1$ and $m=2$. The difference between the structure of $\delta B_{\rm tor}$ and $\delta P_{\rm mhd}$ implies that, unlike in Fig.~\ref{fig:b02_transient-late}(a,b), this is not a compressional response. Instead, $\delta P_{\rm mhd}$ seems to be produced by radial ${\bm E}\times{\bm B}$ flows: $|\delta u_{\rm E,rad}| \approx |\partial_\vartheta\delta\Phi/(r B_{\rm tor})|$. Since the leading-order $m=0$ component of $\delta\Phi$ gives $\partial_\vartheta\delta\Phi = 0$, it is clear that $\delta P_{\rm mhd}$ in Fig.~\ref{fig:b11_transient_416kHz}(b) is dominated by higher-order terms with $m>0$. The same is true for $\delta E_{\rm tor}$ in Fig.~\ref{fig:b11_transient_416kHz}(c): $|\delta E_{\rm tor}| = |-\partial_t\delta A_\parallel - \partial_\zeta\delta\Phi/R| \approx |\partial_\vartheta\delta\Phi/r| + \{\text{higher-order terms}\}$, with the approximations $q \approx 1$ and $E_\parallel\approx 0$. Interestingly, these higher-order Fourier components of the high-frequency quasi-mode are phase-aligned on the high-field side. In contrast, the core and edge components of the low-frequency plasma response studied in the main part of this paper were phase-aligned on the low-field side, as we saw in Figs.~\ref{fig:14_kstar-1.051_phasing} and \ref{fig:b02_transient-late}. The reason for this difference is not known to us.

Finally, the high-$m$ structure of $\delta P_{\rm mhd}$ near the antenna radius in Fig.~\ref{fig:b11_transient_416kHz}(b) can be explained in terms of acoustic waves, which have short poloidal wavelengths at high frequencies like the present $416\,{\rm kHz}$. The initial poloidal spreading of this high-$m$ component in Fig.~\ref{fig:b11_transient_416kHz}(b) indeed proceeds at the same rate as the acoustic spreading of $\delta u_{\rm tor}$ discussed in Appendix~\ref{apdx:modeform_transients} and seen in Figs.~\ref{fig:b01_transient-early} and \ref{fig:b02_transient-late}.

\begin{figure*}
	[tb]
	\centering\vspace{-1.5cm}
	\includegraphics[width=0.96\textwidth]{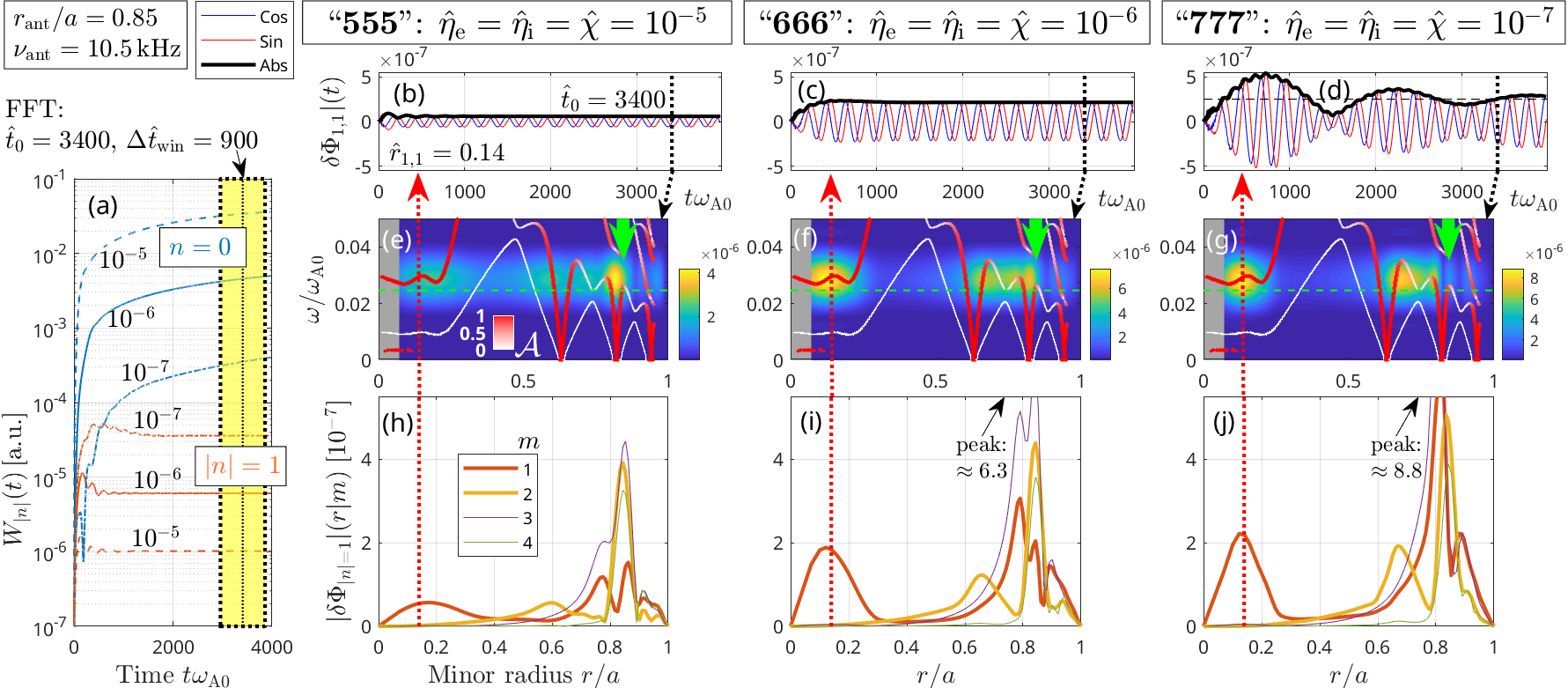}\vspace{0.2cm}
	\includegraphics[width=0.96\textwidth]{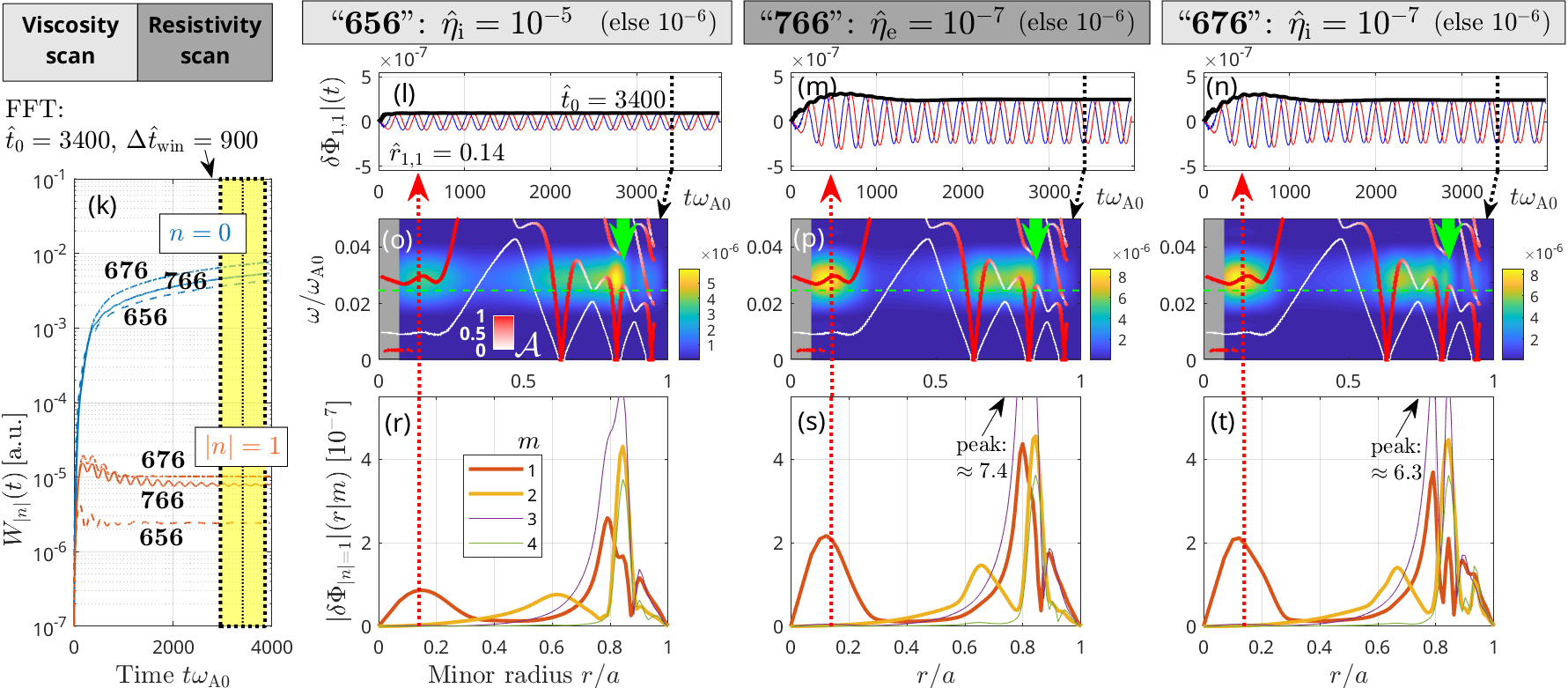}\vspace{-0.3cm}
	\caption{Sensitivity of the core-localized plasma response to a narrow edge-localized $10.5\,{\rm kHz}$ antenna when varying the dissipation coefficients of our MHD model. As a reference, we use KSTAR Model 1 with $\hat{\eta}_{\rm e} = \hat{\eta}_{\rm i} = \hat{\chi} = 10^{-6}$, results for which are shown in panels (c,f,i). The case is similar to that in Fig.~\protect\ref{fig:09_kstar-1.051_scan-f_n1}(d), except that $\nu_{\rm ant}$ was slightly increased from $9\,{\rm kHz}$ to $10.5\,{\rm kHz}$ and the number of grid points was increased to $N_R\times N_z\times N_\zeta = 1024\times 1024\times 16$. The upper half of the figure shows results obtained when using identical values for the normalized electric resistivity $\hat{\eta}_{\rm e}$, ion viscosity $\hat{\eta}_{\rm i}$ and thermal diffusivity $\hat{\chi}$, here $10^{-5}$, $10^{-6}$ and $10^{-7}$. Panel (a) shows the time traces of the total MHD fluctuation energy $W_{|n|}(t)$ from Eq.~(\protect\ref{eq:dg_wn}) for $n=0$ (blue) and $|n|=1$ (orange). Panels (b)--(d) show the time traces of the $|m/n|=1/1$ Fourier component of the electrostatic potential $\delta\hat\Phi_{1,1}(t|r)$ at radius $\hat{r} = 0.14$, measuring the response of the core-localized plateau of the $|m/n|=1/1$ low-frequency Alfv\'{e}nic continuum branch. For the time window highlighted yellow in (a), panels (e)--(g) show the spectrograms and continua for each case (arranged as Fig.~\protect\ref{fig:09_kstar-1.051_scan-f_n1}). Panels (h)--(j) show the radial profiles of the dominant poloidal Fourier components $m=1,2,3,4$ of $|\delta\hat\Phi_{n=1}|(r|m)$ at time $\hat{t}_0 = 2800$ (about $1.2\,{\rm ms}$). The bottom half of the figure shows the results obtained by varying only the viscosity $\hat{\eta}_{\rm i}$ or only the resistivity $\hat{\eta}_{\rm e}$.}
	\label{fig:c01_scan-eta-nu-chi}\vspace{-0.5cm}
\end{figure*}

% =============================================================================
\section{Role of non-ideal effects}
\label{apdx:dissipation}

% -----------------------------------------------------------------------------
\subsection{Sensitivity to resistivity \& viscosity}
\label{apdx:dissipation_sensitivity}

Non-ideal effects enter in the present study primarily in the form of an electric resistivity $\eta_{\rm e}$ in Ohm's law and an ion viscosity $\eta_{\rm i}$ in the MHD momentum balance equation. See Eqs.~(3) and (6) in Ref.~\cite{Bierwage17a}. In addition, the equation of state and the continuity equation each contain a diffusion term with coefficient $\chi$, whose effect on the main results seems to be negligible here, so the sensitivity study in this appendix discusses only the effects of the normalized resistivity $\hat{\eta}_{\rm e} = S^{-1}$ and viscosity $\hat{\eta}_{\rm i} = {\rm Re}^{-1}$ as defined in Eq.~(\ref{eq:dissipation}) in Section~\ref{sec:history}. The results are summarized in Fig.~\ref{fig:c01_scan-eta-nu-chi}.

We use KSTAR Model 1, which is MHD stable for our default settings and driven by a narrow antenna located near the edge at $\hat{r}_{\rm ant} = 0.85$. The case is similar to that in Fig.~\protect\ref{fig:09_kstar-1.051_scan-f_n1}(d), except that the antenna frequency $\nu_{\rm ant}$ was slightly increased from $9\,{\rm kHz}$ to $10.5\,{\rm kHz}$ to better match the central plateau of the low-frequency $|m/n|=1/1$ Alfv\'{e}nic  continuum. The results in Fig.~\ref{fig:c01_scan-eta-nu-chi} were obtained with a relatively fine mesh, consisting of $N_R\times N_z\times N_\zeta = 1024\times 1024\times 16$ grid points, in order to keep numerical dissipation low in simulations with small diffusion coefficients ($10^{-7}$). Similar results (not shown) were obtained with a coarser mesh consisting of $N_R\times N_z = 400\times 400$ grid points in the poloidal plane, except that the mode amplitudes were smaller, presumably because the antenna was not resolved well and thus coupled less efficiently to the plasma. This is, in fact, the reason why all cases shown in Fig.~\ref{fig:c01_scan-eta-nu-chi} were run with the same spatial resolution: A quantitative comparison between the mode amplitudes requires that the antenna profile is resolved equally well in all cases.
 
The default value of the dissipation coefficients was $\hat{\eta}_{\rm e} = \hat{\eta}_{\rm i} = \hat{\chi} = 10^{-6}$, and the corresponding results are shown in Fig.~\ref{fig:c01_scan-eta-nu-chi}(c,f,i). The left and right neighboring panels in the upper half of Fig.~\ref{fig:c01_scan-eta-nu-chi} show results obtained when reducing (b,e,h) or increasing (d,g,j) all diffusion coefficients by a factor $10$. This means that the magnetic Prandtl number is fixed at ${\rm Pr} \equiv \hat{\eta}_{\rm i}/\hat{\eta}_{\rm e} = 1$. The labels ``555'', ``666'' and ``777'' at the top of each panel set are meant to compactly represent the (unsigned) exponents of $\hat{\eta}_{\rm e}$, $\hat{\eta}_{\rm i}$, $\hat{\chi}$.

Panel (a) shows that reducing the diffusion coefficients from $10^{-5}$ via $10^{-6}$ to $10^{-7}$ decreases the energy $W_{|n|}$ of the $n=0$ component and increases that of the $|n|=1$ component, as expected from the reduction of plasma heating. The increase in $W_{|n|=1}$ is primarily due to larger fluctuation amplitudes around the antenna radius $\hat{r} \approx \hat{r}_{\rm ant} = 0.85$, while the amplitude of the core-localized quasi-mode is less sensitive, at least for $10^{-6} \rightarrow 10^{-7}$. This can be seen in panels (h)--(j), where we plot the radial profiles of the electrostatic potential's Fourier components $|\delta\hat\Phi_{|n|=1}|(r|m)$ with $m=1,2,3,4$.

In the case where all diffusion coefficients are $10^{-7}$ (``777''), Fig.~\ref{fig:c01_scan-eta-nu-chi}(d) shows that the amplitude $\delta\hat\Phi_{1,1}(t)$ at $\hat{r}_{1,1}=0.14$ as defined in Eq.~(\ref{eq:philoc}) takes about $5000$ Alfv\'{e}n times ($2\,{\rm ms}$) to settle at a quasi-steady value. Actually, the overall structure of the core-localized quasi-mode is more or less established long before that, so the central peak in Fig.~\ref{fig:c01_scan-eta-nu-chi}(j) is representative also for earlier times. The only exception is a brief period around $\hat{t} \approx 1500$, where $\delta\hat\Phi_{1,1}(t)$ has a deep minimum and the mode structure briefly disintegrates into multiple peaks (not shown). However, the quasi-mode soon recovers. In the end, its amplitude settles near $\delta\hat{\Phi}_{1,1} \approx 2\times 10^{-7}$ as shown in panel (j) for $\hat{\eta}_{\rm e} = \hat{\eta}_{\rm i} = \hat{\chi} = 10^{-7}$, which is similar to the quasi-steady amplitude in panel (i) for $10^{-6}$ (```666'').

In contrast, Fig.~\ref{fig:c01_scan-eta-nu-chi}(b,h) shows that increasing the diffusion coefficients from $10^{-6}$ to $10^{-5}$ (``555'') yields a significantly lowerd quasi-steady amplitude of $\delta\hat{\Phi}_{1,1} \approx 0.6\times 10^{-7}$. Moreover, the peaks of all Fourier components $|\delta\hat\Phi_{n=1}|(r|m)$ in Fig.~\ref{fig:c01_scan-eta-nu-chi}(h) broaden noticeably, so that fluctuations become also more pronounced at intermediate radii $0.3\lesssim\hat{r}\lesssim 0.5$. This broadening is also visible in the spectrogram in Fig.~\ref{fig:c01_scan-eta-nu-chi}(e).

The lower half of Fig.~\ref{fig:c01_scan-eta-nu-chi} shows results obtained when varying only the viscosity $\hat{\eta}_{\rm i}$ (``656'' left, and ``676'' right) or reducing the resistivity $\hat{\eta}_{\rm e}$ (``766'' center) while fixing the remaining diffusion coefficients at $10^{-6}$. Comparison between panels (r) and (h) shows that both higher resistivity and higher viscosity contribute to the broadening of the mode structure and its reduced amplitude, with viscosity perhaps taking a larger share. Panels (m) and (n) show that it suffices to keep either $\hat{\eta}_{\rm e}$ or $\hat{\eta}_{\rm i}$ at $10^{-6}$ (denoted as ``676'' and ``766'', respectively) to facilitate a rapid relaxation and prevent the long-lasting pulsations that were seen for ``777'' in panel (d).\footnote{We emphasize again that $\hat{\chi}$ has no significant effect here.}

With regard to the main theme of this work, namely core-edge coupling, the interpretation of these results is not entirely straightforward. On the one hand, if we consider that the antenna perturbation is subject to less dissipative damping when $\hat{\eta}_{\rm e}$ is reduced, one may think that the core-edge coupling efficiency is reduced because the amplitude of the core-localized quasi-mode does not increase in proportion to the fluctuation amplitude near the antenna in Fig.~\ref{fig:c01_scan-eta-nu-chi}. On the other hand, we did not change the value of $\hat{A}_{\rm ant}$ in Eq.~(\ref{eq:setup_ant}) and the antenna is equally well resolved in all cases, so the similarity of the saturated amplitudes in cases ``666'', ``777'', ``766'' and ``676'' means that, in the end, the core-localized quasi-mode is driven with similar efficiency when $\hat{\eta}_{\rm e,i} \lesssim 10^{-6}$. Meanwhile, diffusion becomes noticeable on a global scale when $\hat{\eta}_{\rm e,i} = 10^{-5}$.

Our current conclusion based on the sensitivity study presented in this appendix is that the simulations with $\hat{\eta}_{\rm e} = \hat{\eta}_{\rm i} = \hat{\chi} = 10^{-6}$ in the main part of the paper provide a reasonable estimate of the plasma response in a weakly collisional plasma, while accelerating convergence by reducing pulsation like those seen for $10^{-7}$ in Fig.~\ref{fig:c01_scan-eta-nu-chi}(d).

The classical (collisional) diffusivities in the central core of the KSTAR plasma that we have modeled here are much smaller. For instance, the normalized Spitzer resistivity is estimated to be about $\hat{\eta}_{\rm e} = S^{-1} \sim 10^{-9}$. The effective diffusivities (such as the fluctuation-induced ``anomalous'' resistivity) could be closer to the values used in our MHD simulations, but verifying this is beyond the scope of this work and currently not on our agenda. Instead, our plan is to find out to what to extent the MHD core-edge coupling phenomena described in the present work can be reproduced with kinetic models. It would then also be interesting to assess the role of kinetic Alfv\'{e}n waves and their radial propagation, although it is possible that their effect is minor (as was speculated in Section 5 of Ref.~\cite{DeChambrier82}).\vspace{-0.3cm}

\begin{figure}
	[tb]%\vspace{-1.5cm}
	\centering
	\includegraphics[width=0.48\textwidth]{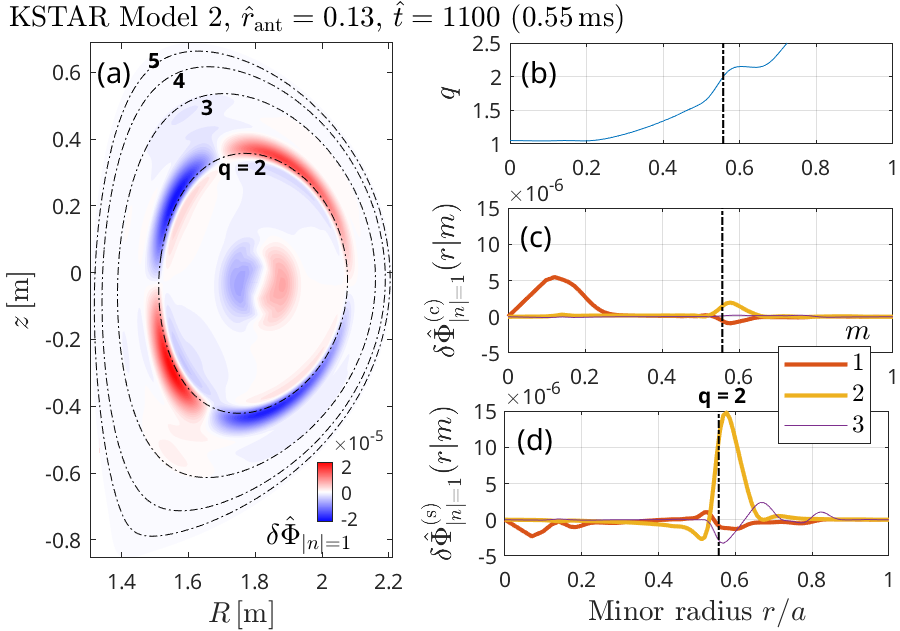}%\vspace{-0.2cm}
	\caption{Structure of the unstable $|m/n|=2/1$ tearing mode in our KSTAR Model 2. This is the same case as in Fig.~\protect\ref{fig:11_kstar-q1.051-flat-q2.15_rant-scan}(d,k,l,p,t), which is driven by a narrow $9\,{\rm kHz}$ antenna located at $\hat{r}_{\rm ant} \approx 0.85$. Panel (a) shows the structure of the electrostatic potential $\delta\hat{\Phi}_{|n|=1}(R,z)$ in the poloidal plane. Underneath the $q$ profile in panel (b), we plot the radial profiles of the poloidal Fourier components $m=1,2,3$ of $\delta\hat{\Phi}_{|n|=1}$; the cosine part in (c) and the sine part in (d). At the time of this snapshot, the antenna-driven $|m/n|=1/1$ response in the central $q\sim 1$ region is still clearly visible, albeit with some distortion.}%\vspace{-0.5cm}
	\label{fig:c02_kstar-q1.051-flat-q2.15_tearing}%
\end{figure}

% -----------------------------------------------------------------------------
\subsection{Tearing instability}
\label{apdx:dissipation_tm}

In Section~\ref{sec:results_scan-r_focus}, we used our KSTAR Model 2 with two continuum plateaus located at different radii to quantify the effect of volumetric focusing. The outer plateau was located just outside of the $q=2$ surface, causing the magnetic shear at the $q=2$ surface to increase beyond the stability threshold of the $|m/n| = 2/1$ resistive tearing mode.

For confirmation, the spatial structure of the tearing mode's fluctuating electrostatic potential $\delta\hat{\Phi}_{|n|=1}$ is shown in Fig.~\ref{fig:c02_kstar-q1.051-flat-q2.15_tearing}. In addition to the tearing mode at the $q=2$ surface, the antenna-driven response at the central $|m/n|=1/1$ continuum plateau is also visible, albeit with some distortion that is presumably caused by the tearing mode.

\begin{figure}
	[tb]%\vspace{-1.5cm}
	\centering
	\includegraphics[width=0.48\textwidth]{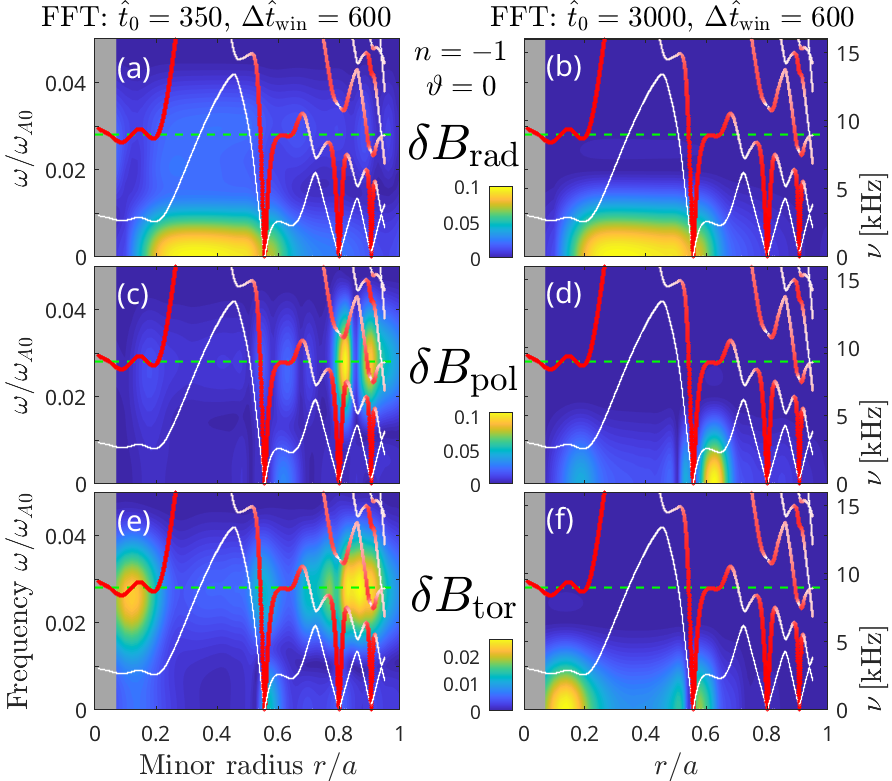}%\vspace{-0.3cm}
	\caption{Early (left) and late (right) spectrograms of the magnetic perturbation $\delta{\bm B}_{|n|=1}$ in KSTAR Model 2, which is subject to an unstable $|m/n|=2/1$ tearing mode at the $q=2$ surface ($\hat{r} \approx 0.55$). Each panel is arranged as Fig.~\protect\ref{fig:11_kstar-q1.051-flat-q2.15_rant-scan}(p) for $|\delta\hat{\Phi}_{|n|=1}|(r,\omega)$. Here we show (a,b) $|[\delta\hat{B}_{\rm rad}]_{|n|=1}|(r,\omega)$, (c,d) $|[\delta\hat{B}_{\rm pol}]_{|n|=1}|(r,\omega)$ and (e,f) $|[\delta\hat{B}_{\rm tor}]_{|n|=1}|(r,\omega)$.}%\vspace{-0.5cm}
	\label{fig:c03_kstar-1.051-flat-q2.15_dB}%
\end{figure}

The spectrogram of $\delta\hat{\Phi}_{|n|=1}$ was shown in Fig.~\ref{fig:11_kstar-q1.051-flat-q2.15_rant-scan}(p) of the main text. For completeness, Fig.~\ref{fig:c03_kstar-1.051-flat-q2.15_dB} shows the spectrograms of the three components of the fluctuating magnetic field $\delta\hat{\bm B}_{|n|=1}$. During the first few hundred Alfv\'{e}n times, the zero-frequency tearing mode already dominates (a) $\delta B_{\rm rad}$, whereas (c) $\delta B_{\rm pol}$ and (e) $\delta B_{\rm tor}$ are still dominated by the $9\,{\rm kHz}$ antenna response. As we have seen in Fig.~\ref{fig:11_kstar-q1.051-flat-q2.15_rant-scan}(d), the tearing mode saturates at about $t\omega_{\rm A0} \approx 2000$ ($\approx 1\,{\rm ms}$), with an amplitude that is several orders of magnitude larger than the antenna-driven fluctuations. The spectrograms of $\delta\hat{\bm B}_{|n|=1}$ in the saturated state are shown in Fig.~\ref{fig:c03_kstar-1.051-flat-q2.15_dB}(b,d,f). One can see that the perturbation reaches all the way into the neighborhood of the magnetic axis, where we observe a peak in $\delta B_{\rm tor}$.

It is possible that the tearing mode's $|m/n|=1/1$ sideband facilitates a helical distortion in the central plasma, similar to what has been proposed in Ref.~\cite{Bando21} as an explanation for helical core observations in JT-60U.

% =============================================================================
\section*{References}%\vspace{-0.3cm}
\addcontentsline{toc}{section}{References}

%\vspace{-0.5cm}
\bibliographystyle{unsrt}
\bibliography{references}

\end{document}